# Lattice QCD Study of Anomalous Transport Phenomena in Strongly Interacting Matter

Eduardo Garnacho Velasco

– PhD thesis –

1$^{\text{st}}$ Supervisor: Dr. Bastian B. Brandt
2$^{\text{nd}}$ Supervisor: Prof. Dr. Gergely Endrődi

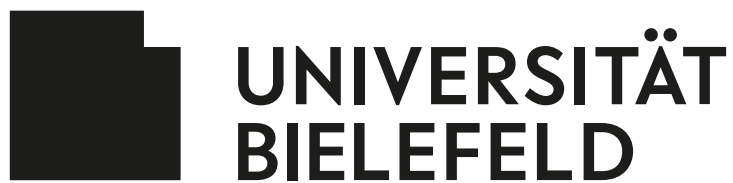

Fakultät für Physik
Universität Bielefeld

# *Abstract*

In this thesis, we study the Chiral Magnetic Effect (CME) and the Chiral Separation Effect (CSE) using lattice QCD simulations. We completely characterize the CSE in QCD using $2+1$ simulations of staggered quarks tuned at the physical point. We find that the CSE conductivity is severely suppressed at low temperatures, and experiences a very sharp increase around the QCD crossover transition, reaching the value of massless non-interacting fermions at high temperatures. This is the first non-perturbative calculation of an anomalous transport conductivity in physical QCD. In the case of the CME, we show that this effect is not present in thermal equilibrium, in accordance with Bloch's theorem, both for free fermions and in QCD. We emphasize the crucial role of using a conserved vector current in the study of anomalous transport phenomena on the lattice, and how the use of non-conserved currents is behind non-vanishing results for the CME in equilibrium that can be found in the literature. Finally, we study the interplay between the CME and inhomogeneous magnetic fields, which leads us to find a novel localized CME signal in equilibrium. This does not contradict Bloch's theorem, since the total signal averages to zero when summed over the full volume. This effect is present in physical QCD, showing that a local equilibrium CME signal is possible in the presence of gluonic interactions.

# Contents

# Chapter 1

# Introduction

## 1.1 Symmetries

For more than 50 years, the theory of high-energy physics has been guided by symmetries. This led to the formulation of the Standard Model of particle physics, arguably the most successful physical theory ever posed. An interesting subset of these driving symmetries is known as *discrete* symmetries. One of the most intuitive among them is **parity**, often denoted as $\mathcal{P}$. In layman's terms, this is just mirror reflection. In the same way our right and left hands reflect differently on the mirror, a concept known as **chirality**, particles can also be chiral: we can have right- or left-handed particles according to their transformation under $\mathcal{P}$. Naively, one could think that parity should be a symmetry of any theory, since the laws of physics should be the same inside and outside a mirror. However, in 1956 T. D. Lee and C. N. Yang, upon examination of different experimental data, pointed out that parity symmetry was untested for one of the interactions explained by the Standard Model: the weak nuclear force [1]. This statement was taken upon by an experimental team led by Chien-Shiung Wu, which that same year performed an analysis of the $\beta$-decay of Cobalt 60, finding a shocking result: the results of the decay had a preferred parity [2]. Therefore, the weak nuclear theory was *not* symmetric under $\mathcal{P}$.

However, theoretical physicists are (in)famous for their persistence. It was already known at that time that some fundamental particles, such as electrons, have partners with similar characteristics but opposite electric charges, called antiparticles. The discrete symmetry that transforms particles into antiparticles and vice versa is called **charge conjugation** ($\mathcal{C}$). The new idea was that parity or charge conjugation were not symmetries of the weak nuclear theory by themselves, but its combination was. This is known as $\mathcal{CP}$**-symmetry**. However, a second surprise was awaiting the particle physics community. In 1964, James Cronin, Val Fitch and collaborators found evidence in the analysis of kaon decay processes that $\mathcal{CP}$-symmetry was also broken by weak interactions [3]. Many other examples of $\mathcal{CP}$-violation have been observed in experiments since then, confirming this groundbreaking result. However, the violation of this symmetry plays an important role in many areas of physics. It is of particular interest in cosmology, where $\mathcal{CP}$-violation is one of the Sakharov conditions [4], that attempt to explain the origin of the imbalance between matter and antimatter in the early universe.

There is a combination of discrete symmetries that is conserved by all physical processes. The last required ingredient is the symmetry under **time reversal** ($\mathcal{T}$), which naively means that processes are invariant under "going backward" in time. The combination of these three symmetries is known as $\mathcal{CPT}$**-symmetry**. All known physical processes respect this combination, and it is a powerful requirement for any theory, since its conservation is related to the relation of spin and statistics or the invariance under Lorentz transformations, see Ref. [5] for a detailed discussion.

## 1.2 Strong interactions

Up until now, we have only referred to the $\mathcal{P}$- and $\mathcal{CP}$-symmetry violation of the weak nuclear interactions. In the case of the other two interactions in the Standard Model, namely the electromagnetic and strong nuclear forces, $\mathcal{P}$ and $\mathcal{CP}$ are indeed a symmetry of all the experimentally analyzed processes involving these theories. It is of special interest for this thesis to analyze the case of strong interactions, which involve the study of quarks and gluons, the fundamental components of atomic nuclei. The quantum field theory that describes these particles and their interactions is called **Quantum Chromodynamics** (QCD). This theory possesses two key characteristics:

- Quarks and gluons are never observed as free states in experiments, but as composite bound states known as hadrons. This phenomenon is known as **confinement**.

- The strength of the interaction between quarks and gluons diminishes with the energy of the system. At infinite energy, QCD becomes a theory of free quarks and gluons, a feature called **asymptotic freedom**.

From a theoretical point of view, no other symmetry forbids the presence of a $\mathcal{CP}$-breaking term in the QCD Lagrangian. This term comes with a parameter usually denoted $\theta$, giving such an addition the name of **theta term**. However, as already mentioned, $\mathcal{CP}$ is experimentally observed to be a symmetry of the strong interactions. In particular, the theta term would give a contribution to the electromagnetic moment of the neutron [6], which has not been detected experimentally [7], giving an upper bound to the theta parameter of $|\theta| < 10^{-10}$. This is usually interpreted as a fine-tuning issue: why is the $\theta$ parameter so small? Is there any physical reason behind this apparent suppression? This is known as the **strong CP-problem**. A possible solution to this issue is the existence of axions, a conjectured particle that can be used to dynamically explain the small value of the theta term. Axions are subject to a large experimental search program, since they are also candidates for dark matter, but have remained elusive so far, see Ref. [8] for a review.

There is another interesting concept that is closely related to $\mathcal{CP}$-symmetry: **quantum anomalies**. In the context of quantum field theories, a symmetry transformation is said to be anomalous if it leaves the classical theory invariant but it is broken by quantum fluctuations. In more formal terms, the action is invariant under the symmetry transformation, but the path integral is not. There is a profound relation between the QCD vacuum and anomalies, in particular through the **axial anomaly**, which makes the axial current in QCD, related to the flow of right- and left-handed quarks, non-conserved even in the massless theory. The conservation law is broken by the **topological charge** $Q_{\text{top}}$, a fundamental object in QCD and precisely the $\mathcal{CP}$-odd operator that appears in the theta term. Therefore the anomaly allows the creation of a chiral imbalance, i.e. a difference between the total number of right- and left-handed quarks, violating parity symmetry. This connection between the axial anomaly and $\mathcal{P}$- and $\mathcal{CP}$-symmetry can have very interesting phenomenological consequences, and one of them is the main topic of study in this thesis: anomalous transport phenomena.

## 1.3   Anomalous transport

Since a non-zero topological charge would lead to $\mathcal{P}$- and $\mathcal{CP}$-symmetry breaking, *global* violations in QCD are currently forbidden by experimental observations, implying that the expectation value of $Q_{\text{top}}$ has to vanish. However, there could be *local* topological fluctuations in the system, leading to local $\mathcal{P}$- and $\mathcal{CP}$-symmetry breaking. When this combines with electromagnetic fields or rotation, novel effects can appear: **anomalous transport phenomena**.

In 1980, Alexander Vilenkin studied a system with a parity-violating term and argued that in the presence of an external magnetic field, an electric current would appear in the system [9]. Unfortunately, this work went unnoticed for almost 30 years, until 2008, when K. Fukushima, D. E. Kharzeev and H. J. Warringa presented in Ref. [10] the **Chiral Magnetic Effect** (CME), the generation of an electric current in presence of a background magnetic field and a chiral imbalance. Although it is closely related to the work of Vilenkin,[1.1] the CME made the novel connection between the axial anomaly and transport effects. Since the introduction of the CME, a whole myriad of related anomalous transport phenomena have appeared. A less famous relative of the CME is the **Chiral Separation Effect** (CSE) [12, 13], which accounts for the generation of an axial current at finite density and magnetic field. Together, they form the **Chiral Magnetic Wave** (CMW) [14], a collective dynamical effect that arises from the interplay of the electric and chiral density waves. As we have mentioned above, these types of phenomena can also appear due to rotation. The prime example is the **Chiral Vortical Effect** [15–17], where angular momentum combined with finite chiral and baryon density create an electromagnetic current. These phenomena and many others constitute a very active area of theoretical research. However, these effects are also highly interesting for the experimental community, since they provide an opportunity to probe the local $\mathcal{CP}$-violations and non-trivial topological nature of QCD. There is an extensive program to detect the CME, CMW and CVE in heavy-ion collision (HIC) experiments and condensed matter systems, see Ref. [18] for a recent review.

Although we have discussed the appearance of these effects due to the axial anomaly in QCD, it is still an open question what is the precise theoretical nature of anomalous transport phenomena in the presence of strong interactions. We now discuss the chosen tool in this thesis to answer this question: lattice gauge theory.

## 1.4  Lattice QCD

One of the main obstacles to studying QCD theoretically is the failure of traditional perturbation theory methods in the low and intermediate energy regimes, where the strong coupling constant is not small. This is why non-perturbative methods have proven to be of extreme use in the study of QCD. Among all the available non-perturbative tools,

---

[1.1] Unfortunately the authors of Ref. [10] were not familiar with the previous work by Vilenkin [9] as explained in Ref. [11]. However, since the explosion in popularity of anomalous transport effects, this work has grabbed much more attention and Ref. [9] has been cited over 500 hundred times from 2012 to 2024, as opposed to the 2 citations it had from 1980 to 2011.

we will concentrate on **lattice QCD**, introduced by Kenneth G. Wilson in his foundational 1974 work [19]. Other techniques include functional methods, like the Functional Renormalization Group [20] and Dyson-Schwinger Equations [21], or holography [22].

The idea behind lattice gauge theories, QCD in particular, is in principle simple. Space-time is discretized, yielding not only a natural ultraviolet regulator to the theory, but also allowing a precise mathematical definition of the path integral. In addition, this integral is then performed in a huge but finite number of dimensions,[1.2] allowing for a numerical solution without relying on perturbation theory. However, as will be clear in Chap. 2, this formulation is full of unexpected issues, profound physical concepts and numerical challenges.

Nevertheless, the success of lattice QCD is undeniable. For example, in the theory at zero temperature, it enables first-principles calculations of hadron masses, hadronic scattering amplitudes, and flavor physics-related quantities like the elements of the Cabibbo–Kobayashi–Maskawa (CKM) matrix, see e.g. Ref. [23] for an extensive overview on these and many other results. It has also become a very important tool to solve the tension between the Standard Model prediction of the anomalous magnetic moment of the muon and its experimental value [24]. In addition, it can be used to study the thermodynamics of QCD. A particularly interesting subject is the phase diagram of strongly interacting matter. Using lattice QCD simulations, it was determined that QCD undergoes a crossover phase transition at high temperatures [25, 26], reaching a new phase of matter where quarks and gluons get deconfined. It has also been used to study the equation of state of QCD [27, 28], which is extensively used by the phenomenological high-energy physics community. These and many other applications have made lattice QCD a very powerful tool to study strong interactions.

## 1.5 Research objectives

In this thesis, we will study two of the aforementioned anomalous transport phenomena, the CME and the CSE, in QCD. To this end, we will make use of lattice QCD simulations, whose theoretical formulation we present in Ch. 2. We continue discussing the most relevant aspects and current theoretical understanding of the CME and the CSE in Ch. 3. After these preliminary chapters, we proceed to present the results obtained in this thesis.

---

[1.2] This is only true when considering space-time to have a finite volume. We will come back to this point in the next chapter.

These are divided into three chapters, which we can associate with the three questions that we will answer:

Q1) What is the conductivity of the CSE in QCD as a function of temperature?

Q2) What is the nature of the CME in global thermal equilibrium?

Q3) How is the CME affected by inhomogeneous magnetic fields?

These questions will be addressed in Chs. 4, 5, 6 respectively. We close with the summary of the results and outlook in Ch. 7.

# Chapter 2

# Lattice QCD

In this chapter, we review the basic aspects of lattice gauge theory, most of which can be found as textbook material in Refs. [29–31]. We will present an introduction to the theoretical concepts of the discretized theory. Since numerical methods also play an important role in lattice QCD calculations, we discuss some of the basic concepts in App. A, focusing particularly on the aspects that are more relevant to the topics of this thesis.

## 2.1  Motivation

In the study of quantum field theories, divergences are an unfortunate ubiquitous companion. In perturbation theory, Feynman diagrams containing loops can involve infinite integrals, both in the high-frequency or ultraviolet (UV) and low-frequency or infrared (IR) regimes. In the case of UV divergences, the usual first step in dealing with these types of integrals is to introduce a **regulator**, which yields a formally finite integral that diverges upon removal of the regulator. The most common examples are a hard momentum cutoff, a Pauli-Villars regulator or dimensional regularization.

Another possibility is to discretize space-time, with a minimum distance between points, considering for example a hypercubic grid.[2.1] This implies that there is a minimum wavelength (or maximum frequency) that can be resolved by the system. In quantum field theory, this imposes a maximum momentum in the integrals, therefore regularizing the UV behavior. This is known as **lattice regularization**. At first glance, this looks like a complicated way of imposing a momentum cutoff. However, as we will discuss later, this formulation has a crucial advantage over a hard momentum cutoff: it preserves gauge invariance. It is important to notice that the regulator is proportional to the (inverse) distance between the lattice points. When this regulator is removed, by sending the lattice distance to zero, the integrals will be divergent again. Hence, the considered theory would in general also need **renormalization**, in the same way as perturbative calculations. The formulation of gauge theories in this discretized formulation is known as **lattice gauge theories**. In principle, we could describe any gauge theory on the lattice. Since we are mostly interested in strong interactions, we will restrict most of our discussion to the case of QCD. Nevertheless, most of the theoretical concepts we will present can be generalized straightforwardly to other theories.

Although we have motivated the use of lattice gauge theory from a regularization point of view, the foundation of this formalism has much more profound roots than just regularizing the theory. For example, the discretized version of the theory can be used to *define* the path integral formulation. It also allows, among other things, to perform non-perturbative calculations in QCD, regardless of how large the coupling is. Furthermore, lattice gauge theory is intimately related to statistical physics models, making this formalism also useful to study phase transitions.

As a first step, we will briefly review the main properties of QCD in the continuum formulation, before discussing the discretized theory.

## 2.2 QCD in the continuum

As we already introduced in Ch. 1, QCD is a quantum field theory that describes the strong nuclear force, one of the three fundamental interactions that are part of the Standard Model of particle physics. In terms of its fundamental particles, QCD models the interactions of quarks and gluons, which are the building blocks of protons and neutrons, as well as all the other hadrons. There are 6 types or "flavors" of quarks, called up ($u$),

---

[2.1] This is the simplest and more widely used geometry to study QCD, hence we will only consider this case in this thesis.

down ($d$), strange ($s$), charm ($c$), bottom ($b$) and top ($t$), which differ in mass and in their electric charges. In addition, each quark can have three different "colors", while gluons come in eight different versions with respect to color, which is a quantum number associated with the strong interactions. For a more extensive introduction to QCD, see for example Ref. [32]. In this section we follow the presentation of Ref. [31].

In more formal terms, QCD is a gauge theory that results from coupling the Dirac theory for free fermions with SU($N_c$) Yang-Mills theory, where $N_c = 3$ is the number of colors. The group SU($N$) denotes the special unitary group of degree $N$, i.e. the Lie group of $N \times N$ unitary matrices with unit determinant. This group has $N^2 - 1$ generators, which we will denote by $\{T^a\}_{a=1}^{N^2-1}$. These generators span the Lie algebra su($N$). Any element $\Omega \in$ SU($N$) can be written in terms of the generators,

$$\Omega = \exp\left(i \sum_{a=1}^{N^2-1} T^a \alpha_a\right) \quad \text{with } \alpha_a \in \mathbb{R}\,. \tag{2.1}$$

We start by considering the Minkowski metric $g^{\mu\nu} = \text{diag}(+1, -1, -1, -1)$, where the Minkowski Dirac matrices $\gamma_{\text{M}}^\mu$ fulfill the relation $\{\gamma_{\text{M}}^\mu, \gamma_{\text{M}}^\nu\} = 2g^{\mu\nu}$. The generating functional of QCD is given by[2.2]

$$\mathcal{Z} = \int \mathcal{D}A \mathcal{D}\bar{\psi} \mathcal{D}\psi \, e^{iS_F^{\text{M}} + iS_G^{\text{M}}}\,. \tag{2.2}$$

The fermionic action $S_F$ can be written as

$$\begin{aligned}
S_F^{\text{M}} &= \sum_f \int \text{d}^4x \, \bar{\psi}_f(x) \left[\sum_{\mu=0}^{3} i\gamma_{\text{M}}^\mu D_\mu - m_f\right] \psi_f(x) \\
&= \sum_f \int \text{d}^4x \, \bar{\psi}_f(x) \left[\sum_{\mu=0}^{3} i\gamma_{\text{M}}^\mu [\partial_\mu - iA_\mu(x)] - m_f\right] \psi_f(x)\,,
\end{aligned} \tag{2.3}$$

where the sum in $f$ runs over the different flavors and $m_f$ is the *bare* (unrenormalized) mass of each quark flavor. For each of them, $\psi_f, \bar{\psi}_f$ have Dirac and color indices, explicitly

$$(\psi)_\alpha^c\,, \ (\bar{\psi})_\alpha^c\,, \qquad c = 1, 2, 3\,; \quad \alpha = 1, 2, 3, 4\,.$$

---

[2.2] During the whole text, we consider natural units $\hbar = c = k_B = 1$.

Each of these components is an anti-commuting Grassmann variable. These objects transform in the fundamental representation of SU(3)

$$\psi(x) \to \psi'(x) = \Omega(x)\psi(x), \qquad \bar{\psi}(x) \to \bar{\psi}'(x) = \bar{\psi}(x)\Omega^\dagger(x), \tag{2.4}$$

with $\Omega(x) \in \mathrm{SU}(3)$. The gauge fields $A_\mu$ are elements of the su(3) Lie algebra, and they can be written in terms of the generators $T_a$ as $A_\mu = \sum_{a=1}^{8} A_\mu^a T_a$, with each $A_\mu^a$ a real-valued field. The object $\slashed{D} \equiv \sum_\mu \gamma^\mu D_\mu$ is the covariant derivative, where the addition of the gauge field to the usual derivative term ensures the gauge invariance of the fermionic action. We will refer to $M = \slashed{D} - m$ as the (massive) **Dirac operator**.

The gauge action $S_G$ is

$$S_G^\mathrm{M} = -\frac{1}{2g^2} \sum_{\mu,\nu=0}^{3} \int \mathrm{d}^4 x \; \mathrm{tr}_C \left[ G^{\mu\nu} G_{\mu\nu} \right], \tag{2.5}$$

with $G_{\mu\nu} = \sum_a G_{\mu\nu}^a T_a = \partial_\mu A_\nu - \partial_\nu A_\mu + i[A_\mu, A_\nu]$ and $g$ the bare coupling.[2.3] Here $\mathrm{tr}_C$ denotes the trace over color space.

In usual perturbative calculations, it is also required to consider two more terms in the QCD action, a gauge fixing part and a Fadeev-Popov ghost part. We will discuss later how these do not play a role in the lattice formulation, so we will also not introduce them in this section.

After discussing the Minkowski formulation, we can continue by introducing the Euclidean formulation, which we will use in most parts of this thesis. We consider a Wick rotation $x_0 \to x_4 = ix_0$, where now $x_4$ is an *imaginary* time. The generating functional then changes to

$$\mathcal{Z} = \int \mathcal{D}A \mathcal{D}\bar{\psi} \mathcal{D}\psi \, e^{-S_F^\mathrm{E} - S_G^\mathrm{E}}, \tag{2.6}$$

with

$$S_F^\mathrm{E} = \sum_f \int \mathrm{d}^4 x \, \bar{\psi}_f(x) \left( \sum_{\mu=1}^{4} \gamma_\mu^\mathrm{E} [\partial_\mu + iA_\mu(x)] + m_f \right) \psi_f(x), \tag{2.7}$$

$$S_G^\mathrm{E} = \frac{1}{2g^2} \sum_{\mu,\nu=1}^{4} \int \mathrm{d}^4 x \, \mathrm{tr}_C \left[ G_{\mu\nu} G_{\mu\nu} \right]. \tag{2.8}$$

---

[2.3] Notice that we consider the convention, widely used in lattice gauge theory, of including the coupling as an overall factor in the gauge action, and not in the Dirac operator. This just accounts for a rescaling of the fields.

After the Wick rotation, we have the Euclidean metric[2.4] $\eta_{\mu\nu} = \mathrm{diag}(+1,+1,+1,+1)$ and $\{\gamma_\mu^{\mathrm{E}}, \gamma_\nu^{\mathrm{E}}\} = 2\eta_{\mu\nu}$. The gamma matrices are now hermitian, and they can be written in terms of their Minkowski counterparts as

$$\begin{aligned}
\gamma_{\mathrm{E}}^4 &= \gamma_{\mathrm{M}}^0\,, \qquad \gamma_{\mathrm{E}}^j = -i\gamma_{\mathrm{M}}^j \quad \text{for } j=1,2,3\,, \\
\gamma_5^{\mathrm{M}} &= i\gamma_0^{\mathrm{M}}\gamma_1^{\mathrm{M}}\gamma_2^{\mathrm{M}}\gamma_3^{\mathrm{M}} = \gamma_1^{\mathrm{E}}\gamma_2^{\mathrm{E}}\gamma_3^{\mathrm{E}}\gamma_4^{\mathrm{E}} = \gamma_5^{\mathrm{E}}\,.
\end{aligned} \tag{2.9}$$

For the rest of the text, we will consider the Euclidean formulation unless stated otherwise, and we drop the M/E subscripts.

The Euclidean formulation has a clear drawback. Since we are now considering an imaginary time, this formalism cannot describe time evolution, only stationary systems. However, it also possesses two main advantages that are crucial for this thesis. Firstly, it enables numerical simulations of quantum field theories, QCD in particular. This is because the exponential factors in the Minkowski path integral (2.2) are complex, oscillating functionals, while after Wick rotating they become real, see Eq. (2.6). This makes the Euclidean generating functional better suited for numerical integration, we will come back to this point in more detail in Sec. 2.7. Secondly, this formalism can describe systems at a finite, global **temperature** $T$. After the Wick rotation, the generating functional can be compared to the partition function of a statistical physics model, and the temperature can be identified with the inverse of the temporal extent, see e.g. Ref. [33] for an introduction to thermal quantum field theory. At finite temperature $T$, the explicit form of space-time integrals reads

$$\int \mathrm{d}^4 x = \int \mathrm{d}^3 x \int_0^{1/T} \mathrm{d}x_4 \,. \tag{2.10}$$

Having reviewed the basic features of QCD in the continuum, we can continue presenting its lattice gauge theory formulation. However, before studying lattice QCD in all its complexity, we will consider the discretization of the Dirac theory of non-interacting fermions. Despite its apparent simplicity, it will give us an insight into some of the challenges we need to face in lattice QCD.

---

[2.4] Since in this metric the distinction between lower and upper indices does not matter, we will usually write both contracted indices as lower indices.

## 2.3 Naive discretization of fermions

Let us consider the Dirac theory for a spin-1/2 non-interacting fermion with mass $m$ in the continuum

$$S_F = \int \mathrm{d}^4 x \, \bar{\psi}(x) \left[ \sum_\mu \gamma_\mu \partial_\mu + m \right] \psi(x) \,. \tag{2.11}$$

To study the lattice version of this theory, we discretize space-time by introducing a four-dimensional hypercubic lattice $\mathcal{N}$,

$$\mathcal{N} = \{ n = (n_1, n_2, n_3, n_4); \quad n_1, n_2, n_3 \in \{0, \ldots, N_s - 1\}, n_4 \in \{0, \ldots, N_t - 1\} \} \,, \tag{2.12}$$

where $N_s$ is the number of points in the spatial directions and $N_t$ is the number of points in the temporal direction. These points are separated by a **lattice spacing** $a$, so that $x = an$,[2.5] and the lattice has a **volume** $V = a^3 N_s^3 \equiv L^3$. Equivalently to the continuum formulation, the temperature is given by the inverse extent of the temporal direction $T = 1/(aN_t)$. The spinors retain the same structure as in the continuum, such as the Dirac structure, or color and flavor indices in QCD. The only difference is that they are now placed on the discrete lattice points $\psi(n), \bar{\psi}(n)$, with $n \in \mathcal{N}$.

This discretization has two direct effects on the action (2.11): the integral over space and time becomes a sum over $\mathcal{N}$ and the derivative is substituted by its symmetric discrete version,[2.6]

$$\partial_\mu \psi(x) \to \frac{1}{2a} [\psi(n + \hat{\mu}) - \psi(n - \hat{\mu})] \,, \tag{2.13}$$

where $\hat{\mu}$ is a unit vector in the direction $\mu$. Therefore, the lattice version of the free Dirac action reads

$$S_F^{\text{naive}} = a^4 \sum_{n \in \mathcal{N}} \bar{\psi}(n) \left[ \sum_\mu \gamma_\mu \frac{\psi(n + \hat{\mu}) - \psi(n - \hat{\mu})}{2a} + m\psi(n) \right] \,. \tag{2.14}$$

This is the most straightforward way of discretizing a fermionic theory, which is usually referred to as **naive fermions**. Since the action is a bilinear in the fermion fields, we can write it as

$$S_F^{\text{naive}} = a^4 \sum_{l,n} \bar{\psi}(l) \, M_{\text{naive}}(l, n) \, \psi(n) \,, \tag{2.15}$$

---

[2.5] The lattice spacing has dimensions of inverse energy, and it is generally used to make quantities dimensionless. We will refer to this as lattice units.

[2.6] Other choices for the discretization of the derivative are also possible, we comment on this point at the end of the section.

where $M_{\text{naive}}(l,n)$ is the (massive) Dirac operator for naive fermions

$$M_{\text{naive}}(l,n) = \sum_\mu \gamma_\mu \frac{1}{2a}(\delta_{l+\hat{\mu},n} + \delta_{l-\hat{\mu},n}) + m\,\delta_{l,n} \equiv \slashed{D}_{\text{naive}}(l,n) + m\,\delta_{l,n}\,. \tag{2.16}$$

To recover the usual Dirac theory, one would need to take the lattice spacing $a$ to zero. This we will refer to as the **naive continuum limit**. It seems trivial to see that in the limit $a \to 0$, the lattice action goes back to Eq. (2.11). However, a closer inspection reveals that the theory is altered in the process. To see this, it is beneficial to work in Fourier space, where it can be readily shown that

$$\widetilde{M}_{\text{naive}}(p,q) = \frac{1}{N_s^3 N_t} \sum_{l,n} e^{-ipla} M_{\text{naive}}(l,n) e^{iqna} = \delta(p-q)\widetilde{M}_{\text{naive}}(p)\,, \tag{2.17}$$

with

$$\widetilde{M}_{\text{naive}}(p) = \frac{i}{a}\sum_\mu \gamma_\mu \sin(ap_\mu) + m = \widetilde{\slashed{D}}_{\text{naive}}(p) + m\,. \tag{2.18}$$

We can compare this expression to the continuum Dirac operator in momentum space

$$\widetilde{M}(p) = \widetilde{\slashed{D}}(p) + m = i\sum_\mu \gamma_\mu p_\mu + m\,. \tag{2.19}$$

The linear dispersion relation in the continuum has been replaced by a sine on the lattice. Hence, the lattice operator is periodic in momentum space. The fundamental domain, which is periodically repeated, is the **Brillouin zone**

$$\text{BZ} = \left\{k\,;\quad -\frac{\pi}{a} < k_\mu \leq \frac{\pi}{a} \quad \forall \mu\right\}\,. \tag{2.20}$$

Notice that when $N_s, N_t$ are finite, as will be the case in numerical simulations, the lattice has a finite volume. This implies the momentum can only take discrete values

$$ap_\mu = \frac{2\pi n_p}{N_\mu} \quad \text{with} \quad n_p \in \mathbb{Z}\,, \tag{2.21}$$

where $N_\mu$ corresponds to $N_s$ for the spatial directions and $N_t$ for the temporal one. When the volume is taken to infinity and the temperature to 0, with $N_s, N_t \to \infty$ but $a$ fixed, the periodicity will remain but the sums in momentum space can be substituted by integrals, and the discretization of the momentum disappears.

The free fermionic propagator is the inverse of the Dirac operator, which can be readily shown to be

$$\widetilde{M}^{-1}_{\text{naive}}(p) = \frac{m - ia^{-1}\sum_\mu \gamma_\mu \sin(ap_\mu)}{m^2 + a^{-2}\sum_\mu \sin^2(ap_\mu)} \ . \tag{2.22}$$

We will now present a heuristic argument to show that the theory gets modified in the trajectory toward the continuum limit. We set now $m = 0$ for simplicity, although the rest of the argument also holds at finite mass. We can now take the limit $a \to 0$ of Eq. (2.22), which again seems to yield the correct propagator in the continuum. In momentum space, the continuum fermionic propagator has a pole at $p = (0, 0, 0, 0)$, corresponding to the single fermion species that it describes. However, the lattice propagator (2.22) has 16 poles ($2^d$ in $d$-dimensions), since $\sin^2(ap_\mu)$ also vanishes at $p_\mu = \pi/a$.[2.7] These 15 extra modes, which were not present in the original theory, are referred to as **doublers**. In an interacting theory, these doublers change the physics of the original formulation non-trivially, and therefore we need a strategy to deal with them. This is the reason behind the wide variety of fermion discretizations in lattice QCD, which correspond to different ways of solving the doubling problem while retaining the symmetries from the continuum, as we will discuss in the next sections.

The origin of the doublers can also be understood by looking at the symmetries of the discretized action and the structure of the Brillouin zone [34]. Let us define the set of ordered indexes for a $d$-dimensional space

$$G = \{\, g; \ g = [\mu_1, \ldots, \mu_h], 1 \leq \mu_1 < \cdots < \mu_h \leq d \,\}, \tag{2.23}$$

where the elements $g$ are tuples of length $h$. The case $h = 0$ corresponds to the empty set $\varnothing$. We define next the set of vectors

$$(\pi_g)_\mu = \begin{cases} \dfrac{\pi}{a} & \text{for } \mu \in g \\ 0 & \text{for } \mu \notin g \end{cases}, \tag{2.24}$$

---

[2.7] This argument is incomplete because a precise calculation is needed to show that the sixteen poles do contribute in the continuum limit and yield a theory with 16 flavors. However, we do not repeat here the necessary steps since they can be found in several textbooks, see for example Ref. [30].

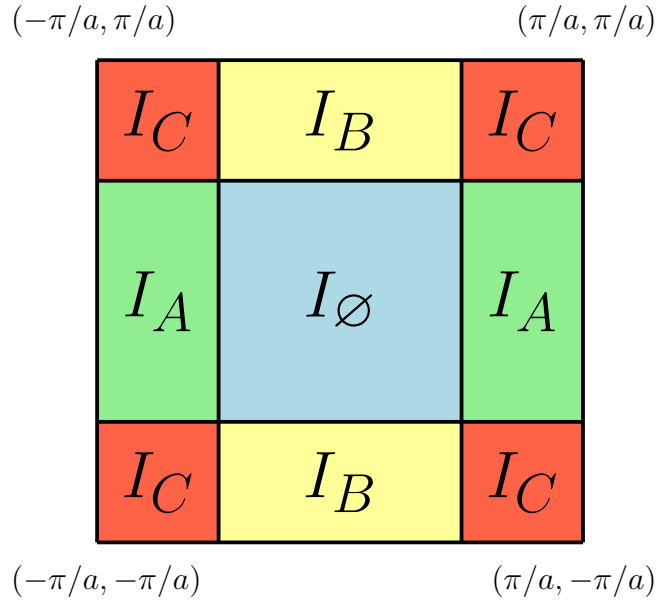

FIGURE 2.1: Division of the Brillouin zone in two dimensions according to the areas defined in Eqs. (2.25), (2.26).

and its dimensionless version $\hat{\pi}_g = a\pi_g$. We can now consider the following regions in momentum space

$$I_g = \left\{ k\,; \quad k = (k_\varnothing + \pi_g) \bmod \frac{2\pi}{a}, \quad k_\varnothing \in I_\varnothing \right\}, \tag{2.25}$$

$$I_\varnothing = \left\{ k\,; \quad -\frac{\pi}{2a} < k_\mu \leq \frac{\pi}{2a}, \, \forall \mu \right\}, \tag{2.26}$$

such as the whole Brillouin zone can be decomposed as

$$\mathrm{BZ} = \bigcup_{g \in G} I_g\,. \tag{2.27}$$

Since this construction is rather difficult to visualize for a four-dimensional lattice, let us briefly consider the simpler case $d = 2$. There are four possible elements in $G$, namely

$$\varnothing\,, \quad A \equiv [1]\,, \quad B \equiv [2]\,, \quad C \equiv [1, 2]\,, \tag{2.28}$$

with associated vectors

$$\hat{\pi}_\varnothing = (0, 0), \quad \hat{\pi}_A \equiv (\pi, 0), \quad \hat{\pi}_B \equiv (0, \pi), \quad \hat{\pi}_C \equiv (\pi, \pi)\,. \tag{2.29}$$

With this, we can build a partition of BZ explicitly, which we show in Fig. 2.1. In the 2-dimensional theory there are 4 doublers, which is suggestively equal to the number of partitions of BZ in this construction. Indeed, each of these areas of the Brillouin zone can be shown to combine to yield a well-defined continuum limit for each of the doublers [30].

Now we are ready to define a symmetry of the naive fermion action that has no analog in the continuum formulation, the **spectrum doubling symmetry**. Considering an element $g \in G$, we define the transformations

$$\begin{aligned}\psi(n) \to \psi'(n) &= e^{in\hat{\pi}_g} P_g \, \psi(n) \,, \\ \bar{\psi}(n) \to \bar{\psi}'(n) &= \bar{\psi}(n) \, P_g^\dagger e^{-in\hat{\pi}_g} \,,\end{aligned} \quad (2.30)$$

with

$$P_g = P_{\mu_1} \dots P_{\mu_h}, \quad g \in G \,, \quad (2.31)$$

and

$$P_\mu = i\gamma_5 \gamma_\mu \quad P_\mu^\dagger = P_\mu^{-1} \,. \quad (2.32)$$

This transformation is a symmetry of the naive fermion action (2.14). In momentum space, it takes the form

$$\begin{aligned}\tilde{\psi}'(k) &= P_g \, \tilde{\psi}(k + \pi_g) \,, \\ \tilde{\bar{\psi}}'(k) &= \tilde{\bar{\psi}}(k + \pi_g) \, P_g^\dagger \,.\end{aligned} \quad (2.33)$$

Now we can give a clearer interpretation of this transformation: it exchanges the corners of the Brillouin zone, i.e. the action is symmetric under the exchange of the doubling modes. This can be seen explicitly, by defining the following doubler fields,

$$\begin{aligned}\tilde{q}^g(k) &= P_g \, \tilde{\psi}(k + \pi_g) \,, \\ \tilde{\bar{q}}^g(k) &= \tilde{\bar{\psi}}(k + \pi_g) P_g^\dagger \,.\end{aligned} \quad (2.34)$$

Now the naive fermion action (2.14) can be factorized into the different doubler contributions as

$$S_F^{\text{naive}} = \sum_{g \in G} \sum_{k \in I_g} \tilde{\bar{q}}^g(k) \, \widetilde{M}_{\text{naive}}(k) \, \tilde{q}^g(k) \,. \quad (2.35)$$

The spectrum doubling symmetry is behind the proliferation of the extra modes in the action, and many of the strategies to deal with the doubling problem are related to modifications of the naive fermion action that break this symmetry.

Let us close this section with two remarks about the doubling problem. Firstly, the appearance of the doublers is an artifact of the use of a *symmetric* discretization of the derivative (2.13). If a non-symmetric version is used, for example a backward one like

$$\partial_\mu \psi(x) \to \frac{1}{a}\left[\psi(n) - \psi(n-\hat{\mu})\right], \tag{2.36}$$

then the theory is free of doublers, as one can check by analyzing the structure of the lattice propagator in this case. Unfortunately, this choice creates more serious problems than doubling modes, for example it yields a non-renormalizable theory [35].

The second comment has to do with the doublers being an issue only in *fermionic* theories. In scalar and vector theories, for example a scalar $\phi^4$ theory or Yang-Mills theory, the doubling problem is not present. This again is related to the type of derivative in the theory, which is a d'Alembert operator (quadratic derivative) instead of the linear derivative in the fermionic case. This changes the structure of the Brillouin zone, such as there is only one pole for the lattice propagator, and hence no doubling problem appears.

## 2.4 QCD on the lattice

### 2.4.1 Gauge theories

Our final goal is to study QCD on the lattice, so we have to discuss the formulation of gauge theories on the lattice. In this section we follow closely Ref. [31]. Let us consider the Lie group SU(3), then Dirac fermion fields transform according to Eq. (2.4). As in the continuum, it is trivial to see that the mass term in Eq. (2.14) is invariant under this transformation. However, the discrete derivative is not. Taking for example the forward term,

$$\bar{\psi}(n)\psi(n+\hat{\mu}) \to \bar{\psi}'(n)\psi'(n+\hat{\mu}) = \bar{\psi}(n)\Omega^\dagger(n)\,\Omega(n+\hat{\mu})\psi(n+\hat{\mu})\,, \tag{2.37}$$

we see that it is not gauge invariant, again analogously to the continuum formulation.

We can introduce a vector field $U_\mu(n)$ connecting the two fields

$$\bar{\psi}'(n)U'_\mu(n)\psi'(n+\hat{\mu}) = \bar{\psi}(n)\Omega^\dagger(n)\,U'_\mu(n)\,\Omega(n+\hat{\mu})\psi(n+\hat{\mu})\,. \tag{2.38}$$

If this object transforms according to

$$U'_\mu(n) = \Omega_\mu(n)U_\mu(n)\Omega^\dagger_\mu(n+\hat{\mu})\,, \tag{2.39}$$

then the combination $\bar{\psi}(n)U_\mu(n)\psi(n+\hat{\mu})$ is gauge invariant. This is exactly how elements of the gauge group transform on its adjoint representation, hence we consider the gauge fields $U_\mu(n) \in \mathrm{SU}(3)$ which we will refer to as **link variables**, since these objects are oriented in the direction $\mu$ and they live on the link connecting the sites $n$ and $n + \hat{\mu}$. Similarly, we can have a link variable going in the *negative* $\mu$ direction, connecting $n$ and $n - \hat{\mu}$. We can relate them to the positively oriented link variables by introducing

$$U_{-\mu}(n) \equiv U_\mu^\dagger(n - \hat{\mu}) . \tag{2.40}$$

The links can be expressed in terms of Lie algebra elements like

$$U_\mu(n) = e^{iaA_\mu(n)}, \tag{2.41}$$

with $A_\mu \in \mathrm{su}(3)$. Putting everything together, we obtain the gauge invariant version of the naive fermion action,

$$S_F^{\text{naive}} = a^4 \sum_n \bar{\psi}(n) \left[ \sum_\mu \gamma_\mu \frac{U_\mu(n)\psi(n+\hat{\mu}) - U_\mu^\dagger(n-\hat{\mu})\psi(n-\hat{\mu})}{2a} + m\psi(n) \right]. \tag{2.42}$$

### 2.4.2 Gauge action

To study QCD, we also need to consider how the gauge action looks like in the lattice formulation, which in the continuum is given by Eq. (2.8). For this, we need a discretized lattice action that is gauge invariant and that yields the correct naive continuum limit. It turns out that this action can be built using the shortest closed loop on the lattice [19], which is commonly known as the **plaquette**,

$$P_{\mu\nu} = U_\mu(n)U_\nu(n+\hat{\mu})U_\mu^\dagger(n+\hat{\nu})U_\nu^\dagger(n). \tag{2.43}$$

The trace over the plaquette is a gauge invariant object, which can be used to define the **Wilson gauge action**,

$$S_G^{\text{Wilson}} = \beta \sum_n \sum_{\mu<\nu} \mathrm{Re}\left[ 1 - \frac{1}{3}\mathrm{tr}_C P_{\mu\nu}(n) \right], \tag{2.44}$$

where we have defined $\beta = 6/g^2$. It can be shown, by expanding the link variables for small $a$, see e.g. Ref. [31], that this action reproduces the correct gauge action in the

$a \to 0$ limit,

$$S_G^{\text{Wilson}} = \beta \sum_n \sum_{\mu<\nu} \text{Re}\left[1 - \frac{1}{3}\text{tr}_C P_{\mu\nu}(n)\right]$$
$$= \frac{a^4}{2g^2} \sum_n \sum_{\mu,\nu} \text{tr}_C[G_{\mu\nu}(n)G_{\mu\nu}(n)] + \mathcal{O}(a^2). \quad (2.45)$$

The expected lattice artifacts for this action are then $\mathcal{O}(a^2)$. However, this is not the only action to yield $S_G$ in the continuum limit, since it can be modified by terms that vanish when $a \to 0$. The lattice action can be chosen differently, such that the quadratic discretization effect terms disappear, and only the terms $\mathcal{O}(a^4)$ survive. This procedure can be iterated to get a better approach to the continuum limit, giving these other formulations of the gauge action the name of **improved actions**.

The first improvement, where the order $\mathcal{O}(a^2)$ is removed, is known as **tree-level improved Symanzik gauge action** [36], and it is given by the second simplest gauge invariant object after a plaquette, a $2 \times 1$ rectangle (see Fig. 2.2). The exact form of this action, for general coefficients $c_0, c_1 \in \mathbb{R}$, is

$$S_G^{\text{Symanzik}} = \beta \sum_n \sum_{\mu<\nu} \text{Re}\left(c_0\left[1 - \frac{1}{3}\text{tr}_C P_{\mu\nu}(n)\right] + c_1\left[1 - \frac{1}{3}\text{tr}_C R_{\mu\nu}(n)\right]\right), \quad (2.46)$$

where $R_{\mu\nu}$ is the path ordered product of the $2 \times 1$ rectangles on the lattice,

$$R_{\mu\nu} = U_\mu(n)U_\mu(n+\hat{\mu})U_\nu(n+2\hat{\mu})U_\mu^\dagger(n+2\hat{\mu}+\hat{\nu})U_\mu^\dagger(n+\hat{\mu}+\hat{\nu})U_\nu^\dagger(n). \quad (2.47)$$

By choosing the coefficients $c_1 = -1/12$ and $c_0 = 1 - 8c_1$, this expression reproduces the continuum gluonic action with discretization effects of $\mathcal{O}(a^4)$.

### 2.4.3 Path integral

Now we are ready to discuss how to formulate the path integral for QCD on the lattice. In the continuum, the Euclidean partition function is given by Eq. (2.6). Let us now consider a generic discretization choice for the fermion and gauge actions $S_F$ and $S_G$. We focus first on the fermionic part, in particular the fermionic measure, which we can define on the lattice as

$$\mathcal{D}\psi = \prod_n d\psi(n), \quad (2.48)$$

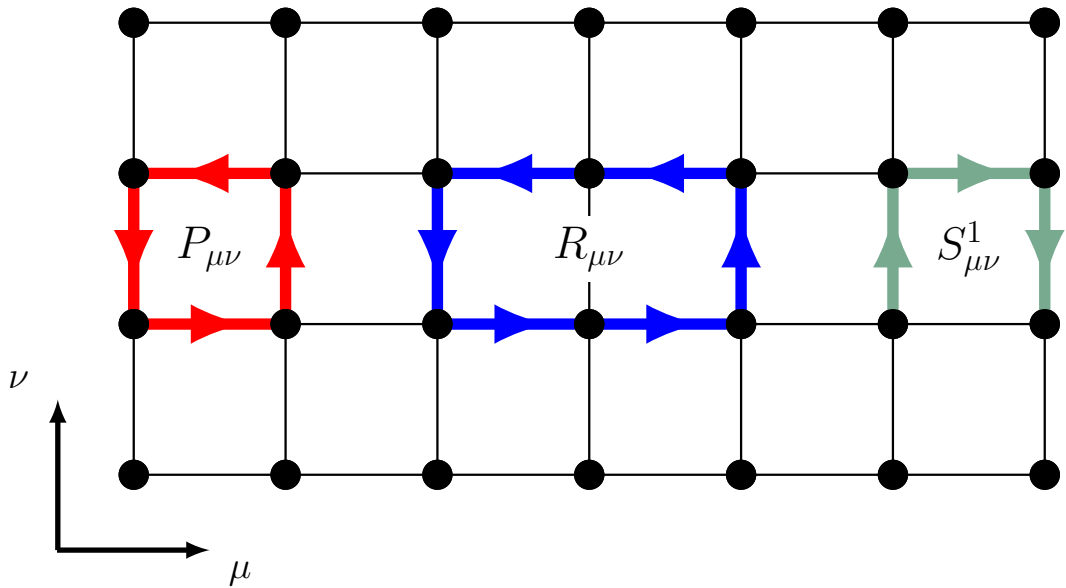

FIGURE 2.2: Graphical representation of a plaquette $P_{\mu\nu}$, a $2 \times 1$ rectangle $R_{\mu\nu}$ and a staple $S^1_{\mu\nu}$ (see Sec. 2.6.2) on the lattice.

for each individual spinor $\psi(n)$. Since the fermionic action is a bilinear in the fields, we can perform the fermionic path integral analytically. Recalling that $\psi(n), \bar{\psi}(n)$ are Grassmann variables, the result of the integration yields the **fermion determinant**

$$\int \mathcal{D}\psi \, \mathcal{D}\bar{\psi} \, e^{-S_F} = \det M \, . \qquad (2.49)$$

We want to emphasize that this determinant is still a function of the links $U$, so it does not factor out from the total path integral.

In the case of the gluonic variables, the definition of the measure requires a bit more work. Following Ref. [31], we consider first the integration measure for a single group element $U \in \mathrm{SU}(3)$, which can be parameterized by 8 real numbers $\alpha^a$ with $a = 1, \ldots, 8$. We can derive an integral measure $\mathrm{d}U$ by requiring that it fulfills two conditions:

1) Invariance under left- and right-side multiplication by another group element $V \in \mathrm{SU}(3)$: $\mathrm{d}U = \mathrm{d}(UV) = \mathrm{d}(VU)$ .

2) Normalization: $\int \mathrm{d}U = 1$ .

The measure defined in this way is known as **Haar measure**. It takes the explicit form

$$\mathrm{d}U = c \, \sqrt{\det[g(\alpha)]} \, \prod_{a=1}^{8} \mathrm{d}\alpha^a \, , \qquad (2.50)$$

up to a normalization factor $c$ to fulfill condition 2), with

$$g(\alpha)_{nm} = \text{tr}\left[\frac{\partial U}{\partial \alpha^n}\frac{\partial U^\dagger}{\partial \alpha^m}\right]. \tag{2.51}$$

Then the total gluonic measure can be defined as

$$\mathcal{D}U = \prod_{n,\mu} \mathrm{d}U_\mu(n)\,. \tag{2.52}$$

Finally, we can write the **partition function of lattice QCD**

$$\mathcal{Z} = \int \mathcal{D}U\,\mathcal{D}\psi\,\mathcal{D}\bar{\psi}\, e^{-S_G - S_F} = \int \mathcal{D}U\, e^{-S_G}\det M\,. \tag{2.53}$$

There is an important detail that we would like to discuss now. All the SU($N$) groups, SU(3) in particular, are *compact* groups. This is reflected in condition 2), since the integral over a constant element yields a finite result. This is opposed to the usual continuum formulation in terms of the *algebra* elements $A$. The su($N$) Lie algebras are not compact and hence this integration over the unit element would diverge, a behavior that can be ultimately traced back to the gauge redundancy. This is solved in perturbation theory[2.8] by fixing the gauge, which for non-abelian theories also results in the appearance of the Fadeev-Popov ghosts. On the lattice, this procedure is not required, and we do not need to think about gauge fixing or ghosts. It is nevertheless possible to fix the gauge in the lattice formulation, which can be beneficial in certain calculations, but we do not consider these procedures in this thesis.

Now that we have a path integral formalism on the lattice, we can calculate observables analogously as in the continuum. If we consider a purely gluonic observable $O_G$, its expectation value can be written as

$$\langle O_G \rangle = \frac{T}{V}\int \mathcal{D}U\, e^{-S_G}\det M\, O_G\,. \tag{2.54}$$

For a fermionic observable $O_F$, we need to use **Wick's theorem**, similarly as in usual perturbation theory calculations. For example, for a one-point function,

$$\langle \bar{\psi}\psi \rangle \equiv \frac{T}{V}\frac{1}{\mathcal{Z}}\int \mathcal{D}U\mathcal{D}\psi\,\mathcal{D}\bar{\psi}\, e^{-S_G - S_F}\int \mathrm{d}^4 x\, \bar{\psi}(x)\psi(x) = \frac{T}{V}\left\langle \text{Tr}\, M^{-1}\right\rangle, \tag{2.55}$$

---

[2.8] Up to the Gribov ambiguity problem [37].

and for a two-point function,

$$\langle\bar{\psi}\psi\bar{\psi}\psi\rangle \equiv \frac{T}{V}\frac{1}{\mathcal{Z}}\int \mathcal{D}U\mathcal{D}\psi\,\mathcal{D}\bar{\psi}\, e^{-S_G-S_F}\int \mathrm{d}^4x\,\mathrm{d}^4y\,\bar{\psi}(x)\psi(x)\bar{\psi}(y)\psi(y)$$
$$= -\frac{T}{V}\left\langle \mathrm{Tr}\big(M^{-1}M^{-1}\big)\right\rangle + \frac{T}{V}\left\langle \mathrm{Tr}\,M^{-1}\,\mathrm{Tr}\,M^{-1}\right\rangle - \frac{T}{V}\left\langle \mathrm{Tr}\,M^{-1}\right\rangle^2, \quad (2.56)$$

where Tr denotes the sum over all the lattice points and trace over other internal spaces, like color or Dirac ($\mathrm{tr}_D$),

$$\mathrm{Tr} \to \sum_n \mathrm{tr}_C\,\mathrm{tr}_D\,. \quad (2.57)$$

The first term in Eq. (2.56) is usually referred to as the **connected** part, while the other two terms are called **disconnected** terms.

Let us inspect Eq. (2.55) closer,

$$\langle\bar{\psi}\psi\rangle = \frac{T}{V}\frac{1}{\mathcal{Z}}\int \mathcal{D}U\,\det M\,\mathrm{Tr}\,M^{-1}\,. \quad (2.58)$$

In this expression, we can distinguish two fermionic contributions: one coming from the determinant, usually referred to as **sea contribution**, and another coming from the Dirac operator in the trace, called **valence contribution**. Both the sea and valence Dirac operators are supposed to be the same. However, on the lattice one may choose different setups for each operator. For example, one can employ different discretizations for each operator or choose different fermion masses, $m_{\mathrm{sea}}$ for the sea Dirac operator and $m_{\mathrm{val}}$ for the valence one. A particular approximation, which is relevant for this thesis, where $m_{\mathrm{sea}} \to \infty$, is known as the **quenched approximation**. If we consider this limit of QCD, the quarks decouple from the theory and we get a pure SU(3) gauge theory. In this case, the fermionic determinant approaches a constant, that drops when taking expectation values. Therefore, we can consider the case where the quarks have a very heavy but finite mass, so that taking the determinant to be a constant is a good approximation. In this setup, expectation values are taken with respect to the partition function of the pure gauge theory, even for fermionic observables

$$\langle O_F\rangle_{\mathrm{quenched}} = \frac{T}{V}\int \mathcal{D}U\,e^{-S_G}\,O_F\,. \quad (2.59)$$

In the language of perturbation theory, this approximation accounts for neglecting the virtual fermionic loops in Feynman diagrams. In the study of physical QCD, where the $u$ and $d$ quarks are lighter than most of the relevant scales in the theory, neglecting the fermionic determinant is in general not justified.

The reason why this is a popular approximation is that the calculation of expectation values in lattice QCD is performed numerically, see App. A for a brief introduction to this topic. The most demanding part of the calculation is the fermionic determinant, so the quenched approximation can be a very useful tool to reduce the computational cost of simulations. When the first simulations of lattice QCD started in the 1990s, the lack of computational power forced the use of the quenched approximation. In the present days, where full dynamical simulations are possible, the quenched theory is still a practical way to study certain problems in a first approximation. We will make use of the quenched approximations as a test case for full QCD in Chs. 4 and 5.

As a last remark, in real computer simulations the number of lattice points has to be finite, therefore it is necessary to impose boundary conditions. For the gluon fields, we use periodic boundary conditions in every direction, while for the fermion fields we impose periodic boundary conditions in the spatial directions and anti-periodic ones in the temporal direction, as dictated by the Fermi-Dirac statistics,

$$U_\mu(N_s \hat{i}, n_t) = U_\mu(0), \quad U_\mu(\vec{n}, N_t) = U_\mu(\vec{n}, 0), \tag{2.60}$$

$$\psi(N_s \hat{i}, n_t) = \psi(\vec{n}, 0), \quad \psi(\vec{n}, N_t) = -\psi(\vec{n}, 0), \tag{2.61}$$

for any $\hat{i} \in \{(1,0,0), (0,1,0), (0,0,1)\}$.

## 2.5 Global symmetries of QCD

Up to this point, we have only discussed the naive fermion discretization. This formalism is rarely used in lattice calculations, since the 16 doublers make simulations for an arbitrary number of flavors unfeasible. If we want to describe physical QCD on the lattice, we need to eliminate the contributions of the doublers. However, this turns out to be a very delicate issue, due to the connection of the doublers with the global symmetries of the theory. To understand this crucial point, we will now explain how the global symmetries of the QCD action appear on the lattice. To this end, we first review some of these symmetries in the continuum theory. In this section, we follow Refs. [31] and [38].

### 2.5.1 Continuum formulation

Let us first briefly consider the free Dirac action (2.11). A Dirac spinor $\psi$ can be decomposed into two Weyl spinors $\varphi_L, \varphi_R$ that transform according to the $(1/2, 0)$, $(0, 1/2)$ irreducible representations of the Lorentz group respectively,

$$\psi = \begin{pmatrix} \varphi_L \\ \varphi_R \end{pmatrix}. \tag{2.62}$$

A Weyl spinor $\varphi_{L/R}$ is said to be left/right-handed, a property referred to as **chirality**. Dirac spinors can be projected into a definite handedness using the chiral projectors $P_{L/R}$,

$$P_L \psi \equiv \psi_L = \begin{pmatrix} \varphi_L \\ 0 \end{pmatrix}, \quad P_R \psi \equiv \psi_R = \begin{pmatrix} 0 \\ \varphi_R \end{pmatrix}, \tag{2.63}$$

with

$$P_{L/R} = \frac{\mathbb{1} \pm \gamma_5}{2}. \tag{2.64}$$

The Dirac action is invariant under global *vector* transformations, which transform equally left- and right-handed spinors

$$\begin{aligned} \psi'(x) &= e^{i\alpha} \psi(x), \\ \bar{\psi}'(x) &= \bar{\psi}(x) e^{-i\alpha}. \end{aligned} \tag{2.65}$$

In addition, for massless fermions, it is also invariant under global *axial* (or *chiral*) transformations, i.e. different rotations for the left- and right-handed components, which can be written as

$$\begin{aligned} \psi'(x) &= e^{i\alpha\gamma_5} \psi(x), \\ \bar{\psi}'(x) &= \bar{\psi}(x) e^{i\alpha\gamma_5}. \end{aligned} \tag{2.66}$$

This can be summarized in the following property of the Dirac operator

$$\{\slashed{D}, \gamma_5\} = 0. \tag{2.67}$$

The mass term breaks this symmetry explicitly, since it mixes right- and left-handed spinors.

These symmetry concepts can be generalized for QCD with $N_f$ flavors. The fermionic action is given by

$$S_F = \int d^4x \, \bar{\Psi}(x) \left[ \mathbb{1}_F \sum_\mu \gamma_\mu (\partial_\mu + i A_\mu(x)) + \mathcal{M} \right] \Psi(x) \,, \tag{2.68}$$

where $\mathbb{1}_F$ is the identity matrix in flavor space, $\mathcal{M}$ is the mass matrix

$$\mathcal{M} = \mathrm{diag}\,(m_1, \ldots, m_{N_f}) \,, \tag{2.69}$$

and $\Psi, \bar{\Psi}$ are vectors in flavor space

$$\Psi = \begin{pmatrix} \psi_1 \\ \vdots \\ \psi_{N_f} \end{pmatrix}, \qquad \bar{\Psi} = \left( \bar{\psi}_1, \ldots, \bar{\psi}_{N_f} \right) \,. \tag{2.70}$$

Now the chiral transformations are given by the group $\mathrm{SU}(N_f)$. We can define **vector transformations** as

$$\Psi' = e^{i\alpha T_i} \Psi \,, \qquad\qquad \bar{\Psi}' = \bar{\Psi} e^{-i\alpha T_i} \,, \tag{2.71}$$

$$\Psi' = e^{i\alpha \mathbb{1}_F} \Psi \,, \qquad\qquad \bar{\Psi}' = \bar{\Psi} e^{-i\alpha \mathbb{1}_F} \,. \tag{2.72}$$

Similarly, we can define a set of **axial transformations**

$$\Psi' = e^{i\alpha \gamma_5 T_i} \Psi \,, \qquad\qquad \bar{\Psi}' = \bar{\Psi} e^{i\alpha \gamma_5 T_i} \,, \tag{2.73}$$

$$\Psi' = e^{i\alpha \gamma_5 \mathbb{1}_F} \Psi \,, \qquad\qquad \bar{\Psi}' = \bar{\Psi} e^{i\alpha \gamma_5 \mathbb{1}_F} \,. \tag{2.74}$$

This set of transformations leaves the action (2.68) invariant when all the flavors are massless. Altogether, the massless (classical) action has the following global symmetry

$$\mathrm{SU}(N_f)_L \times \mathrm{SU}(N_f)_R \times \mathrm{U}(1)_V \times \mathrm{U}(1)_A \,. \tag{2.75}$$

The notation $\mathrm{SU}(N_f)_L \times \mathrm{SU}(N_f)_R$ emphasizes that left- and right-handed components of the fermion fields transform independently. In the case of non-zero but degenerate masses, the symmetry breaks to

$$\mathrm{SU}(N_f)_V \times \mathrm{U}(1)_V \times \mathrm{U}(1)_A \,, \tag{2.76}$$

where $\mathrm{SU}(N_f)_V$ is the diagonal vector group, where left- and right-handed spinors transform equally. Finally, for non-zero and non-degenerate masses only the symmetry group

$$\underbrace{\mathrm{U}(1)_V \times \cdots \times \mathrm{U}(1)_V}_{N_f} \times \mathrm{U}(1)_A \tag{2.77}$$

remains.

We can derive the currents associated with these symmetries. For simplicity, we consider the case of $N_f = 2$ (up and down quarks in QCD) with degenerate masses $m_u = m_d = m_{ud}$. The generators of SU(2) are $\tau^j/2$ with $j = 1, 2, 3$ and $\tau$ the Pauli matrices. By applying the transformations (2.71), (2.72) to the action (2.68), we can derive, see e.g. Ref. [38], the singlet and non-singlet **Ward identities** (WIs)

$$\sum_\mu \partial_\mu \mathcal{J}_\mu(x) = 0 \qquad \text{(singlet)}, \tag{2.78}$$

$$\sum_\mu \partial_\mu \mathcal{J}_\mu^j(x) = 0 \qquad \text{(non-singlet)}, \tag{2.79}$$

with the singlet and non-singlet vector currents

$$\mathcal{J}_\mu(x) = \bar{\Psi}(x)\gamma_\mu \frac{\mathbb{1}_F}{2}\Psi(x) \qquad \text{(singlet)}, \tag{2.80}$$

$$\mathcal{J}_\mu^j(x) = \bar{\Psi}(x)\gamma_\mu \frac{\tau^j}{2}\Psi(x) \qquad \text{(non-singlet)}. \tag{2.81}$$

Similarly, using the transformations (2.73), (2.74), the singlet and non-singlet **axial Ward identities** (AWIs) are given by[2.9]

$$\begin{aligned}\sum_\mu \partial_\mu \mathcal{J}_{\mu 5}(x) &= 2m_{ud}\,\bar{\Psi}(x)\gamma_5\frac{\mathbb{1}_F}{2}\Psi(x) \\ &+ \frac{1}{16\pi^2}\sum_{\alpha,\beta,\mu,\nu}\epsilon_{\alpha\beta\mu\nu}\,\mathrm{tr}_C[G_{\alpha\beta}(x)G_{\mu\nu}(x)] \qquad \text{(singlet)},\end{aligned} \tag{2.82}$$

and

$$\sum_\mu \partial_\mu \mathcal{J}_{\mu 5}^j(x) = 2m_{ud}\,\bar{\Psi}(x)\gamma_5\frac{\tau^j}{2}\Psi(x) \qquad \text{(non-singlet)}, \tag{2.83}$$

---

[2.9] More precisely, the last term in the singlet AWI (2.82) does not appear from the transformation (2.73), it is an explicit breaking of the symmetry that arises from quantum effects. We will explain this point in detail below.

where the singlet and non-singlet axial currents are

$$\mathcal{J}_{\mu 5}(x) = \bar{\Psi}(x)\gamma_\mu \gamma_5 \frac{\mathbb{1}_F}{2}\Psi(x) \qquad \text{(singlet)}, \tag{2.84}$$

$$\mathcal{J}_{\mu 5}^j(x) = \bar{\Psi}(x)\gamma_\mu \gamma_5 \frac{\tau^j}{2}\Psi(x) \qquad \text{(non-singlet)}. \tag{2.85}$$

Although in QCD the quarks are massive, one would expect that the chiral symmetry $\mathrm{SU}(N_f)_L \times \mathrm{SU}(N_f)_R$ is still a good approximate symmetry for $N_f = 2$, and also in part for $N_f = 3$, since the explicit breaking due to the masses is small for the lighter quarks. However, these symmetries turn out to have a more complicated structure. Let us consider again the case of $N_f = 2$ in QCD, with degenerate up and down quarks. As we just mentioned, the fact that these particles are very light, just a few MeV, would naively point to chiral symmetry being still an approximate symmetry of QCD. This has some experimental consequences, in particular it implies that the nucleon[2.10] $N$ must have the same mass as his negative parity-partner $N^*$. However, the observed mass for the nucleon is around 940 MeV, while $N^*$ possesses a mass of 1535 MeV. This difference cannot be solely explained by the masses of the up and down quarks, there has to be another source of explicit breaking. Since the action is invariant under the symmetry, this violation has to come from a **spontaneous breaking of chiral symmetry**, where the ground state is not invariant under this symmetry. The symmetry-breaking pattern is

$$\mathrm{SU}(2)_L \times \mathrm{SU}(2)_R \to \mathrm{SU}(2)_V. \tag{2.86}$$

The order parameter for this spontaneous symmetry breaking is the **chiral condensate**,

$$\left\langle \bar{\Psi}(x)\frac{\mathbb{1}_F}{2}\Psi(x) \right\rangle \tag{2.87}$$

and the theory has 3 quasi-Goldstone bosons which correspond to the pions. This vacuum expectation value transforms like a mass term, and therefore is not invariant under chiral rotations. We will discuss later how chiral symmetry can be restored by increasing the temperature, which leads to a very active research topic in lattice QCD, the phase diagram of strongly interacting matter.

Finally, there is yet another source of symmetry breaking, which is the most relevant one for this thesis. The flavor-singlet axial symmetry $\mathrm{U}(1)_A$ is a symmetry of the *classical* action, however, it is broken explicitly in the *quantum* theory. This is known as the **axial**

---

[2.10] For degenerate $u$ and $d$ quarks and without taking into account QED effects, the proton and the neutron are the same particle, which is usually called nucleon.

**anomaly**, and in the path integral formalism it can be understood as a non-invariance of the fermionic integration measure under U(1)$_A$ transformation. This is reflected in the last term in Eq. (2.82), which breaks the conservation law explicitly even in the $m_{ud} = 0$ case. This term is proportional to an object known as topological charge density, and its integral over the whole space is the **topological charge** $Q_{\text{top}}$

$$Q_{\text{top}} = \frac{1}{32\pi^2} \sum_{\mu,\nu,\alpha,\beta} \int d^4x \, \epsilon_{\mu\nu\alpha\beta} \operatorname{tr}_C [G_{\mu\nu}(x) G_{\alpha\beta}(x)]. \tag{2.88}$$

The topological charge possesses very interesting properties, for example it can be shown to be an integer $Q_{\text{top}} \in \mathbb{Z}$. In **Quantum Electrodynamics** (QED), where the gauge group is U(1), this anomaly is also present, and it is again generated by the topological charge,[2.11] that in this case reads

$$Q_{\text{top}}^{\text{U(1)}} = \frac{1}{32\pi^2} \sum_{\mu,\nu,\alpha,\beta} \int d^4x \, \epsilon_{\mu\nu\alpha\beta} [F_{\mu\nu}(x) F_{\alpha\beta}(x)], \tag{2.89}$$

with the electromagnetic field tensor $F_{\mu\nu}(x) = \partial_\mu a_\nu(x) + \partial_\nu a_\mu(x)$ and $a_\mu(x) \in \mathfrak{u}(1)$. This can be rewritten in terms of the usual electric and magnetic fields $\vec{E}$ and $\vec{B}$ as

$$Q_{\text{top}}^{\text{U(1)}} = \frac{1}{4\pi^2} \int d^4x \, \vec{E} \cdot \vec{B}(x). \tag{2.90}$$

This means that a parallel electric and magnetic field generates the anomaly in QED, a concept that will be relevant in the next chapter. The axial anomaly has several physical consequences, e.g. the mass difference of the $\eta$ and $\eta'$ mesons or the enhancement decay rate of the process $\pi^0 \to \gamma\gamma$, see for example Ref. [39] for an introduction. We will discuss more closely the specifics of this anomaly in Ch. 3.

### 2.5.2 Lattice formulation

Having revisited the symmetries of QCD in the continuum, we will now focus on the equivalent of these symmetries in the lattice formulation. The first consideration is about chiral symmetry, which as we discussed above can be expressed in terms of the Dirac operator as the condition

$$\{\gamma_5, \slashed{D}\} = \gamma_5 \slashed{D} + \slashed{D} \gamma_5 = 0. \tag{2.91}$$

---

[2.11] In a more precise way, the topological charge is zero in QED for every gauge configuration, since the integral of the operator in Eq. (2.89) vanish. Only when considering fields $a_\mu(x)$ which are non-zero in the boundary $|\vec{x}| \to \infty$, like in the case of homogeneous background electromagnetic fields, a non-trivial topology can arise. We will concentrate on this case for the rest of the text.

Unfortunately, the implementation of this symmetry on the lattice encounters an unavoidable obstacle, a no-go theorem for chiral symmetry on the lattice proved in 1981 by H. B. Nielsen and M. Ninomiya, now known as the **Nielsen-Ninomiya theorem** [40, 41]:[2.12]

> Let $D$ be a lattice Dirac operator which is a translationally invariant, hermitian operator. Then one of the following properties has to be violated:
>
> - $\widetilde{\slashed{D}}(p)$ is local, i.e. it decays exponentially fast, such that its Fourier transform exists and its derivatives are continuous.
> - $\widetilde{\slashed{D}}(p)$ is invertible everywhere except at $p = 0$, i.e. it is free of doublers.
> - The Dirac operator fulfills $\{\gamma_5, \slashed{D}\} = 0$, i.e. it possesses chiral symmetry.

At first glance, it seems that the locality condition should be the first candidate to be sacrificed. Unfortunately, this turns out to have catastrophic consequences in gauge theories, as shown for a non-local fermionic discretization called SLAC fermions [42], where it leads to violations of Lorentz and gauge invariance [42]. This means that we have to choose between having doublers in the theory or breaking chiral symmetry. However, if we choose the latter, we can try to break chiral symmetry in a controlled fashion. In particular, we can introduce a violation of $\mathcal{O}(a)$, in the following way

$$\slashed{D}\gamma_5 + \gamma_5\slashed{D} = a\slashed{D}\gamma_5\slashed{D}. \tag{2.92}$$

This is known as the **Ginsparg-Wilson equation** [43]. Lattice Dirac operators fulfilling this equation can be shown to possess desirable chiral properties, very close to the ones in the continuum theory, hence they are usually referred to as chiral discretizations. Two examples of discretizations in this category are overlap fermions [44] or domain-wall fermions [45].

How do the Ward identities appear on the lattice? Let us analyze in detail the derivation of the singlet Ward identities,[2.13] since they play a central role in this thesis. We will see that, while the U(1)$_V$ Ward identity is straightforward to construct on the lattice, the appearance of the axial anomaly is a more complicated topic. Let us first consider the vector U(1) transformations. We can derive the Ward identity on the lattice

---

[2.12] This is the lattice version of the theorem. There is also a generalization to the continuum theory that we will discuss in Ch. 3.
[2.13] We will briefly come back to the non-single axial Ward identity on the lattice in Sec. 2.6.1.

for naive fermions by performing a local vector transformation

$$\begin{aligned} \delta\psi(n) &= i\alpha(n)\,\psi(n)\,, \\ \delta\bar\psi(n) &= -i\alpha(n)\,\bar\psi(n)\,. \end{aligned} \qquad (2.93)$$

to the action (2.42). Defining the lattice backward derivative as

$$\nabla_\mu f(n) = f(n) - f(n-\hat\mu)\,, \qquad (2.94)$$

it is straightforward to arrive at the **lattice Ward identity**

$$\sum_\mu \nabla_\mu J_\mu(n) = 0\,, \qquad (2.95)$$

with the **conserved vector** current

$$J_\mu(n) = \frac{1}{2a}\bar\psi(n)\gamma_\mu[U_\mu(n)\psi(n+\hat\mu) + U_\mu^\dagger(n-\hat\mu)\psi(n-\hat\mu)]\,. \qquad (2.96)$$

Notice that this current depends on the two nearest neighbors of $n$ in every direction. We will refer to these types of current as point-split or 1-link operators.

In the case of the $U(1)_A$ symmetry, one could naively consider the following local axial transformation

$$\begin{aligned} \delta\psi(n) &= i\alpha(n)\,\gamma_5\psi(n)\,, \\ \delta\bar\psi(n) &= i\alpha(n)\,\bar\psi(n)\,\gamma_5\,, \end{aligned} \qquad (2.97)$$

analogously to the vector case. However, this can be shown, see Ref. [46], to lead to an axial current

$$J^0_{\mu 5}(n) = \frac{1}{2a}\bar\psi(n)\gamma_\mu\gamma_5\bigl[U_\mu(n)\psi(n+\hat\mu) + U_\mu^\dagger(n-\hat\mu)\psi(n-\hat\mu)\bigr]\,, \qquad (2.98)$$

and a pseudoscalar density

$$P^0_5(n) = \bar\psi(n)\gamma_5\psi(n)\,, \qquad (2.99)$$

which fulfill a conservation equation of the form

$$\sum_\mu \nabla_\mu J^0_{\mu 5}(n) = 2m P^0_5(n)\,. \qquad (2.100)$$

Surprisingly, there is no sign of the anomaly in this equation.[2.14] This fact was explained by L. H. Karsten and J. Smit in their seminal 1981 work [46]. The key aspect is to consider how the transformation (2.97) acts on the doubler fields $\tilde{q}^g$ defined in Eq. (2.34). In momentum space, the fermion fields transform as

$$\delta\tilde{\psi}(k) = i\tilde{\alpha}(k)\,\gamma_5\tilde{\psi}(k)\,, \tag{2.101}$$

implying that $\tilde{q}^g$ changes as

$$\delta\tilde{q}^g(k) = i\tilde{\alpha}(k)\,M_g\gamma_5\,\tilde{\psi}(k+\pi_g) = \pm i\tilde{\alpha}(k)\,\gamma_5\,\tilde{q}^g(k)\,, \tag{2.102}$$

where the sign depends on $g$, it is $+1$ when $\pi_g$ has an even number of non-zero components and $-1$ when it has an odd number. We can formalize this by defining the set of vectors[2.15]

$$\Xi = \{\xi;\ \xi_\mu = \pm 1 \quad \forall \mu\}\,. \tag{2.103}$$

Then the axial transformation for the doublers reads

$$\delta\tilde{q}^g(k) = i\tilde{\alpha}(k)\,e^{i\hat{\pi}_g\xi}\gamma_5\,\tilde{q}^g(k) \qquad \text{for any } \xi \in \Xi\,. \tag{2.104}$$

Physically, this means that half of the doublers have opposite axial charges as the other half, and this produces a cancellation between them that eliminates the anomaly.

However, it is possible to obtain the correct form of the anomaly with naive fermions. This was calculated in Ref. [34], by modifying the axial transformation so there is no cancellation between the doublers. The modified transformation is

$$\begin{aligned}\delta\tilde{\psi}(n) &= i\alpha(n)\,\gamma_5\,\psi(n+\xi)\,,\\ \delta\tilde{\bar{\psi}}(n) &= i\alpha(n)\,\bar{\psi}(n+\xi)\,\gamma_5\,,\end{aligned} \tag{2.105}$$

---

[2.14] An important point is that in the continuum, the anomaly stems from the fact that the fermionic measure is divergent [47]. Since the lattice regularizes the measure, the anomaly has to appear as an *explicit* breaking of the axial symmetry.

[2.15] This particular choice of the vectors $\xi$ turns out to be advantageous in the derivation of the anomaly [34]. In principle other choices for the set of vectors are possible, for example considering only the vectors with all components equal to $+1$,

$$\Xi = \{\xi;\ \xi_\mu = +1 \quad \forall \mu\}\,.$$

with $\xi \in \Xi$. This transformation in momentum space acts as

$$\delta\psi(k) = i\tilde{\alpha}(k)\,\gamma_5\,e^{-iak\xi}\,\tilde{\psi}(k)\,. \tag{2.106}$$

Now we can show that all the doublers transform equally

$$\begin{aligned}\delta\tilde{q}^g(k) &= i\tilde{\alpha}(k)\,P_g\gamma_5 e^{-i\xi a(k+\pi_g)}\tilde{\psi}(k+\pi_g) = i\tilde{\alpha}(k)\,\gamma_5 P_g e^{i\hat{\pi}\xi}e^{-i\xi a(k+\pi_g)}\tilde{\psi}(k+\pi_g) \\ &= i\tilde{\alpha}(k)\,\gamma_5 P_g e^{-ia\xi k}\tilde{\psi}(k+\pi_g) = i\tilde{\alpha}(k)\,e^{-ia\xi k}\gamma_5\tilde{q}^g(k)\,, \end{aligned} \tag{2.107}$$

since the extra phase is independent of $g$.

We can generalize this transformation by averaging over all the possible values of $\xi$ and by inserting gauge links to obtain a gauge invariant object. The transformation law is then

$$\begin{aligned}\delta\psi(n) &= i\alpha(n)\,\gamma_5\,\psi(n,\Xi)\,, \\ \delta\bar{\psi}(n) &= i\alpha(n)\,\bar{\psi}(n,\Xi)\,\gamma_5\,, \end{aligned} \tag{2.108}$$

where

$$\psi(n,\Xi) = \frac{1}{16}\sum_{\xi\in\Xi}U(n,\xi)\psi(n+\xi)\,, \tag{2.109}$$

and $U(n,\xi)$ is the product of links in a given path from $n$ to $n+\xi$. In principle, one could also average over these paths, since the choice is not unique. We will come back to this point in Ch. 4. Using this transformation, the correct AWI can be written as

$$\sum_\mu \nabla_\mu J_{\mu 5}(n) = 2mP_5(n) + Q(n)\,, \tag{2.110}$$

with the current

$$J_{\mu 5}(n) = \frac{1}{2a}\bar{\psi}(n)\bigl[U_\mu(n)\gamma_\mu\gamma_5\psi(n+\mu,\Xi) + U_\mu^\dagger(n-\hat{\mu})\gamma_\mu\gamma_5\psi(n-\hat{\mu},\Xi)\bigr]\,, \tag{2.111}$$

the pseudoscalar density

$$P_5(n) = \bar{\psi}(n)\gamma_5\psi(n,\Xi)\,, \tag{2.112}$$

and a remaining term

$$Q(n) = \frac{1}{2a}\sum_{\xi\in\Xi}\frac{1}{16}\sum_\mu \bar{\psi}(n)\gamma_\mu\gamma_5\Bigl[U_\mu(n)U(n+\hat{\mu},\xi) \\ - U(n+\hat{\mu},\xi)U_\mu(n+\hat{\mu})\Bigr]\psi(n+\hat{\mu}+\xi)\,. \tag{2.113}$$

In the continuum limit, this last term can be shown to yield the topological contribution coming from the anomaly [34, 48].

In this section, we have explained the global symmetries of the discretized theory, and the delicate balance between eliminating the doublers and retaining the symmetries of the continuum theory. In the next section, we give two concrete examples of discretizations that tackle this issue from two different angles.

## 2.6 Lattice fermionic formulations

We now present the two discretizations that have been used to obtain the results of this thesis: Wilson and staggered quarks. We mainly follow the presentation in Refs. [30] and [31].

### 2.6.1 Wilson fermions

The first strategy to remove the contributions from the doublers was proposed by Wilson in 1977 [49]. The idea is to lift the edges of the Brillouin zone by an amount proportional to the inverse lattice spacing, to modify the dispersion relation of the doublers. In particular, the Dirac operator in momentum space is modified like

$$\begin{aligned}\widetilde{M}(p) &= m + \frac{i}{a}\sum_\mu \gamma_\mu \sin(ap_\mu) + \frac{r}{a}\sum_\mu [1 - \cos(ap_\mu)] \\ &\equiv m + \frac{i}{a}\sum_\mu \left[\gamma_\mu \sin(ap_\mu) + r\widetilde{X}(ap_\mu)\right],\end{aligned} \qquad (2.114)$$

where $X$ is known as the **Wilson term** and $r$ is the Wilson parameter, which is usually given the value $r = 1$. Notice that $r = 0$ corresponds to the Dirac operator of naive fermions (2.18). For the physical particle, for which all momentum components are $p_\mu = 0$, the extra term vanishes. However, the doublers contain $N > 0$ momentum components with $p_\mu = \pi/a$, hence the newly added term will modify the mass of the doublers like

$$m + \frac{2rN}{a}. \qquad (2.115)$$

When $a \to 0$, the mass of the doublers will tend to infinity, making them decouple in the continuum limit. Another way to understand how the Wilson term eliminates the doublers is via the spectrum doubling symmetry defined in momentum space in Eq. (2.33). The Wilson term explicitly violates this symmetry, breaking the degeneracy.

In position space, the Wilson term corresponds to a discretized d'Alembert operator $-(a/2)\Box_\mu$,

$$X(l,n) = -\frac{a}{2} \sum_\mu \frac{\delta_{l+\hat{\mu},n} - 2\delta_{l,n} + \delta_{l-\hat{\mu},n}}{a^2}\,. \tag{2.116}$$

The total action, including the gauge links, reads

$$S_F^{\text{Wilson}} = a^4 \sum_{l,n} \bar{\psi}(l)\, M_{\text{Wilson}}(l,n)\, \psi(n)\,, \tag{2.117}$$

with the Dirac operator for Wilson fermions

$$\begin{aligned}M_{\text{Wilson}}(l,n) = &-\frac{1}{2a} \sum_\mu \left[ (r-\gamma_\mu)\delta_{l+\hat{\mu},n} U_\mu(n) + (r+\gamma_\mu)\delta_{l-\hat{\mu},n} U_\mu^\dagger(n-\hat{\mu}) \right] \\ &+ \left(m + \frac{4r}{a}\right)\delta_{l,n}\,,\end{aligned} \tag{2.118}$$

which describes a single fermion in the continuum limit. The Wilson fermions action is usually rewritten in terms of the **hopping parameter** $\kappa$

$$M_{\text{Wilson}} = C(1 - \kappa H)\,, \tag{2.119}$$

where

$$\kappa = \frac{1}{8 + 2am}\,, \quad C = m + \frac{4r}{a}\,, \tag{2.120}$$

and $H$ is the **hopping operator**,

$$H(l,n) = \frac{1}{2a} \sum_\mu \left[ (r-\gamma_\mu)\delta_{l+\hat{\mu},n} U_\mu(n) + (r+\gamma_\mu)\delta_{l-\hat{\mu},n} U_\mu^\dagger(n-\hat{\mu}) \right]\,. \tag{2.121}$$

This action has lattice artifacts of order $\mathcal{O}(a)$, as opposed to the $\mathcal{O}(a^2)$ in the case of naive fermions. This can be improved in a similar way as discussed for the gauge action: an irrelevant term can be added to the action to cancel the lower order, making the new action converge faster to the continuum limit. In the Wilson case, the $\mathcal{O}(a)$ improvement consists of adding the Sheikholeslami–Wohlert term [50]. However, for this thesis it is enough to only consider unimproved Wilson fermions.

Unfortunately, removing the doublers comes with a price, as the Nielsen-Ninomiya theorem taught us. Since the Wilson term acts like a mass term, it means that this action lacks chiral symmetry even in the $m \to 0$ limit. This has serious consequences, in particular for the renormalization of the Wilson action. To see this, it is useful to consider the Ward identities in the Wilson formulation. In the case of the U(1)$_V$, we obtain a very similar result as in the naive fermions formulation. Considering again the transformation (2.93), we obtain the WI [46]

$$\sum_\mu \nabla_\mu J_\mu(n) = 0\,, \tag{2.122}$$

with

$$J_\mu(n) = \frac{1}{2a} \sum_\mu \bar\psi(n)\left[(\gamma_\mu - r)U_\mu(n)\psi(n+\hat\mu) + (\gamma_\mu + r)U_\mu^\dagger(n-\hat\mu)\psi(n-\hat\mu)\right]. \tag{2.123}$$

Notice that the Wilson term enters the definition of the conserved vector current.

For the U(1)$_A$ anomaly, the situation compared to the naive fermions case has changed. Now that the doubler degeneracy has been broken, if we consider again the local axial transformation (2.97), the cancellation of the anomaly no longer takes place. Therefore, we can use this transformation to derive the singlet AWI [46]

$$\sum_\mu \nabla_\mu J_{\mu 5}(n) = 2m\bar\psi(n)\gamma_5\psi(n) + rX_5(n)\,, \tag{2.124}$$

with

$$X_5(n) = -\frac{1}{2a}\sum_\mu \Big\{ \left[\bar\psi(n)U_\mu(n)\gamma_5\psi(n+\hat\mu) + \bar\psi(n+\hat\mu)U_\mu^\dagger(n)\gamma_5\psi(n+\hat\mu)\right] \\ + [n \to n - \hat\mu] - 4\bar\psi(n)\gamma_5\psi(n) \Big\}, \tag{2.125}$$

and the anomalous axial current

$$J_{\mu 5}(n) = \frac{1}{2a}\sum_\mu \bar\psi(n)\gamma_\mu\gamma_5[U_\mu(n)\psi(n+\hat\mu) + U_\mu^\dagger(n-\hat\mu)\psi(n-\hat\mu)]. \tag{2.126}$$

In the continuum limit, it can be shown that the term $X_5$ generates the anomaly [46]. Therefore, in the Wilson formulation, the anomaly arises from the Wilson term. This is an important concept of the Wilson fermions discretization that we will refer to below.

It is instructive to also consider the non-singlet axial current in this case, since it is related to the explicit breaking of flavor chiral symmetry. For the case of two degenerate flavors with mass $m_{ud}$, we can consider the transformation

$$\begin{aligned}\delta\Psi(n) &= i\alpha(n)\,\gamma_5 T_j \Psi(n)\,,\\ \delta\bar{\Psi}(n) &= i\alpha(n)\,\bar{\Psi}(n)\,\gamma_5 T_j\,,\end{aligned} \qquad (2.127)$$

with $T_j = \tau_j/2$. The non-singlet AWI can be derived using this transformation, which yields [38]

$$\sum_\mu \nabla_\mu \mathcal{J}_{\mu 5}^j(n) = 2m_{ud}\bar{\Psi}(n)\gamma_5\frac{\tau^j}{2}\Psi(n) + rX_5^j(n)\,, \qquad (2.128)$$

with

$$X_5^j(n) = -\frac{1}{2a}\sum_\mu \left\{ \left[\bar{\Psi}(n)U_\mu(n)\gamma_5\frac{\tau^j}{2}\Psi(n+\hat{\mu}) + \bar{\Psi}(n+\hat{\mu})U_\mu^\dagger(n)\gamma_5\frac{\tau^j}{2}\Psi(n+\hat{\mu})\right] \right.\\ \left. + [n\to n-\hat{\mu}] - 4\bar{\Psi}(n)\gamma_5\frac{\tau^j}{2}\Psi(n) \right\}, \qquad (2.129)$$

and the current

$$\mathcal{J}_{\mu 5}^j(n) = \frac{1}{2a}\sum_\mu \bar{\Psi}(n)\gamma_\mu\gamma_5[U_\mu(n)\frac{\tau^j}{2}\Psi(n+\hat{\mu}) + U_\mu^\dagger(n-\hat{\mu})\frac{\tau^j}{2}\Psi(n-\hat{\mu})]\,. \qquad (2.130)$$

Now the effect of the chiral symmetry breaking is transparent, since setting $m_{ud} = 0$ does not give a conserved current. Although the term is very similar to the singlet case, this current is not anomalous due to its flavor structure and vanishes in the continuum limit. However, it has another effect. In the continuum and in lattice discretizations where some subset of the chiral symmetry is retained at $m = 0$, the bare quark mass only renormalizes multiplicatively $m_R = Z_S m$. However, when there is no chiral symmetry at all, like in the Wilson fermions case, the mass also **renormalizes additively**. This can be understood by looking at the explicit form of $X_5^j$. The last term in Eq. (2.129) is the pseudoscalar density, meaning that this operator mixes with $X_j^5$. Therefore the renormalization of the axial current and the pseudoscalar condensate also mixes, which results in the additive renormalization for the mass [51]. The renormalized mass takes the form

$$m_R = Z_S[m - m_{\text{cr}}]\,, \qquad (2.131)$$

where $m_{\text{cr}}$ is the **critical mass**. Equivalently, a critical hopping parameter $\kappa_{\text{cr}}$ can be defined. These critical values can be determined in several ways, a possibility is to calculate the value of $\kappa$ where the pion is massless, i.e. the chiral limit. This then corresponds to the additive mass renormalization $\kappa_{\text{cr}}$.

### 2.6.2 Staggered fermions

As we have seen in the formulation of Wilson fermions, modifying the dispersion relation is a possible path to follow to deal with the doublers. Another possible strategy consists of reducing the Brillouin zone by effectively doubling the lattice spacing, eliminating some of the doublers. This can be accomplished by redistributing the Dirac degrees of freedom on the lattice, with a formulation known as **staggered fermions**. It was first developed in the Hamiltonian formulation by J. B. Kogut and L. Susskind in 1975 [52] (this is the reason why staggered fermions are also known as Kogut-Susskind fermions) and later adapted to the Euclidean formalism [53].

Let us consider the free naive fermion action (2.14), and make the Dirac indices explicit

$$S_F^{\text{naive}} = a^4 \sum_{\alpha,\beta=1}^{4} \sum_{n} \bar{\psi}_\alpha(n) \left[ \sum_\mu (\gamma_\mu)_{\alpha\beta} \frac{\psi_\beta(n+\hat{\mu}) - \psi_\beta(n-\hat{\mu})}{2a} + m\psi_\alpha(n) \right]. \tag{2.132}$$

The most straightforward way of deriving the staggered formalism is through the spin diagonalization of this action. We can consider the following transformation

$$\begin{aligned} \psi(n) &= T(n)\psi'(n)\,, \\ \bar{\psi}(n) &= \bar{\psi}'(n)T^\dagger(n)\,, \end{aligned} \tag{2.133}$$

where $T(n)$ are $4 \times 4$ spin matrices which fulfill

$$T^\dagger(n)T(n) = \mathbb{1}_D\,, \tag{2.134}$$

$$T^\dagger(n)\gamma_\mu T(n \pm \hat{\mu}) = \mathbb{1}_D\, \eta_\mu(n)\,, \tag{2.135}$$

where $\eta_\mu$ are complex phases and $\mathbb{1}_D$ is the identity in Dirac space, i.e. a $4 \times 4$ identity matrix. Applying this transformation to the action yields

$$S_F^{\text{naive}} = a^4 \sum_\alpha \sum_n \bar{\psi}'_\alpha(n) \left[ \sum_\mu \eta_\mu(n) \frac{\psi'_\alpha(n+\hat{\mu}) - \psi'_\alpha(n-\hat{\mu})}{2a} + m\psi'_\alpha(n) \right]. \tag{2.136}$$

Now the action is diagonal in Dirac space, so we have four copies of the same action for $\alpha = 1, 2, 3, 4$. Hence we can select only one of the four spinor components, for example $\psi'_1(n) \equiv \chi(n)$. This is the **staggered action**, which in the presence of gauge links can

be written as

$$S_F^{\text{staggered}} = a^4 \sum_n \bar{\chi}(n) \left[ \sum_\mu \eta_\mu(n) \frac{U_\mu(n)\chi(n+\hat{\mu}) - U_\mu^\dagger(n-\hat{\mu})\chi(n-\hat{\mu})}{2a} \right. \\ \left. + m\chi(n) \right]. \quad (2.137)$$

Notice that the objects $\chi$ are no longer 4-spinors but 1-component spinors, hence the degrees of freedom are reduced by a factor of 4. By removing these degrees of freedom, we have also reduced the number of doubles from 16 to 4. A possible (but not unique) choice for the $T(n)$ matrices is

$$T(n) = \gamma_1^{n_1} \gamma_2^{n_2} \gamma_3^{n_3} \gamma_4^{n_4}, \quad (2.138)$$

which yields the following phases

$$\eta_1(n) = 1,\ \eta_2(n) = (-1)^{n_1},\ \eta_3(n) = (-1)^{n_1+n_2},\ \eta_4(n) = (-1)^{n_1+n_2+n_3}. \quad (2.139)$$

As we anticipated, we have reduced the number of doublers by redistributing the Dirac degrees of freedom on the lattice, although this is not completely transparent from our derivation. A clearer view of what we have achieved with the staggered transformation can be seen by expressing the action in what is known as the taste basis. We will reconstruct the 4-spinors by grouping the degrees of freedom within a hypercube, which will also expose some of the drawbacks of the staggered formulation.

Let us relabel the lattice sites in terms of a coordinate $h_\mu$ labeling to which hypercube the points belong and another coordinate $s_\mu$ that indicates to which corner of the hypercube the point corresponds to

$$n_\mu = 2h_\mu + s_\mu \quad \text{with} \quad h_\mu \in \{0, 1, \ldots, N_\mu/2 - 1\}, \quad s_\mu \in \{0, 1\}. \quad (2.140)$$

We can define the new fields[2.16] $\zeta$ that live within the hypercube as

$$\zeta_{\alpha a}(h) = \frac{1}{8} \sum_s T_{\alpha a}(s)\chi(2h+s), \quad \bar{\zeta}_{\alpha a}(h) = \frac{1}{8} \sum_s \bar{\chi}(2h+s) T^*_{a\alpha}(s). \quad (2.141)$$

---

[2.16] These fields have to be understood in the same spirit as the doubler fields $q^g$ defined in Eq. (2.34) since they have Dirac and doubler indexes. A very similar presentation to Sec. 2.3 can be carried out for staggered fermions using the fields $q^g$ and the set $G$ defined in Eq. (2.23), see Ref. [34] for further details. However, we follow here a different presentation, which can be found in most textbooks, for simplicity.

We have made the matrix indexes of $\zeta$ explicit, which we will identify in the end with a Dirac index $\alpha = \{1, 2, 3, 4\}$ and a doubler index $a = \{1, 2, 3, 4\}$. Note that the matrices $T$ mix doubler and Dirac components, therefore we assign them both types of indices. Using the following properties of $T(s)$

$$\frac{1}{4}\operatorname{tr}\left[T^\dagger(s)T(s')\right] = \delta_{s,s'}, \quad \frac{1}{4}\sum_s \left[T(s)^*_{b\beta}T_{a\alpha}(s)\right] = \delta_{\alpha,\beta}\delta_{a,b}, \quad (2.142)$$

the definitions can be inverted,

$$\chi(2h+s) = 2\operatorname{tr}\left[T(s)\zeta(h)\right], \quad \bar{\chi}(2h+s) = 2\operatorname{tr}\left[\bar{\zeta}(h)T^\dagger(s)\right]. \quad (2.143)$$

With these ingredients, and the explicit representation of $T(n)$ given in Eq. (2.138), it can be shown (see for example Ref. [48]) that the staggered action takes the form

$$S_F^{\text{staggered}} = b^4 \sum_h \bar{\zeta}_{a\alpha}(h) \left[\sum_\mu \left((\gamma_\mu)_{\alpha\beta}\delta_{ab}\nabla^b_\mu + \frac{1}{2}b\,(\gamma_5)_{\alpha\beta}\,(\gamma_\mu^*\gamma_5)_{ab}\,\Box^b_\mu\right) \right. \\ \left. + m\,\delta_{\alpha\beta}\delta_{ab}\right]\zeta_{\beta b}(h), \quad (2.144)$$

with

$$\nabla^b_\mu f(h) = \frac{f(h+\hat{\mu}) - f(h-\hat{\mu})}{2b}, \quad (2.145)$$

$$\Box^b_\mu f(h) = \frac{f(h+\hat{\mu}) - 2f(h) + f(h-\hat{\mu})}{b^2}, \quad (2.146)$$

and $b = 2a$ is the lattice spacing between hypercubes. Notice that the doubling of the lattice spacing is natural, since the fields $\zeta, \bar{\zeta}$ live in the hypercubes. Although we will not perform the derivation, it can be shown by analyzing the staggered action in momentum space, that this is the effective doubling of the lattice spacing that leads to a reduction of the Brillouin zone, reducing the number of doubler modes in the theory [30].

Now we try to make the aforementioned identification of the indexes with Dirac and flavor spaces. Considering the first and last terms in Eq. (2.144), it is natural to identify $\alpha$ and $\beta$ as Dirac indexes. This leaves us with $a, b$ as flavor indexes. However, the second term is not diagonal in flavor space, breaking chiral symmetry explicitly. This has no analogy in QCD in the continuum, so we will refer to the remaining doublers in the staggered formulation as **tastes**. Notice that the taste-breaking term, which allows for taste-changing processes, vanishes in the continuum limit, where we recover a four flavor theory. With this presentation, it might seem like staggered fermions are just a worse version of the Wilson formulation: chiral symmetry is lost *and* some of the doublers remain. We can make the comparison with a Wilson term more explicit. Let us consider the direct product of Dirac space ⊗ taste space, where the usual $\gamma$-matrices act on the former and the operators $t_\mu = \gamma_\mu^*$, $t_5 = \gamma_5$ on the latter. We can then rewrite the staggered action as

$$S_F^{\text{staggered}} = b^4 \sum_h \bar{\zeta}(h) \left[ \sum_\mu \left( (\gamma_\mu \otimes \mathbb{1}) \nabla_\mu^b + \frac{1}{2} b (\gamma_5 \otimes t_\mu t_5) \Box_\mu^b \right) \right. \\ \left. + m (\mathbb{1} \otimes \mathbb{1}) \right] \zeta(h) \,. \tag{2.147}$$

Similarly, we can express the Wilson fermions action (2.117) in Dirac ⊗ taste space as

$$S_F^{\text{Wilson}} = a^4 \sum_n \bar{\psi}(n) \left[ \sum_\mu \left( (\gamma_\mu \otimes \mathbb{1}) \nabla_\mu^a - \frac{r}{2} a (\mathbb{1} \otimes \mathbb{1}) \Box_\mu^a \right) \right. \\ \left. + m (\mathbb{1} \otimes \mathbb{1}) \right] \psi(n) \,. \tag{2.148}$$

Now we see explicitly that both actions contain a Wilson term, which in the case of staggered fermions is not diagonal in taste space. Then why are staggered fermions interesting? The answer lies in the degree of chiral symmetry breaking. While the usual Wilson term breaks it completely, the staggered action retains a subgroup. This turns out to be crucial, since the staggered action is protected against additive mass renormalization, and they possess some advantages over Wilson fermions when studying the spontaneous breaking of chiral symmetry, for example the phenomenon of the Aoki phase [54].

The remnant chiral symmetry arises from the simple observation that the massless staggered action (2.137) only connects even and odd points, where a lattice site $n$ is said to be even/odd if $(n_1 + n_2 + n_3 + n_4) \bmod 2 = 0/1$. Therefore, the action is invariant under the global transformation

$$\delta \chi(n) \to i\alpha \, \eta_5(n) \chi(n) \,, \\ \delta \bar{\chi}(n) \to i\alpha \, \bar{\chi}(n) \eta_5(n) \,, \tag{2.149}$$

with

$$\eta_5(n) = (-1)^{n_1 + n_2 + n_3 + n_4} \,. \tag{2.150}$$

This corresponds to an independent U(1) rotation for even and odd sites, which is precisely the subgroup of chiral symmetry that is retained. The symmetry generator $\eta_5(n)$ mixes Dirac and taste space, in particular, it is unitarily equivalent to $(\gamma_5 \otimes t_5)$.

Concerning the singlet vector and axial Ward identities, their derivation is equivalent to the naive formulation we discussed in Sec. 2.5.2, after spin diagonalizing the associated currents applying the transformation (2.133). From the naive version of the conserved vector current (2.96), we obtain

$$J_\mu(n) = \frac{1}{2a} \bar{\chi}(n)\,\eta_\mu(n)\left[U_\mu(n)\chi(n+\hat{\mu}) + U_\mu^\dagger(n-\hat{\mu})\chi(n-\hat{\mu})\right]. \tag{2.151}$$

The anomalous axial current, using Eq. (2.111), yields

$$J_{\mu 5}(n) = \frac{1}{2a} \bar{\chi}(n) \left[\prod_\nu \eta_\nu(n)\right] \eta_\mu(n) \left[U_\mu(n)\chi(n+\hat{\mu},\Xi) + U_\mu^\dagger(n-\hat{\mu})\chi(n-\hat{\mu},\Xi)\right], \tag{2.152}$$

with

$$\chi(n,\Xi) = \frac{1}{16} \sum_{\xi\in\Xi} U(n,\xi)\chi(n+\xi). \tag{2.153}$$

And finally, the pseudoscalar density from Eq. (2.112), reads

$$P_5(n) = \left[\prod_\nu \eta_\nu(n)\right] \bar{\chi}(n)\chi(n,\Xi). \tag{2.154}$$

Maintaining the subgroup of chiral symmetry came with a drawback, which is the four remaining doublers in the theory. We still have to address this fact to simulate an arbitrary number of flavors and describe QCD. Is there a way of removing the three extra doublers? Unfortunately, the answer is not known. However, staggered fermions are a very attractive discretization, since simulations are much less computationally expensive than with Wilson, overlap or domain-wall fermions. This led the lattice community to formulate what is known as the **rooting trick**. The idea is relatively simple, one takes the 4th root of the staggered determinant, hoping to get a theory with only one flavor

$$\det M \to (\det M)^{1/4}. \tag{2.155}$$

The problem lies in the taste-breaking term, which implies that the doublers are not real flavors at finite lattice spacing. Therefore, rooting has a priori no reason to work, since the order between taking the 4th root and the continuum limit does not necessarily commute, see Ref. [55] for a detailed review of the topic. We will not revisit here the arguments against rooting and how the evidence points to rooting being valid, but we will emphasize two important arguments that are relevant to this thesis.

Firstly, the rooting trick is *exact* for non-interacting fermions, since taste-breaking processes cannot occur without gluon exchanges. Therefore, free fermion results with rooted staggered fermions describe the Dirac theory in the continuum limit. The results that we will show in Chs. 4, 5 and 6 will be performed in physical QCD using rooted staggered fermions. However, these results can already be understood at a qualitative level by looking at the free theory, which serves as an important hint towards the validity of the rooting trick in our setup. In addition, in Chs. 4 and 5 there will be a crosscheck with a different discretization, Wilson fermions, which will further lessen any possible doubt about issues due to rooting in our results.

Secondly, one of the most common concerns about the rooted theory has to do with the implications for the axial anomaly, see e.g. Ref. [56]. Indications of the anomaly being correctly described by rooted staggered fermions have been carried out, for example, in the Schwinger model through the study of the $\eta'$ mass [57]. More recently, an important milestone has been achieved, by showing that rooted staggered fermions reproduce the experimental values for the $\eta - \eta'$ mass splitting in QCD at the physical point [58], a remarkable result that reinforces the current view of the staggered community on the rooting trick. This and many other successful comparisons of rooted staggered fermion results with experimental measurements constitute strong (but empirical) hints pointing toward the validity of the rooting trick. Nevertheless, the lack of a complete proof of the correctness of rooting should also make staggered practitioners alert, and always analyze possible issues that can arise from the use of this method.

To finish this section, we will discuss a general lattice procedure that also helps in dealing with taste-breaking effects: **smearing**. In general, smearing refers to a family of techniques where an average of the links in its vicinity is introduced, smoothing the UV behavior of the theory without altering the large range correlations. In the staggered formalism, taste-breaking processes are generated by gluons with momentum close to the edge of the Brillouin zone, see e.g. Ref. [48]. Therefore, smearing techniques can be used to reduce these effects, since they smooth out the contributions from high momentum gluons [59]. In particular, we will focus on a type of smearing called **stout smearing** [60]. The smeared links $U^S$, called fat links, are defined as

$$U_\mu^S(n) = e^{iQ_\mu(n)} U_\mu(n) \,, \tag{2.156}$$

with

$$Q_\mu(n) = \frac{i}{2}\left(L_\mu(n)^\dagger - L_\mu(n) - \frac{1}{3}\operatorname{tr}\bigl[L_\mu(n)^\dagger - L_\mu(n)\bigr]\right), \qquad (2.157)$$

and

$$L_\mu(n) = \left(\sum_{\nu \neq \mu} \rho_{\mu\nu}\, C_{\mu\nu}(n)\right) U_\mu^\dagger(n). \qquad (2.158)$$

The objects $C_{\mu\nu}$ are defined as

$$\begin{aligned}
C_{\mu\nu}(n) &= U_\nu(n) U_\nu(n+\hat\nu) U_\nu^\dagger(n+\hat\mu) + U_\nu^\dagger(n-\hat\nu) U_\mu(n-\hat\nu) U_\nu(n-\hat\nu+\hat\mu) \\
&\equiv S^1_{\mu\nu} + S^2_{\mu\nu},
\end{aligned} \qquad (2.159)$$

where $S^1_{\mu\nu}$, $S^2_{\mu\nu}$ are referred to as **staples** (see Fig. 2.2). The smearing procedure can be iterated over several smearing steps. The particular choice of $\rho_{\mu\nu}$ also needs to be tuned. In particular, in our staggered fermions simulations, we will use the staggered action with two stout-smearing steps and $\rho_{\mu\nu} \equiv \rho = 0.15\ \forall \mu, \nu$, which has been shown to reduce significantly the taste breaking effects [61].

With this we conclude the discussion of the discretizations that we have used to obtain the results of this thesis, which we will present in Chs. 4, 5 and 6. Another two important ingredients that appear in the investigation of the CME and the CSE, as we will discuss in depth in the next chapter, are the study of QCD thermodynamics and magnetic fields. In the next two chapters, we present how these two concepts can be introduced on the lattice.

## 2.7 Lattice QCD thermodynamics

We start by discussing the study of the thermodynamics of QCD on the lattice. As we introduced in Sec. 2.3, we are considering the Euclidean thermal field theory, where the temperature is given by the inverse number of lattice points in the temporal direction $T = 1/(aN_t)$. It is important to emphasize that in this imaginary time formalism, lattice QCD in particular, only systems in **global thermodynamic equilibrium** can be described. Therefore, we do not have access to out-of-equilibrium or real-time quantities directly.[2.17] This plays a very important role in the discussion of anomalous transport phenomena, as we will see in the next chapter. Nevertheless, the study of QCD thermodynamics in equilibrium is already a very interesting topic, with relevant phenomenological

---

[2.17] There are indirect ways to study some out-of-equilibrium quantities from Euclidean correlators, we came back to this point in Ch. 7.

applications, like the study of phase transitions in QCD. Before succinctly reviewing some relevant observables for these transitions, we will discuss an important part of the study of thermodynamics on the lattice: how to introduce a quark chemical potential in the discretized theory.

### 2.7.1 Quark chemical potential on the lattice

Considering a grand canonical ensemble, we can study QCD matter in the presence of finite temperature and a **quark chemical potential** $\mu$. The introduction of this type of chemical potential on the lattice carries some important issues and subtleties that we will present now.

The first problem is a conceptual one, and it has to do with the introduction of a chemical potential in the lattice formulation. The chemical potential enters the Lagrangian coupled to the conserved U(1)$_V$ charge

$$\mu \bar{\psi} \gamma_4 \psi \,.$$

It is then tempting to introduce it in the same way on the lattice, i.e. as a linear term in the Dirac operator. For example, for naive fermions

$$S_F^{\text{naive}} = a^4 \sum_n \bar{\psi}(n) \left[ \sum_\nu \gamma_\nu \frac{U_\nu^\dagger(n)\psi(n+\hat{\nu}) - U_\nu(n-\hat{\nu})\psi(n-\hat{\nu})}{2a} + m\psi(n) \right. \\ \left. + \mu \gamma_4 \psi(n) \right] \,. \tag{2.160}$$

However, this can be shown to lead to divergences in the free energy in the continuum limit [62]. The reason behind this is a violation of gauge invariance.[2.18] A chemical potential $\mu$ can be thought of as a constant u(1) gauge field in the temporal direction. In the continuum, the chemical potential enters in the same way as a gauge field in the covariant derivative. In the lattice formulation (2.160), this is not true anymore, a linear chemical potential is not equivalent to a U(1) gauge link. This violation of gauge invariance affects the Ward identity, leading to new divergences which are an artifact of this formulation. After realizing this fact, the solution is more apparent: introduce the

---

[2.18] This is only true in the interacting theory, where the breaking of gauge symmetry leads to the violation of the Ward identity. However, this issue is also present in the free theory, where it can be understood as a violation of the conservation law for the vector current.

chemical potential *exponentially*, in the same way as a gauge field

$$S_F^{\text{naive}} = a^4 \sum_n \bar{\psi}(n) \left[ \sum_\nu \gamma_\nu \frac{U_\nu^\dagger(n) e^{\mu \delta_{\nu 4}} \psi(n+\hat{\nu}) - U_\nu(n-\hat{\nu}) e^{-\mu \delta_{\nu 4}} \psi(n-\hat{\nu})}{2a} \right. \\ \left. + m\psi(n) \right]. \tag{2.161}$$

This is now free of any divergences that are also not present in the continuum formulation, so the lattice gauge theory at non-zero $\mu$ is well-defined.

Unfortunately, the largest obstacle is a numerical one. In general, lattice QCD simulations rely on Monte Carlo methods to solve the path integral, see App. A for a brief introduction to the topic. This implies that we need to be able to identify the fermion determinant with a probability distribution, i.e. it has to be real and positive definite. The reality condition is guaranteed if the Dirac operator has the property of $\gamma_5$-hermiticity

$$\gamma_5 M \gamma_5 = M^\dagger. \tag{2.162}$$

The Dirac operator (2.11) fulfills this condition, also in QCD. However, the continuum Dirac operator in the presence of a quark chemical potential is modified to

$$M(\mu) = \slashed{D} + m + \mu \gamma_4. \tag{2.163}$$

The new term violates $\gamma_5$-hermicity[2.19]

$$\gamma_5 M(\mu) \gamma_5 = M^\dagger(-\mu), \tag{2.164}$$

and thus the determinant is complex. This is known as **sign problem**, and forbids the usage of usual Monte Carlo techniques. There are other examples of parameters that could in principle be added to the simulations but suffer from the same problem, like a background electric field or a $\theta$-term, as well as real-time path integral simulations.

### 2.7.2 Phase transitions in QCD

One of the main goals that makes studying thermodynamics on the lattice interesting is understanding the phase diagram of strongly interacting matter. As we mentioned

---

[2.19] Although we use the continuum version for this derivation, the discretized action suffers the same issue. We also emphasize that the sign problem is completely independent of the linear or exponential introduction of the chemical potential.

in Sec. 2.5.1, chiral symmetry is spontaneously broken at $T = 0$ by the chiral condensate (2.87). It was conjectured that by increasing the temperature, the chiral condensate could be eroded until it finally disappears, restoring chiral symmetry. As a function of temperature, it has been established by lattice simulations that QCD undergoes a crossover phase transition [25] at around $T_c = 155$ MeV [26, 63].

Another important transition that can be studied is the **deconfinement transition**. As we discussed in Ch. 1, color-singlets (quarks and gluons) are never observed in nature, but confined inside hadrons. Since at high energies, asymptotic freedom implies the quarks become quasi-free, one would expect that at high temperatures the quarks and gluons get deconfined. A precise mathematical definition of confinement is a challenging task, but we can define an order parameter that relates the confined and deconfined phases in the pure SU(3) theory, the **Polyakov loop**

$$P = \frac{1}{V} \sum_n P(n) = \frac{1}{V} \sum_n \mathrm{tr}_C \left[ \prod_{n_4=0}^{N_t-1} U_4(n) \right]. \tag{2.165}$$

Let us consider two, static, test charge quarks $q, \bar{q}$ in the pure SU(3) Yang-Mills theory. It can be shown, see e.g. Ref. [31], that the correlator of two Polyakov loops is related to the potential between the static quark-antiquark pair $q\bar{q}$ separated by distance $r = a|\vec{n} - \vec{l}|$

$$\left\langle P(n) P^\dagger(l) \right\rangle = e^{-F_{\bar{q}q}(r)/T}. \tag{2.166}$$

At large distances, the decorrelation of the observables implies the factorization

$$\lim_{a|\vec{n}-\vec{l}|\to\infty} \left\langle P(n) P^\dagger(l) \right\rangle = |\langle P \rangle|^2, \tag{2.167}$$

where in the last term we have used translational invariance to omit the spatial dependence. In the confined phase, the potential has to grow indefinitely with the distance, which is only possible if $|\langle P \rangle|$ vanishes. This indicates that the Polyakov loop is an order parameter for deconfinement: $\langle P \rangle = 0$ corresponds to the confined phase, while $\langle P \rangle \neq 0$ signals the deconfined phase. In pure SU(3) gauge theory and quenched QCD, this is a first-order phase transition that occurs at a temperature around $T_c^q = 270$ MeV [64]. This deconfinement transition can be understood in terms of the **center symmetry** of SU(3). The center of a group is composed of the elements of the group that commute with all the elements. In the case of SU(3), the center group is $Z_3$, defined as the cubic roots of the identity, namely

$$Z_3 = \{\mathbb{1}, \mathbb{1}e^{2\pi i/3}, \mathbb{1}e^{-2\pi i/3}\}. \tag{2.168}$$

Using this, we can define the **center transformations** for any given time slice $\tau = n_4$ as

$$U_4(\vec{n}, \tau) \to z\, U_4(\vec{n}, \tau), \quad \text{with } z \in Z_3. \tag{2.169}$$

The discretized gauge action, consider for example the Wilson action (2.44), is invariant under this transformation, since the plaquette always closes trivially in the temporal direction. However, the Polyakov loop is not invariant

$$P \to zP. \tag{2.170}$$

Since the action is invariant, it would naively seem that the expectation value of the Polyakov loop has to vanish. However, as we have discussed before, in the deconfined phase $\langle P \rangle \neq 0$, which is only possible if the $Z_3$ symmetry disappears. From this, we conclude that the center symmetry needs to be *spontaneously broken* at high temperatures in pure SU(3) theory.[2.20] The theory chooses one of the three $Z_3$ symmetric vacua, where the Polyakov loop will have a non-zero imaginary part $\arg P = 0, \pm 2\pi/3$, corresponding to the elements of $Z_3$. The Polyakov loop and its $Z_3$ vacuum sectors will play an important role in Ch. 4, where it will affect the study of the Chiral Separation Effect in quenched QCD. As a last remark, in QCD the Polyakov loop is no longer an exact order parameter for the deconfinement transition, since the fermionic action breaks the center symmetry explicitly.

Finally, we also mention that there are many other axes in the phase diagram of strong interacting matter, that can be studied directly or indirectly using lattice QCD, for example finite baryon chemical potential [65], isospin chemical potential [66] or background magnetic fields [67], among others. Since we will mostly concentrate on observables defined at vanishing magnetic fields and chemical potentials, or in the weak $B$ regime, none of these have a direct impact on our study of anomalous transport, so we refer to the cited literature for further discussions on the topic.

---

[2.20] More precisely, the Polyakov loop expectation value is zero $\langle P \rangle = 0$ in any finite volume simulation, also at high temperatures. In order to investigate the spontaneous symmetry-breaking pattern, one can either use the module as the order parameter $\langle |P| \rangle$, or, more rigorously, introduce an external symmetry-breaking field, in order to take the thermodynamic limit of $\langle P \rangle$ and then the external field to zero.

## 2.8 Magnetic fields on the lattice

Now we continue with the other ingredient required to study anomalous transport on the lattice: the introduction of background electromagnetic fields. Let us consider QED with a single fermion flavor of electric charge $q_f = \tilde{q}_f e$,[2.21]

$$\mathcal{L}_{\text{QED}} = \bar{\psi}(x)\gamma_\mu[\partial_\mu + iq_f a_{f\mu}(x)]\psi(x) + m\bar{\psi}(x)\psi(x) + \frac{1}{4}F_{\mu\nu}(x)F_{\mu\nu}(x). \quad (2.171)$$

The discretization of this theory is completely analogous to the SU(3) case. In addition, we are ultimately interested in *background* magnetic fields, hence we will treat the photon as a classical field and we neglect the photonic kinetic term $F_{\mu\nu}F_{\mu\nu}$. This also entails that there is not a path integral over the U(1) fields. The discretized version of this action, which is just the Dirac theory coupled with classical electromagnetism, yields

$$S_{\text{EM}} = a^4 \sum_n \bar{\psi}(n) \left[ \sum_\mu \gamma_\mu \frac{u_{f\mu}(n)\psi(n+\hat{\mu}) + u_{f\mu}^\dagger(n-\hat{\mu})\psi(n-\hat{\mu})}{2a} + m\psi(n) \right], \quad (2.172)$$

where we have introduced the links $u_{f\mu}(n) \in$ U(1). Therefore, when combining QCD with electromagnetism, we just need to make the substitution

$$U_\mu(n) \to u_{f\mu}(n)\, U_\mu(n). \quad (2.173)$$

However, there is another important detail that we need to address. Actual lattice QCD simulations are always performed in a finite volume, which leads to the **quantization of the electromagnetic fluxes**. To prove this, let us consider for simplicity a 3-dimensional torus and a constant magnetic field pointing in the third spatial direction $\vec{B} = B\,\hat{e}_3$, with $L_1, L_2$ the length of the orthogonal spatial directions. The total flux is given by the surface integral of the vector field $\vec{A}$ over a given path $\mathcal{C}_1$ of area $\mathcal{A}$. Using Stokes theorem,

$$\oint_{\mathcal{C}_1} \mathrm{d}S \cdot \vec{A} = \mathcal{A}B. \quad (2.174)$$

But it is also possible to choose the different path $\mathcal{C}_2$, which encloses the complementary surface,

$$\oint_{\mathcal{C}_2} \mathrm{d}S \cdot \vec{A} = (\mathcal{A} - L_1 L_2)B. \quad (2.175)$$

---

[2.21] Notice that now the charge is introduced in the covariant derivative. This is important to treat several flavors with different electric charges.

The total flux, calculated for the two choices, differs. This apparent inconsistency is only resolved if the fields fulfill

$$\exp\left(iqB\mathcal{A}\right) = \exp\left(iqB[\mathcal{A} - L_1 L_2]\right), \tag{2.176}$$

implying

$$qB = \frac{2\pi N_b}{L_1 L_2} \qquad N_b \in \mathbb{Z}. \tag{2.177}$$

We want to emphasize that this quantization is a consequence of the finite volume, it is completely unrelated to the discretization. When we consider a theory with more than one charge, for example QCD with $u, d, s$ quarks and $\tilde{q}_u = 2/3$, $\tilde{q}_d = \tilde{q}_s = -1/3$, it is natural to define the quantization condition in terms of the (absolute value of the) smallest charge in the system. In this example we have

$$eB = \frac{6\pi N_b}{L_1 L_2} \qquad N_b \in \mathbb{Z}. \tag{2.178}$$

In Chs. 4 and 5, we will focus on lattice simulations with a background magnetic field pointing in the third spatial direction $\vec{B} = B\hat{e}_3$. We can choose the gauge $A_2 = Bx_1$, which yields the following prescription for the links [67],

$$\begin{aligned} u_{f1}(n) &= \begin{cases} e^{-ia^2 N_1 q_f B n_2} \text{ if } n_2 = N_2 \\ 1 \text{ otherwise} \end{cases} \\ u_{f2}(n) &= e^{ia^2 q_f B n_1} \\ u_{f3}(n) &= 1 \\ u_{f4}(n) &= 1 \end{aligned} \tag{2.179}$$

Notice that the links in the $x_1$-direction also obtained a spatial dependence, which originates from a gauge transformation required to fulfill boundary conditions, see Ref. [67] for the complete derivation. In Ch. 6 we will discuss a generalization of this formalism to the case of an *inhomogenous* magnetic field.

Simulations at finite magnetic field are currently a well-understood subject, where numerous studies have been performed, see Ref. [68] for a recent review. However, background *electric* fields are a more complicated variation. Not only do simulations at $E \neq 0$ suffer from a sign problem, but also the definition of thermal equilibrium has to be carefully analyzed, since naively the Lorentz force would accelerate particles indefinitely.

Since we will only require simulations at finite $B$ for this thesis, we will not discuss further the case of background electric fields and refer to Refs. [69, 70] for a detailed discussion.

## 2.9 Continuum limit

We will close this chapter by briefly discussing one of the most important concepts in lattice gauge theory: the **continuum limit**. So far, we have referred to the limit $a \to 0$ as the *naive* continuum limit, implying that there are more details hidden in this limit than is obvious at first glance. And there are indeed two important points to consider when sending the lattice spacing to zero.

The first one has to do with external parameters like temperature, volume or magnetic field. When simulating at different values of $a$, we should keep these external parameters constant, maintaining the physics of the system unchanged. This means that $L, T$ and $eB$ have to be fixed for all choices of $a$, so for example the dimensionless combination $LT = N_s/N_t$ known as **aspect ratio** has to be constant. Similarly, $qBL^2 = 2\pi N_b$ has to remain fixed, so if the volume is constant, the same $N_b$ for different values of the lattice spacing will yield the same physical magnetic field. In addition, since simulations are performed in a finite volume, they will suffer from **finite volume effects**, which are usually exponentially suppressed like $\exp(-mL)$, where $m$ is the mass gap (lowest energy state) of the theory [31]. These are only completely removed in the infinite volume limit, usually called the **thermodynamic limit**. Since this procedure can be computationally very demanding, it is usual to consider only large enough volumes, where the finite size effects are already exponentially suppressed. A usual rule of thumb for QCD simulation is that the condition $m_\pi L > 4$ needs to be fulfilled, with $m_\pi$ the mass of the pion, to have finite size effects under control [31].

The second concept has deeper roots. At the beginning of the chapter, we mentioned how lattice gauge theory is equivalent to a statistical physics model. In this spirit, this system has a correlation length $\hat{\xi}$ (in lattice units), which is controlled by the lowest energy state $m$ in the system

$$\hat{\xi} = \frac{1}{am} \equiv \frac{1}{\hat{m}}. \tag{2.180}$$

Since $m$ is a physical observable, then it should be finite in the continuum limit. Therefore, $\hat{m}$ has to vanish so that $\xi$ diverges. The divergence of the correlation length in the continuum is an indication of a second-order phase transition, implying that the continuum quantum field theory is realized at the critical point of a statistical physics model.

This is to be expected, since the divergence of the correlation length implies that the UV details of the lattice model, for example the lattice spacing, the precise discretization of the action, etc. are irrelevant in the continuum limit.

Let us consider the concrete example[2.22] of the pure SU(3) gauge theory, where the only free parameter is the bare gauge coupling $g$. Since we have established the analogy with critical phenomena, we can follow it to the end. In the vicinity of a critical point, the parameters of the theory have to be tuned to reach it. In the pure SU(3) theory, we have to tune $g(a)$ to reach the point where the correlation length $\hat{\xi}$ diverges, keeping the physical observables fixed. Let us consider a general physical observable $O(g(a), a)$ on the lattice, which is a function of the bare coupling and the lattice spacing. In the continuum limit, it yields its physical value

$$\lim_{a \to 0} O(g(a), a) = O_{\text{phys.}} \,. \tag{2.181}$$

Close enough to the continuum limit, i.e. to the critical point, the observable has to become independent of the lattice spacing. This is equivalent to satisfy the following differential equation,

$$\left[ a \frac{\partial}{\partial a} - \beta(g) \frac{\partial}{\partial g} \right] O(g, a) = 0 \,, \tag{2.182}$$

which defines the $\beta$-**function**[2.23]

$$\beta(g) = -a \frac{\partial g}{\partial a} \,. \tag{2.183}$$

This equation sets the relation between the bare coupling $g$ and the lattice spacing $a$, and how the former flows to the continuum limit. In the weak coupling regime $g \ll 1$, the $\beta$-function can be calculated perturbatively. The 1-loop result in SU(3) Yang-Mills theory theory is [30]

$$\beta(g) = -\frac{33}{48\pi^2} g^3 + \mathcal{O}(g^5) \,. \tag{2.184}$$

The negative sign reflects the asymptotic freedom property, and it shows that there is an attractive fixed point at $g = 0$. This is the critical point in the statistical physics model that corresponds to the continuum limit of the theory. In theories like QCD, asymptotic freedom guarantees that the continuum limit can be reached, see Ref. [29] for a detailed discussion.

---

[2.22] We follow here the simplified presentation in Refs. [30, 31], for a more formal analysis of the continuum limit in several theories see Ref. [29].

[2.23] This function should not be confused with the inverse gauge coupling $\beta = 6/g^2$.

What happens in theories that are not asymptotically free? The best-known examples are a scalar $\phi^4$ theory and Quantum Electrodynamics (QED), where the presence of a Landau pole raises questions about whether the cutoff can be removed or not. Lattice simulations are a great tool to tackle the fate of the continuum limit in these theories. The current consensus is that the limit $a \to 0$ cannot be taken due to the concept of quantum triviality, and they have to be regarded as effective theories where the cutoff needs to be finite but higher than any physical scale of the theory. Although a very interesting topic, quantum triviality is far beyond the scope of this thesis, see Refs. [71–73] for a discussion of the $\phi^4$ theory and Ref. [74] for an analysis of QED.

We still need a practical way of finding a mapping between the bare lattice parameters and physical observables, i.e. calculating the $\beta$-function non-perturbatively, as well as the dependence of the bare quark masses in QCD with the lattice spacing $m(a)$. This is known as **scale setting**. Although there are many different ways of doing this, we concentrate here only on a particular example using hadron masses with two degenerate $u, d$ quarks and a heavier $s$, a setup usually referred to as "$2+1$ simulations", following Ref. [27]. For a review of other methods, like the Sommer parameter or the Wilson flow, see e.g. Ref. [75]. We consider quantities whose experimental values are known, in particular the mass of the pion[2.24] and its decay constant, as well as the mass of the charged kaon and its respective decay constant [76]. Due to its quark content, pion-related quantities can be used to set the (degenerate mass) of the $u$ and $d$ quarks $m_{ud}$, while the kaonic ones can be used for setting the $s$ quark mass $m_s$. We start by defining the dimensionless ratio

$$\mathcal{R}_{ud} = \frac{m_\pi^{\text{exp}}}{f_\pi^{\text{exp}}} \approx 1.069 \,. \tag{2.185}$$

This quantity can also be calculated on the lattice at different values of $am_{ud}$ at a fixed value of $\beta$,

$$R_{ud}(am_{ud}) = \frac{am_\pi}{af_\pi} \,. \tag{2.186}$$

By extrapolating to the point where $R_{ud}(am_{ud}^*) = \mathcal{R}_{ud}$, we find the physical value $am_{ud}^*$ for a given $\beta$. Having tuned the light quark masses, we can turn to the strange quark. In this case, we define the ratio

$$\mathcal{R}_s = \frac{m_\pi^{\text{exp}}}{f_K^{\text{exp}}} \approx 0.868 \,. \tag{2.187}$$

---

[2.24] With two degenerate light quarks, the charged and neutral pions are degenerate.

We can again calculate this ratio on the lattice, using the value $am_{ud}^*$ determined before,

$$R_s(am_s) = \frac{am_\pi(m_{ud}^*)}{af_k(m_{ud}^*, m_s)} \ . \tag{2.188}$$

The physical value $am_s^*$ for the $s$-quark mass in lattice units is given by $R_s(am_s^*) = \mathcal{R}_s$. Once both quantities have been set to their physical values in lattice units, we can determine the value of the lattice spacing $a$, for example using the kaon decay constant determined on the lattice $af_K$, and taking the ratio with its experimental value $a = f_K/f_K^{\text{exp}}$.

As a last comment, one could consider the case of simulations at finite magnetic field, which will be the main focus of this thesis. Do we need an LCP at finite $B$? This seems challenging, if not impossible, since experimental measurements of the masses of the particles are not known in a finite magnetic field. However, it can be shown using staggered fermions that the scale setting procedure, using for example the Sommer parameter, is not affected by using background magnetic fields (given that they are weak enough) and theoretical expectations like the dependence of the mass of the pion with the magnetic field can be reproduced with a $B = 0$ LCP [67]. In the Wilson fermions case, the situation is more challenging, since the additive renormalization of the mass depends on $B$ [38]. For unimproved Wilson quarks, the $\kappa$ has to be tuned for different values of the magnetic field, to reduce lattice artifacts. We note that in Chs. 4 and 5, in our quenched Wilson simulations, we checked that this tuning does not have a noticeable impact on our observables, and hence we will not consider this case when presenting the results.

# Chapter 3

# Anomalous transport phenomena

In this chapter we will present in detail the concept of anomalous transport phenomena. We will start by reviewing the chiral anomaly and its subtleties, which are of particular relevance for the posterior study of anomalous transport effects. We will then move on to discuss the two particular phenomena we have studied throughout this thesis, the Chiral Magnetic Effect (CME) and the Chiral Separation Effect (CSE).

## 3.1 A closer look into the chiral anomaly

In Chap. 2, we succinctly reviewed the symmetries of QCD, and we discussed that the $U(1)_A$ symmetry is anomalous. Let us now dive into the details of this statement. Both global and gauge symmetries can be anomalous, although in this thesis we will only focus on the case of global symmetries.[3.1] In particular, we will focus on the case of the axial $U(1)_A$ anomaly.

Historically, this anomaly has its roots in an apparent paradox. In 1967, D. G. Sutherland [77] and M. Veltman [78] showed that the decay of neutral pions into two photons $\pi^0 \to \gamma\gamma$ should be forbidden if *both* the vector and the axial currents are conserved. This was in contradiction with the experimental evidence available at that time, since it was known that the process does indeed occur in nature. Since the $u$ and $d$ quarks were not heavy enough to explain the violation of the axial symmetry, this came to be known as the Sutherland-Veltman paradox. The only plausible solution was that

---

[3.1] For a gauge theory to be well-defined, the gauge symmetry itself cannot be anomalous. This leads to the topic of anomaly cancellations, which plays a central role in the Standard Model.

the axial U(1) was not a symmetry of the quantum theory, leading to the formulation of the axial anomaly.

There are famously two ways of deriving this anomaly, among many other possibilities. One is via the non-invariance under axial transformations of the fermionic integration measure, derived by Kazuo Fujikawa in 1979 [47]. The other method, which was chronologically the first, is considering the so-called **triangle diagrams** which was developed independently by S. L. Adler [79] and J. S. Bell and R. Jackiw [80] in 1969 (for this reason the axial anomaly is sometimes referred to as the ABJ anomaly). We will focus on this last method, since it will give us a useful insight to later discuss anomalous transport.

### 3.1.1 Triangle diagram

In this section, we consider the Minkowski signature and, to simplify the calculation, the case of the abelian anomaly in QED. Following the work by ABJ, we will consider the following three-point function

$$iT^{\alpha\mu\nu} = \int \mathrm{d}^4x\,\mathrm{d}^4y\,\mathrm{d}^4z\,e^{-ipx}e^{iq_1y}e^{iq_2z}\,\langle J_5^\alpha(x)J^\mu(y)J^\nu(z)\rangle\,. \tag{3.1}$$

The benefit of studying this general expression is twofold, since it will allow us to investigate the conservation equations of both the U(1)$_V$ and U(1)$_A$ currents, as well as its interplay, since

$$\text{Ward Identity:} \quad \partial_\mu J^\mu \quad \leftrightarrow \quad q_\mu^1 T^{\alpha\mu\nu}\,, \tag{3.2}$$

$$\text{Axial Ward Identity:}\, \partial_\alpha J_5^\alpha \quad \leftrightarrow \quad p_\alpha T^{\alpha\mu\nu}\,, \tag{3.3}$$

There are two Feynman diagrams contributing to this amplitude at the one-loop level:[3.2]

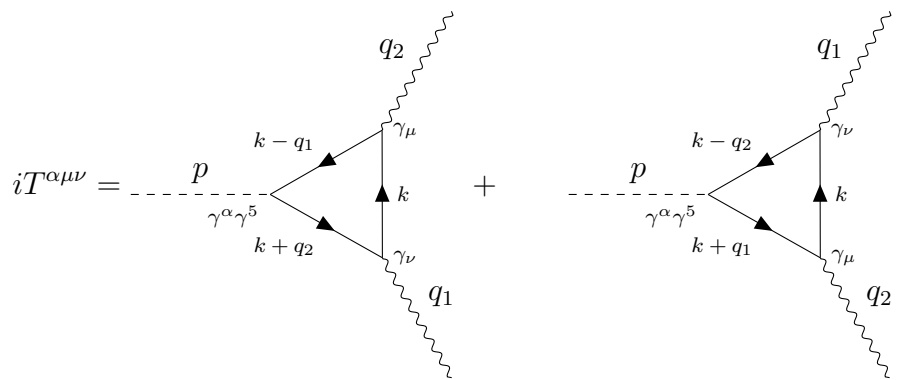

---
[3.2] One of the most interesting features of the axial anomaly is that it does not receiver higher order corrections, i.e. it is a 1-loop *exact* result.

where we have used momentum conservation $p = q_1 + q_2$. Using the Feynman rules of QED, we obtain

$$\begin{aligned} iT^{\alpha\mu\nu} &= -\int \frac{\mathrm{d}^4 k}{(2\pi)^4} \, \mathrm{tr}_D \left[ \gamma^\mu \frac{i}{\slashed{k}-m} \gamma^\nu \frac{i}{\slashed{k}+\slashed{q}_2-m} \gamma^\alpha \gamma^5 \frac{i}{\slashed{k}-\slashed{q}_1-m} \right. \\ &\quad \left. + \begin{pmatrix} \mu \leftrightarrow \nu \\ q_1 \leftrightarrow q_2 \end{pmatrix} \right] \\ &= i\int \frac{\mathrm{d}^4 k}{(2\pi)^4} \left[ \frac{\mathrm{tr}_D\left(\gamma^\mu[\slashed{k}+m]\gamma^\nu[\slashed{k}+\slashed{q}_2+m]\gamma^\alpha\gamma^5[\slashed{k}-\slashed{q}_1+m]\right)}{(k^2-m^2)([k+q_2]^2-m^2)([k-q_1]^2-m^2)} \right. \\ &\quad \left. + \begin{pmatrix} \mu \leftrightarrow \nu \\ q_1 \leftrightarrow q_2 \end{pmatrix} \right]. \end{aligned} \tag{3.4}$$

We will now illustrate a well-known yet sometimes forgotten aspect of this integral: **it requires a careful regularization despite being finite**. To see this, we consider the simplified case $m = 0$, although the argument also holds in the massive case as well. Our exposition follows closely Refs. [39, 81]. The amplitude at $m = 0$ reads

$$T^{\alpha\mu\nu} = \int \frac{\mathrm{d}^4 k}{(2\pi)^4} \left[ \frac{\mathrm{tr}_D\left(\gamma^\mu\slashed{k}\gamma^\nu[\slashed{k}+\slashed{q}_2]\gamma^\alpha\gamma^5[\slashed{k}-\slashed{q}_1]\right)}{k^2[k+q_2]^2[k-q_1]^2} + \begin{pmatrix} \mu \leftrightarrow \nu \\ q_1 \leftrightarrow q_2 \end{pmatrix} \right]. \tag{3.5}$$

Let us first consider the case of the Ward identity. For that, we contract with $q^1_\mu$ yielding

$$\begin{aligned} q^1_\mu T^{\alpha\mu\nu} &= \int \frac{\mathrm{d}^4 k}{(2\pi)^4} \left[ \frac{\mathrm{tr}_D\left(\slashed{q}_1\slashed{k}\gamma^\nu[\slashed{k}+\slashed{q}_2]\gamma^\alpha\gamma^5[\slashed{k}-\slashed{q}_1]\right)}{k^2[k+q_2]^2[k-q_1]} \right. \\ &\quad \left. + \frac{\mathrm{tr}_D\left(\gamma^\nu\slashed{k}\slashed{q}_1[\slashed{k}+\slashed{q}_1]\gamma^\alpha\gamma^5[\slashed{k}-\slashed{q}_2]\right)}{k^2[k+q_1]^2[k-q_2]^2} \right]. \end{aligned} \tag{3.6}$$

Calculating the traces and simplifying the expression, we arrive at the following integral

$$q^1_\mu T^{\alpha\mu\nu} = -4i\epsilon^{\alpha\nu\rho\sigma} \int \frac{\mathrm{d}^4 k}{(2\pi)^4} \left[ \frac{(k-q_1)_\rho(k+q_2)_\sigma}{(k-q_1)^2(k+q_2)^2} - \frac{(k-q_2)_\rho(k+q_1)_\sigma}{(k-q_2)^2(k+q_1)^2} \right]. \tag{3.7}$$

Naively, it looks like we can just shift both terms under the integral, $k \to k + q_1$ in the first term and $k \to k + q_2$ in the second one, and get a vanishing result. So far, this is in accordance with our expectation, since the vector current should be conserved. Let us now turn to the AWI,

$$p_\alpha T^{\alpha\mu\nu} = \int \frac{d^4k}{(2\pi)^4} \left[ \frac{\text{tr}_D \left( q_1^\mu \slashed{k} \gamma^\nu [\slashed{k} + \slashed{q}_2] \slashed{p} \gamma^5 [\slashed{k} - \slashed{q}_1] \right)}{k^2 [k+q_2]^2 [k-q_1]} \right. \\ \left. + \frac{\text{tr}_D \left( \gamma^\nu \slashed{k} \slashed{q}_1 [\slashed{k} + \slashed{q}_1] \gamma^\alpha \gamma^5 [\slashed{k} - \slashed{q}_2] \right)}{k^2 [k+q_1]^2 [k-q_2]^2} \right]. \tag{3.8}$$

Again this can be simplified to

$$p_\alpha T^{\alpha\mu\nu} = 4i\epsilon^{\mu\nu\rho\sigma} \int \frac{d^4k}{(2\pi)^4} \left[ \frac{k_\rho (q_2)_\sigma}{k^2(k+q_2)^2} + \frac{k_\rho (q_1)_\sigma}{k^2(k-q_1)^2} \right] + \binom{\mu \leftrightarrow \nu}{q_1 \leftrightarrow q_2}, \tag{3.9}$$

where have used $p = q_1 + q_2$. This integral is zero, due to Lorentz invariance. For example, the first term in the integral needs to be proportional to $(q_2)_\rho (q_2)_\sigma$, since it has to be contracted with $\epsilon^{\mu\nu\rho\sigma}$ and $q_2$ is the only momentum present in the term, and thus vanishes. This argument holds for the other three terms. Then, surprisingly, we obtain that the axial current is also conserved. This goes against the anomalous nature of the axial current we discussed above, and it points towards an issue in this calculation.

A careful inspection reveals that the problem resides in the divergent nature of these integrals when considered individually. Let us consider again Eq. (3.7). Each of the terms seems to diverge quadratically by power counting, however the piece proportional to $k_\rho k_\sigma$ vanishes when contracted with the Levi-Civita symbol. Then we are left with

$$q^1_\mu T^{\alpha\mu\nu} = -4i\epsilon^{\alpha\nu\rho\sigma} \int \frac{d^4k}{(2\pi)^4} \left[ \frac{k_\rho(q_1+q_2)_\sigma - (q_1)_\rho (q_2)_\sigma}{(k-q_1)^2(k+q_2)^2} - \frac{k_\rho(q_1+q_2)_\sigma + (q_1)_\rho (q_2)_\sigma}{(k-q_2)^2(k+q_1)^2} \right] \\ \equiv -4i\epsilon^{\alpha\nu\rho\sigma} \int \frac{d^4k}{(2\pi)^4} [H_{\rho\sigma}(k) - H_{\rho\sigma}(k+q_2-q_1)]. \tag{3.10}$$

This integral is a difference of two *linearly* divergent integrals. Although it seems that the difference is finite, zero in particular, shifting one of the terms is a very delicate issue, that we will illustrate with a 1-dimensional example. Let us consider the function

$$f(x) = \frac{1}{1+e^{-x}}. \tag{3.11}$$

We calculate the following integral

$$I(a) = \int_{-\infty}^{\infty} dx \left[ f(x+a) - f(x) \right] = \int_{-\infty}^{\infty} dx \left[ \frac{1}{1+e^{-x-a}} - \frac{1}{1+e^{-x}} \right]. \tag{3.12}$$

The individual integrals are divergent. If we shift the first term, it looks like the total integral vanishes. However, a careful evaluation, using the definition of improper integrals via two cut-offs, reveals that the integral is non-zero

$$I(a) = \lim_{\Lambda_1 \to \infty} \lim_{\Lambda_2 \to \infty} \int_{-\Lambda_1}^{\Lambda_2} dx \left[ \frac{1}{1 + e^{-x-a}} - \frac{1}{1 + e^{-x}} \right] = a. \quad (3.13)$$

The result is not only finite but proportional to the shift. Now we see that shifting before integrating is not a legitimate operation, since it gives an incorrect result.

We can further understand this fact by considering a generic function $g(x)$ and the integral

$$\Delta(a) = \int_{-\infty}^{\infty} dx \left[ g(x+a) - g(x) \right]. \quad (3.14)$$

Let us Taylor expand $\Delta(a)$ around $a = 0$

$$\Delta(a) = \int_{-\infty}^{\infty} dx \left[ a g'(x) + \frac{a^2}{2} g''(x) + \ldots \right], \quad (3.15)$$

and integrate

$$\Delta(a) = \lim_{\Lambda_1 \to \infty} \lim_{\Lambda_2 \to \infty} \left( a[g(\Lambda_2) - g(-\Lambda_1)] + \frac{a^2}{2} [g'(\Lambda_2) - g'(-\Lambda_1)] + \ldots \right). \quad (3.16)$$

If the integral $\int_{-\infty}^{\infty} dx\, g(x)$ converges or diverges logarithmically at most, then $0 = g(\pm\infty) = g'(\pm\infty) = \ldots$ and then $\Delta(a) = 0$. However, for linearly divergent integrals, the first term does not vanish $0 \neq g(\pm\infty)$, while the higher derivatives do. Therefore the final result depends on a "surface" term

$$\Delta(a) = a \lim_{\Lambda_1 \to \infty} \lim_{\Lambda_2 \to \infty} [g(\Lambda_2) - g(-\Lambda_1)]. \quad (3.17)$$

We have purposefully displayed the formal expression with the limits, to emphasize the role of the cutoff as a regulator to arrive at this expression. In four dimensions, something similar occurs. For example the result of a linearly divergent integral like

$$\Delta^\mu(a) = \int \frac{d^4k}{(2\pi)^4} \left[ \frac{k^\mu + a^\mu}{(k+a)^4} - \frac{k^\mu}{k^4} \right], \quad (3.18)$$

is proportional to the shift $a$ and a factor coming from a surface integral [81]

$$\Delta^\mu(a) = \frac{i}{32\pi^2} a^\mu. \quad (3.19)$$

In the same way as in the 1-dimensional case, the surface term can only arise because of an implicit hard momentum cutoff.

We have established that shifting the integrals in Eq. (3.7) is not allowed under the integral and that a careful evaluation using a cutoff makes the final result proportional to the shift. The problem in the triangle diagram is that the required shift depends on the parameterization chosen for the momenta in the loop, which is arbitrary. Therefore we have to consider the most general combination to keep track of all the possibilities, which is

$$k^\mu \to k^\mu + b_1 q_1^\mu + b_2 q_2^\mu, \tag{3.20}$$

with $b_1, b_2 \in \mathbb{R}$. Now this change affects both the calculation of the WI and the AWI. It can be shown that the result for the WI after taking into account the generalized shift, see e.g. Ref. [39], is

$$q_\mu^1 T^{\alpha\mu\nu} = \frac{1}{4\pi^2} \epsilon^{\mu\nu\rho\sigma} (q_1)_\rho (q_2)_\sigma (1 - b_1 + b_2), \tag{3.21}$$

and for the AWI

$$p_\alpha T^{\alpha\mu\nu} = \frac{1}{4\pi^2} \epsilon^{\mu\nu\rho\sigma} (q_1)_\rho (q_2)_\sigma (b_1 - b_2), \tag{3.22}$$

where in both expressions the use of a cutoff is implicitly required to get a result proportional to the shift.

This already proves that, regardless of the values of $b_1$ and $b_2$, the vector and axial currents cannot be conserved at the same time. However, it also shows an inherent ambiguity in the triangle diagram. Even with a careful evaluation taking into account the shift and using an implicit hard momentum cutoff, the result is only defined up to the choice of parameterization.

Why does a cutoff not yield directly a well-defined unambiguous result? This shouldn't come as a surprise, since a hard momentum cutoff violates gauge invariance. This can already be seen in the calculation of the vacuum polarization diagram in QED, where the use of a cutoff provides an effective mass to the photon that breaks gauge invariance explicitly [32]. Hence the cutoff does not guarantee fulfilling the WI, nor reproducing the correct axial anomaly.

To preserve the WI, we need a *gauge invariant* regulator. This provides an extra constraint in the theory that eliminates the ambiguity, recovering the correct form of the anomaly. The most used regulator in this category is dimensional regularization. However, this option comes with some technical drawbacks, since the $\gamma_5$ is inherently a 4-dimensional object. Although these issues can be circumvented, for this particular case it is more convenient to use the Pauli-Villars (PV) regularization. As a reminder, in the PV regularization we add three extra fields to the theory[3.3] with coefficient $c_s$ and masses $m_s$ [82]. Reserving $s = 0$ for the physical field with mass $m$, a possible choice of the parameters is

$$c_0 = c_1 = 1, \quad c_2 = c_3 = -1, \\ m_0^2 = m^2, \quad m_1^2 = m^2 + 2\Lambda^2, \quad m_2^2 = m_3^2 = m^2 + \Lambda^2. \tag{3.23}$$

The regulator is removed by taking the limit $\Lambda \to \infty$, when the extra fields get infinitely heavy and decouple from the theory.

This regulator[3.4] automatically chooses the combination $b_1 - b_2 = 1$, satisfying the WI and the anomalous character of the axial current

$$q_\mu^1 T^{\alpha\mu\nu} = 0 \tag{3.24}$$

$$p_\alpha T^{\alpha\mu\nu} = \frac{1}{4\pi^2} \epsilon^{\mu\nu\rho\sigma} (q_1)_\rho (q_2)_\sigma, \tag{3.25}$$

Although we will not reproduce here the details of the calculation, see e.g. Refs. [81, 82], it is worth mentioning a detail about the anomaly in the PV regularization. In this case, the anomalous term arises from a contribution coming solely from the regulator

$$\sum_{s=1}^{3} p_\alpha T^{\alpha\mu\nu}(m_s) \xrightarrow{\Lambda \to \infty} \frac{1}{4\pi^2} \epsilon^{\mu\nu\rho\sigma} (q_1)_\rho (q_2)_\sigma. \tag{3.26}$$

This is analogous to the case of Wilson fermions that we discussed in Sec. (2.6.1), where the anomaly appears from the Wilson term in the continuum limit, i.e. when the doublers get infinitely heavy and decouple.

We present now a summary of the axial anomaly via the triangle diagram:

- The triangle diagram yields a finite but ill-defined integral that needs to be regularized.

---

[3.3] This choice is enough to regularize the Dirac theory or QED, but the precise number of required extra fields depends on the theory and the number of dimensions.

[3.4] The Pauli-Villars regularization breaks chiral symmetry explicitly, hence it should only be applied to the massive theory, where this symmetry is already broken by the physical fermion mass. To connect to the $m = 0$ case, we drop the term proportional to the physical fermion mass in Eq. (3.25).

- A hard momentum cutoff violates gauge invariance and therefore is not suited for anomaly-related quantities since it does not eliminate the ambiguities and does not reproduce the correct vector and axial Ward identities.

- A gauge invariant regulator, like Pauli-Villars, reproduces the correct WI and AWI without any ambiguities.

Although everything we have presented so far has been known for a very long time, even in the original work by ABJ [79, 80] these ambiguities were already discussed, this turns out to play a very important role in the context of anomalous transport effects, in particular for the CME, contributing to some of the confusion surrounding this effect, as we will discuss below.

We will close this section with a technical comment. In the theories we are going to focus on, namely Dirac theory and physical QCD in the presence of background electromagnetic fields, it is possible to choose a gauge invariant regulator that respects the symmetries of the theory, hence resolving any possible ambiguity for the axial anomaly. However, this is not always the case in other theories. In particular, in chiral gauge theories, the choice of a regulator is a much more delicate issue. This is reflected in the *continuum*[3.5] version of the Nielsen-Ninomiya theorem [84]:

> It is not possible to construct a regularized theory that satisfies all the following properties:
>
> - Invariance under the global part of the gauge group.
> - The number of right- and left-handed Weyl fermions is different.
> - The theory reproduces the correct axial anomaly.
> - The action is bilinear in the Weyl fields.

An example of a theory where the Ward identity cannot be satisfied, making the theory inconsistent, is QED with one Weyl fermion species [39].

---

[3.5] This and the lattice theorem we presented in Sec. 2.5.2 are complementary, since a doubler-free theory can have a different number of left- and right-handed fermions. A corollary of these two theorems is that chiral and doubler-free but non-local lattice discretizations cannot reproduce the correct anomaly. An example of this is the SLAC discretization, where it can be shown that the anomaly vanishes [83].

## 3.1.2 Anomaly equation

Before discussing anomalous transport phenomena, we will analyze in more detail the axial Ward identity and some of its properties, since they will play an important role in the next section.

Let us consider the AWI in the continuum Euclidean formalism for QCD with one flavor, which takes the form

$$\partial_\mu J_{\mu 5}(x) = 2m\bar{\psi}(x)\gamma_5\psi(x) + \frac{1}{16\pi^2}\epsilon_{\mu\nu\alpha\beta}\operatorname{tr}_C\left[G_{\mu\nu}(x)G_{\alpha\beta}(x)\right]. \tag{3.27}$$

For a given gluonic configuration, which we can treat as a background field, we can take the fermionic expectation value of the fermionic operators $\langle O \rangle_F = \int \mathcal{D}\bar{\psi}\mathcal{D}\psi\, O$ and integrate the AWI over the entire volume.[3.6] The left-hand side term vanishes in the Euclidean space due to boundary conditions, yielding

$$Q_{\text{top}} = -m\int d^4x\, \langle\bar{\psi}(x)\gamma_5\psi(x)\rangle_F = -m\int d^4x\, \operatorname{tr}\left[\gamma_5\left(\slashed{D}+m\right)^{-1}(x)\right]. \tag{3.28}$$

We can now consider the spectrum of $\slashed{D}$, which is an anti-hermitian operator, therefore its eigenvalues $\lambda_n$ and eigenvectors $\phi_n$ are of the form

$$\slashed{D}\phi_n = i\lambda_n\phi_n. \tag{3.29}$$

We can then write the integrated AWI as

$$Q_{\text{top}} = -m\int d^4x \sum_n \frac{\phi_n^\dagger(x)\gamma_5\phi_n(x)}{i\lambda_n + m}. \tag{3.30}$$

Furthermore, the Dirac operator fulfills $\{\slashed{D},\gamma_5\} = 0$, implying

$$\slashed{D}(\gamma_5\phi_n) = -\gamma_5\slashed{D}\phi_n = -i\lambda_n\gamma_5\phi_n. \tag{3.31}$$

This implies that for $\lambda_n \neq 0$, the vectors $\phi_n$ and $\gamma_5\phi_n$ are orthonormal to each other, since they are eigenvectors with different eigenvalues. Therefore

$$\int d^4x\, \phi_n^\dagger(x)\gamma_5\phi_n(x) = 0 \quad \text{if} \quad \lambda_n \neq 0 \quad \forall n. \tag{3.32}$$

---

[3.6] For simplicity, we consider a finite volume, so that the spectrum is discrete.

So in the integrated anomaly, we are only left with the modes with $\lambda_n = 0$, whose eigenvectors $\phi_n$ are eigenstates of $\slashed{D}$ and $\gamma_5$ simultaneously. For obvious reasons, these states are called **zero modes**. Since $\gamma_5^2 = \mathbb{1}$, the associated eigenvalues can only be $\pm 1$. Considering that there are $n_\pm$ zero modes with eigenvalue $\pm 1$, we obtain

$$Q_{\text{top}} = n_- - n_+ \,. \tag{3.33}$$

This equality is known as the **index theorem** (or Atiyah-Singer theorem) [85]. The index theorem is a profound mathematical relation, that connects the zero modes of the Dirac operator with the topology of the gluonic fields.

Finally, we would like to connect the concept of chirality with the **helicity**, which corresponds to the projection of the spin of a particle onto its momentum. A particle where spin and momentum align is said to be right-handed and left-handed if they anti-align. This notation coincides with the chirality of a spinor, since solving the massless Dirac equation shows that a particle chirally left/right-handed is helically left/right-handed. However, this implies that an antiparticle chirally left/right-handed is helically right/left-handed, see e.g. Ref. [32]. Now we can give a precise interpretation to the operators $J_{45} = N_5 = \bar{\psi}\gamma_4\gamma_5$ and $P_5 = \bar{\psi}\gamma_5\psi$. Considering a particular representation of the Dirac matrices,

$$\gamma_4 = \begin{pmatrix} 0 & \mathbb{1}_{2\times 2} \\ \mathbb{1}_{2\times 2} & 0 \end{pmatrix}, \qquad \gamma_5 = \begin{pmatrix} -\mathbb{1}_{2\times 2} & 0 \\ 0 & \mathbb{1}_{2\times 2} \end{pmatrix}, \tag{3.34}$$

where here $\mathbb{1}_{2\times 2}$ is the $2 \times 2$ identity, and Eq. (2.63), we obtain

$$N_5 \to \bar{\psi}\gamma_4\gamma_5\psi = \psi_R^\dagger \psi_R - \psi_L^\dagger \psi_L \,, \tag{3.35}$$

$$P_5 \to \bar{\psi}\gamma_5\psi = \psi_L^\dagger \psi_R - \psi_R^\dagger \psi_L \,. \tag{3.36}$$

In terms of *helicity*, this means:

- Finite axial imbalance $N_5 > 0$ implies the creation of helically right-handed particles and antiparticles.

- Finite pseudoscalar density $P_5 > 0$ implies the creation of helically right-handed particles and helically left-handed antiparticles.

As a clarifying note, sometimes $P_5$ is referred to as chirality. To avoid any confusion, we will always refer to it as pseudoscalar density and to $J_{45} = N_5$ as axial or chiral density.

Having reviewed all the required preliminary concepts, we are ready now to present the details of anomalous transport effects.

## 3.2 Anomalous transport

The response of materials to electromagnetic fields is one of the most studied effects in physics. Let us consider the simplest case possible: the Ohmic conductivity. When a homogeneous electric field $E$ is turned on, which we can consider to be pointing in the third spatial direction $\vec{E} = E\hat{e}_3$, an electric current parallel to the field is created. The Ohmic conductivity $\sigma_E$ is defined as

$$\vec{J} = \sigma_E \vec{E} \,. \tag{3.37}$$

We can analyze the discrete symmetries of this conductivity. Considering that $\vec{J}$ is $\mathcal{P}$-even and $\mathcal{T}$-odd and $\vec{E}$ is $\mathcal{P}$-even and $\mathcal{T}$-even, we have that $\sigma_\mathrm{E}$ is $\mathcal{P}$-even and $\mathcal{T}$-odd. This is to be expected, since the conduction of electricity is a dissipative process and therefore thermodynamically irreversible.

Can a homogeneous magnetic field $\vec{B} = B\hat{e}_3$ transport charges in the same way? Let us write the equivalent expression,

$$\vec{J} = \sigma_B \vec{B} \,. \tag{3.38}$$

In this case the magnetic field $\vec{B}$ is $\mathcal{P}$-odd and $\mathcal{T}$-odd, therefore the conductivity is $\mathcal{P}$-odd. This would be forbidden in a theory with no $\mathcal{P}$- or $\mathcal{CP}$-violations, like QCD or QED. Again this is not a surprise, since this can already be understood from Maxwell's equations: a homogeneous magnetic field does not create a current. However, let us take a closer look at the Lorentz structure of this expression [68]. We can write it as

$$J_\mu \propto \epsilon_{\mu\nu\alpha\beta}\, eF_{\nu\alpha} W_\beta \,, \tag{3.39}$$

where $W_\beta$ needs to be a pseudovector, which we can identify with the axial current $W_\beta = J_{\beta 5}$. Specifying $F_{21} = B$, we obtain

$$J_3 \propto \epsilon_{321\beta} eF_{21} J_{\beta 5} \,. \tag{3.40}$$

Now we see that this current is possible if we have a non-zero axial density $J_{45} = N_5$, i.e. an axial imbalance between right- and left-handed particles. However, is it possible to have this imbalance in QCD?

In the Minkowski formalism, where there is real-time evolution, the axial imbalance can be generated *dynamically* through the anomaly equation. Let us consider a particular gluonic configuration with non-trivial topology, which again can be thought of as a background field. Neglecting the fermion mass contribution, the AWI implies

$$\frac{\mathrm{d}N_5(t)}{\mathrm{d}t} = \int \mathrm{d}^3x \, \epsilon_{\mu\nu\alpha\beta} \, \mathrm{tr}_C[G_{\mu\nu}(x)G_{\alpha\beta}(x)] \,. \tag{3.41}$$

Since the considered fields have non-trivial topology, the right-hand side is non-zero and it generates an axial imbalance over the time evolution of the system. However, as we discussed in Ch. 1, strong interactions conserve both $\mathcal{P}$ and $\mathcal{CP}$ symmetries in expectation value. This implies that these topological fluctuations have a zero expectation value averaged over all the space. However, we can have *localized* violations of these symmetries, i.e. local topological fluctuations that get converted into axial imbalance through the anomaly [86].

However, in the Euclidean formalism, the situation is different. As pointed out in Sec. 3.1.2, the integral of the left-hand side in the AWI vanishes when integrated over all the space. This is to be expected, since there is no time evolution in this setup in the Minkowskian sense. Therefore, the anomaly cannot transfer axial imbalance from the gauge fields dynamically. In this case, the imbalance has to be introduced *ad hoc*,[3.7], by considering a **chiral chemical potential** $\mu_5$, that couples to the axial density $J_{45}$

$$\mu_5 \bar{\psi} \gamma_4 \gamma_5 \psi \,.$$

We will discuss further details about $\mu_5$ in the next section.

From this discussion we learn that it is in principle possible to create these $\mathcal{P}$- and $\mathcal{CP}$-odd environments in QCD due to the axial anomaly, and it is in these systems where novel currents can appear, which are referred to as **anomalous transport effects**. In the example that we presented in Eq. (3.40), the electric current created at finite magnetic

---

[3.7] We want to emphasize that, following the analysis of the Lorentz structure in Eq. (3.40) it is not possible to create a vector current in a static setup at finite $B$ and $Q_{\mathrm{top}}$, since the topological charge is a pseudoscalar.

field and chiral imbalance is called **Chiral Magnetic Effect** (CME) [10]

$$\vec{J} = \sigma_{\text{CME}}(\mu_5)\,\vec{B}\,. \tag{3.42}$$

We could continue our quest and wonder further: is it also possible to create an *axial* current from a magnetic field? Let us write down again the Lorentz expression,

$$J_{35} \propto \epsilon_{321\beta} e F_{21} W_\beta\,. \tag{3.43}$$

Now we need $W_\beta$ to be a vector, so we would need finite quark density, i.e. a quark chemical potential $\mu$. When combined with a magnetic field, it yields an axial current which is known as **Chiral Separation Effect** (CSE) [12, 13]

$$\vec{J}_5 = \sigma_{\text{CSE}}(\mu)\,\vec{B}\,. \tag{3.44}$$

The relation of this phenomenon with the anomaly is slightly more obscure and will be clarified later. However, the CSE can be seen as a complimentary effect to CME, a relation that can be summarized as follows

$$\begin{pmatrix}\vec{J}\\ \vec{J}_5\end{pmatrix} = \begin{pmatrix}\sigma_{\text{E}} & \sigma_{\text{CME}}\\ ? & \sigma_{\text{CSE}}\end{pmatrix}\begin{pmatrix}\vec{E}\\ \vec{B}\end{pmatrix}\,. \tag{3.45}$$

An attentive reader probably has realized that there is one combination missing, the generation of an axial current by an electric field. This is called the Chiral Electric Separation Effect (CESE) [87, 88], and it requires finite quark and chiral density. The CESE is not directly related to the anomaly, it arises similarly to the usual Ohmic conductivity.

Although in this thesis we will concentrate on the CME and the CSE, which are effects that appear due to a magnetic field, it is also possible to have anomalous transport arising from the vorticity $\vec{\omega} \sim \vec{\nabla} \times \vec{v}$, where $\vec{v}$ is the flow velocity of the field. The most famous example is the Chiral Vortical Effect (CVE) [15–17], the generation of an electric current in the presence of vorticity and both chiral imbalance and finite density. This effect is the equivalent of the CME for rotating systems, where $\vec{\omega}\mu$ plays the role of the magnetic field in the Chiral Magnetic Effect. Following the same logic as before, we can guess that there is an axial equivalent of the CVE, the Axial Vortical Effect (AVE) [89, 90]. However, this effect carries a surprise: it exists at finite vorticity and chiral or quark density, but it has a thermal contribution that depends solely on the vorticity. This term has been linked to a different anomaly, the *gravitational* anomaly [91], whose detailed discussion is beyond the scope of this thesis.

Another possibility is to have collective excitations that arise from anomalous transport phenomena. For example, the interplay of CME and CSE can create the Chiral Magnetic Wave (CMW) [14] and the CVE can create the Chiral Vortical Wave (CVW) [92]. These effects arise from the mutually induced fluctuations of the axial and quark density, yielding propagating collective waves.

One of the most appealing features of these effects is that they are suited for experimental detection, making them candidates to probe the topological nature of QCD. Many of these phenomena are the subject of an intense experimental search, see Ref. [18] for a recent review. A particular set of experiments where the search for anomalous transport is relevant is heavy-ion collision experiments. These environments serve as a laboratory to test our understanding of the strong interactions, and they are expected to possess the conditions for anomalous transport to appear. Several of these experiments have aimed at the detection of the CME: the STAR collaboration in RHIC [93–100], and CMS [101] and ALICE [102–104] in the LHC. Unfortunately, a conclusive signal has remained elusive so far.

Another interesting system to look for the CME is Weyl-semimetals, an exotic type of new material whose excitations correspond to Weyl fermions. It has been observed in these materials that the Ohmic conductivity gets enhanced at finite magnetic field, in the direction parallel to it. This has been interpreted as an indirect signal of the CME [105].

Unfortunately, the CSE is a more challenging phenomenon to detect in heavy-ion collisions, since it does not generate a conserved current. However, its study is relevant for the detection of the CMW in HICs [106]. See also Ref. [107] for an idea to detect the CSE in Weyl-semimetals.

As one can already intuit from this presentation, there is a whole zoo of anomalous transport phenomena that include the discussed effects and many more. However, the main goal of this thesis is to study the CME and the CSE, already a challenging task, so we turn now to discuss the more precise aspects of the theoretical foundations of these effects.

## 3.3 Chiral Magnetic Effect

We will start with the formulation of the Chiral Magnetic Effect. Since its introduction in Ref. [10], the CME has been treated within a wide variety of theoretical approaches in the presence of strong interactions, including QCD models and effective theories [108] (see the review [109]), holography [110–112], hydrodynamics [113, 114], kinetic theory [115] and lattice simulations [116–120], see also the lattice studies looking for indirect effects of the CME [121–127]. It has also been argued that the CME plays a role in cosmology [128], e.g. for neutrino transport in supernovae [129].

The CME can be intuitively understood using a very simplified system, depicted in Fig. 3.1. Let us explain the different steps in the picture:

(0) - We consider a system of massless up and down quarks, with a defined helicity. The spin of the particles is depicted with a blue arrow, while the momentum is shown as a red one.

(1) - We turn on a homogeneous and static magnetic field, pointing without loss of generality in the third spatial direction. The spin will then (anti-align) align with the magnetic field for (negative) positively charged particles.

(2) - If the system is chirally imbalanced, which here we denote by a finite $\mu_5$, there will be more right-handed particles than left-handed particles (or vice versa).

(3) - The positive charged particles flow in the direction of the magnetic field, while the negatively charged particles flow in the opposite direction, thus separating charges and creating an electric current. This is the Chiral Magnetic Effect.

This argument is of course very naive, since it is a classical argument: quarks are considered to be massless billiard balls directed solely by the magnetic fields. In reality, for example, quarks are massive particles. Also quantum fluctuations have to be taken into account, using a quantum field theory, which requires renormalization. These points lead to many subtleties, both in the continuum and on the lattice, that we will present below.

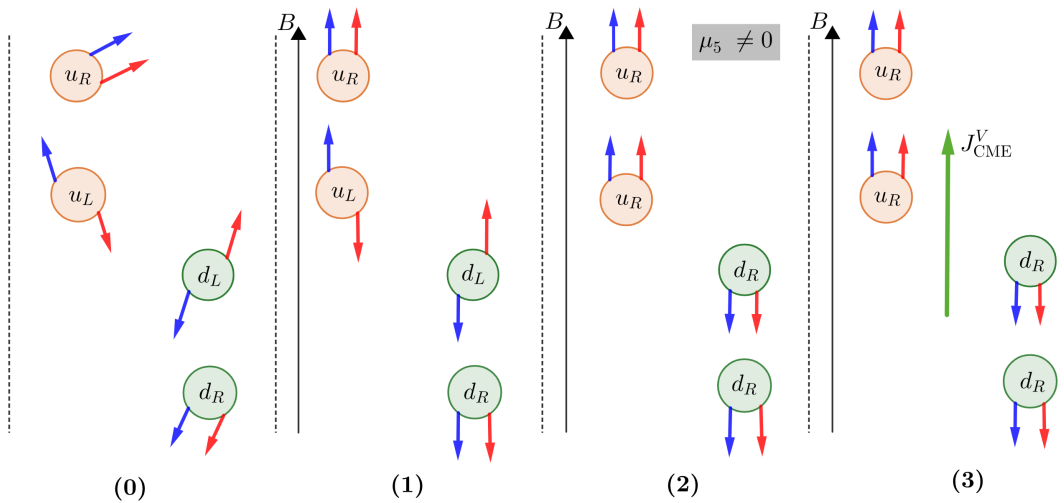

FIGURE 3.1: Illustration of the Chiral Magnetic Effect for a system of *u* and *d* quarks. The chiral imbalance is depicted by a chiral chemical potential $\mu_5$, which is taken to be positive so the number of right-handed particles is enhanced. When a magnetic field is present, particles with different charges get separated creating the CME.

### 3.3.1  CME with non-interacting fermions

The CME can be expressed in terms of its conductivity, which we define as the linear term with respect to the magnetic field

$$J_3 = \sigma_{\text{CME}}(\mu_5)\, eB_3 + \mathcal{O}(B^3)\,. \tag{3.46}$$

The full conductivity can be Taylor-expanded in the chiral chemical potential,

$$J_3 = C_{\text{CME}}\, \mu_5\, eB_3 + \mathcal{O}(B^3, \mu_5^3)\,. \tag{3.47}$$

We will concentrate our analysis on the coefficient $C_{\text{CME}}$, which is defined at $\mu_5 = 0$, since this also will turn out to be convenient for the lattice simulations.[3.8] We can define this coefficient by taking a derivative with respect to $\mu_5$,

$$\left.\frac{\partial J_3}{\partial \mu_5}\right|_{\mu_5 = 0} = C_{\text{CME}}\, eB_3\,. \tag{3.48}$$

Let us now rederive the CME using a simple argument given in the seminal work by Fukushima, Kharzeev and Warringa in Ref. [10]. We consider a system of free massless fermions in a background electric $\vec{E}$ and magnetic $\vec{B}$, taken to be parallel. The presence

---

[3.8] Although lattice simulations at finite $\mu_5$ are possible, its implementation in the staggered formulation carries technical difficulties. We will discuss this in detail in Ch. 5.

of the electric field creates an electric current $\vec{J}$ parallel to $\vec{E}$. We consider the anomaly equation

$$\frac{\mathrm{d}N_5(\vec{x},t)}{\mathrm{d}t\,\mathrm{d}^3x} = \frac{e^2}{2\pi^2}\vec{E}\cdot\vec{B}\,, \tag{3.49}$$

which dictates the rate of change in the chiral imbalance. The energy cost of this process is given by the chiral chemical potential $\mu_5$, therefore the total energy per unit of time is $\mu_5 \mathrm{d}N_5$. If we assume no losses in the system, the energy for the chiral rate change has to come from the power delivered by the current, so we can equate both

$$e\int \mathrm{d}^3x\,\vec{J}\cdot\vec{E} = \mu_5\,\frac{\mathrm{d}N_5(t)}{\mathrm{d}t}\,. \tag{3.50}$$

Using the anomaly equation, we can rewrite this expression as

$$e\int \mathrm{d}^3x\,\vec{J}\cdot\vec{E} = \mu_5\frac{e^2}{2\pi^2}\int \mathrm{d}^3x\,\vec{E}\cdot\vec{B}\,. \tag{3.51}$$

In the limit $\vec{E}\to 0$, we have

$$\vec{J} = \frac{1}{2\pi^2}\,\mu_5\,e\vec{B}\,, \tag{3.52}$$

which is precisely the CME current. From here we can read that

$$C_{\mathrm{CME}} = \frac{1}{2\pi^2}\,, \tag{3.53}$$

a very well-known number that is ubiquitously found in the literature. This value is usually mentioned to be fixed by the anomaly, as the derivation suggests, and therefore topologically protected against corrections due to interactions, meaning that this is a 1-loop exact result as the axial anomaly. However, we want to stress here that this derivation refers to the **out-equilibrium** effect, where the chiral imbalance is dynamically generated over time through the anomaly, as discussed above.

In **global thermal equilibrium**, as described by the Euclidean formalism, where we consider a static $\mu_5$, this effect is very different. It is not uncommon to find examples in the literature where the value $C_{\mathrm{CME}} = 1/(2\pi^2)$ is derived in the thermal equilibrium, see e.g. Refs. [130, 131] and some of the derivations in Ref. [10]. This adds to a still reigning confusion concerning the nature of the CME in equilibrium. However, as we will try to convey as one of the most important points in this thesis, **the Chiral Magnetic Effect does not exist in global thermal equilibrium**. We will discuss below the mathematical aspects that lead to this conclusion, but there is a more physical way of understanding this fact. In 1949, David Bohm published a summary [132] of an unpublished result derived by Felix Bloch in the 1930s, where he proved, in the context of quantum mechanics, what is now known as **Bloch's theorem**[3.9]

> Global conserved currents cannot flow in the equilibrium ground state of a system in the thermodynamic limit.

This result was (much) later generalized to the context of quantum field theory by Naoki Yamamoto [134]. Since the CME current is the electric one, it is a conserved current and therefore cannot flow in global thermal equilibrium, at least globally. This result will be of central importance in Ch. 5, where we will study the CME in QCD using lattice simulations. However, the theorem contains several keywords that allow us to circumvent this no-go theorem. The first one is *global*, since it implies that local currents which are not volume-averaged are allowed. We will explore the possibility of the existence of a localized CME in Ch. 6. The second important word is *conserved*. If the current flowing is a *non-conserved* current, it would be again not forbidden by the theorem. This is relevant for the CSE, which corresponds to an axial current. We will discuss the consequences of this in the next section and the study of the CSE in QCD in Ch. 4. Lastly, there is another "loophole" in the theorem which is the *thermodynamic limit*. This means, for example, that finite-volume effects could make the CME non-zero. It is also important for effects involving vorticities like the CVE, since rigidly-rotating systems are only defined in a finite volume to preserve causality.

The absence of the CME in equilibrium for non-interacting fermions can be shown in many other ways, for example Refs. [118, 120, 135–137]. The crucial point is regularization, which is of central importance for the anomaly as we have discussed at the beginning of the chapter. The conductivity coefficient $C_{\text{CME}}$ can be expressed in terms of Feynman diagrams in 1-loop perturbation theory, which is exact for non-interacting fermions. For example, one could consider the triangle diagram that arises from taking derivatives of the partition function with respect to the gauge field, the magnetic field and the chiral chemical potential, corresponding to diagram (C) in Fig. 3.2. This makes transparent the relation between the CME and the anomaly through the triangle diagram. It is also possible to calculate the vacuum polarization diagram at finite $\mu_5$ [135], corresponding to diagram (A). Another possibility, corresponding to diagram (B), is the calculation of the axial vector vacuum polarization in a background magnetic field. Let us briefly discuss this last case at $T = 0$. For a more detailed discussion and the finite temperature case,

---

[3.9] Not to be confused with the other Bloch's theorem about the periodicity of the wavefunction [133].

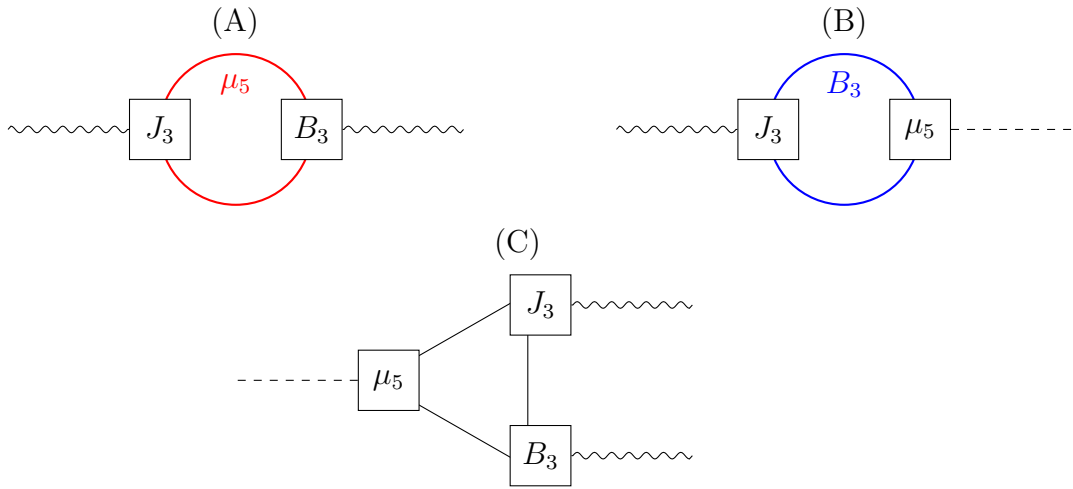

FIGURE 3.2: Feynman diagrams that can be used to calculate $C_{\text{CME}}$ in 1-loop perturbation theory (free fermions).

we refer to Ref. [138]. The calculation of the unregularized diagram yields the following integral

$$C_{\text{CME}} = \frac{1}{4\pi^3} \int_{-\infty}^{\infty} \mathrm{d}p_3 \, \mathrm{d}p_4 \, \frac{m^2 + p_4^2 - p_3^2}{(m^2 + p_4^2 + p_3^2)^2} \,. \tag{3.54}$$

It can be separated as

$$C_{\text{CME}} = (f_m + f_p) \,, \tag{3.55}$$

with

$$f_m = \frac{1}{4\pi^3} \int_{-\infty}^{\infty} \mathrm{d}p_3 \, \mathrm{d}p_4 \frac{m_s^2}{(m_s^2 + p_4^2 + p_3^2)^2} \,, \tag{3.56}$$

and

$$f_p = \frac{1}{4\pi^3} \int_{-\infty}^{\infty} \mathrm{d}p_3 \, \mathrm{d}p_4 \frac{p_4^2 - p_3^2}{(m_s^2 + p_4^2 + p_3^2)^2} \,. \tag{3.57}$$

The first term $f_m$ is a well-behaved integral, whose result is

$$f_m = \frac{1}{4\pi^2} \,. \tag{3.58}$$

However, the integral $f_p$ contains an ambiguity, it depends on the parameterization in the $p_3 - p_4$ plane. For example, performing first the $p_4$ integral yields

$$f_{p,34} = \frac{1}{4\pi^3} \int_{-\infty}^{\infty} dp_3 \left( \int_{-\infty}^{\infty} dp_4 \frac{p_4^2 - p_3^2}{(m_s^2 + p_4^2 + p_3^2)^2} \right)$$
$$= \frac{1}{4\pi^3} \int_{-\infty}^{\infty} dp_3 \frac{\pi}{2(m_s^2 + p_3^2)^{3/2}} = \frac{1}{4\pi^2}, \qquad (3.59)$$

while the other order gives a different result

$$f_{p,43} = \frac{1}{4\pi^3} \int_{-\infty}^{\infty} dp_4 \left( \int_{-\infty}^{\infty} dp_3 \frac{p_4^2 - p_3^2}{(m_s^2 + p_4^2 + p_3^2)^2} \right)$$
$$= \frac{1}{4\pi^3} \int_{-\infty}^{\infty} dp_4 \frac{-\pi}{2(m_s^2 + p_4^2)^{3/2}} = -\frac{1}{4\pi^2}. \qquad (3.60)$$

In particular, we can rotate the integration variables to parameterize the integral by an arbitrary angle $\alpha$

$$\begin{pmatrix} p_3 \\ p_4 \end{pmatrix} = \begin{pmatrix} \cos\alpha & \sin\alpha \\ -\sin\alpha & \cos\alpha \end{pmatrix} \begin{pmatrix} u \\ v \end{pmatrix}. \qquad (3.61)$$

Then the integral yields

$$f_{p,\alpha} = -\frac{\cos 2\alpha}{4\pi^2}. \qquad (3.62)$$

Taking for example $\alpha = \pi/2$, i.e. considering the ordering $f_{p,34}$, results in the value $C_{\text{CME}} = 1/(2\pi^2)$. This discussion is completely analogous to our previous analysis of the axial anomaly: the result is ambiguous without a proper regulator.

To solve this issue, we can follow the steps of the calculation of the axial anomaly and consider a gauge invariant regulator. Using the PV regulator, it can be shown that the integral gives a well-defined result,

$$C_{\text{CME}} = \frac{1}{4\pi^3} \sum_{s=0}^{3} c_s \int_{-\infty}^{\infty} dp_3\, dp_4 \frac{m_s^2 + p_4^2 - p_3^2}{(m_s^2 + p_4^2 + p_3^2)^2} \propto \sum_{s=0}^{3} c_s = 0. \qquad (3.63)$$

The gauge invariant regulator yields the value $C_{\text{CME}} = 0$, which is consistent with Bloch's theorem. This derivation shows that the CME suffers from the same issues as the axial anomaly, and that regularization plays a central role in the theoretical study of the Chiral Magnetic Effect.

Before continuing to the discussion of the CME in QCD, we will briefly comment on a technical issue that is especially relevant for the holography community studying the CME. The form of the anomaly we have derived in Sec. 3.1.1 is known as the **consistent anomaly** of the abelian theory. In simplified terms, this just means that the currents can be obtained through variations of the partition function with respect to gauge vector and axial gauge transformations.[3.10] However, it can be shown that the Ward identities derived in this way do not transform covariantly under vector and axial gauge transformations. Although there is nothing wrong with this, it is just a consequence of the anomaly, covariant expressions can be achieved if desired by adding local terms to the currents. These are known as **covariant anomalies**. The price to pay is that the new currents cannot be obtained through derivatives of the partition function. The covariant anomaly has an interesting consequence for the CME. It can be shown that in a chiral gauge theory, the covariant form of the anomaly yields an equilibrium CME current with $C_{\rm CME} \neq 0$ [90, 140]. However, this vector current is *not* conserved anymore. We emphasize that here we do not consider these types of covariant anomalies, and we follow the physical definition of the vector current that is conserved and can be obtained as a derivative with respect to the vector gauge field of the partition function.

### 3.3.2 CME in QCD

Having clarified the CME for non-interacting fermions, we can turn on the strong interactions. One would expect that Bloch's theorem still holds in QCD, since its generalization also works in the presence of interactions. Some studies find that this is indeed the case in certain interacting systems, see Ref. [141]. However, there are lattice simulation results pointing in a different direction. As opposed to the case of a quark chemical potential, simulations at finite $\mu_5$ are free of a sign problem, since the Dirac operator retains its $\gamma_5$-hermiticity. We can easily show this, since

$$\gamma_5(\mu_5\gamma_4\gamma_5)\gamma_5 = -\mu_5\gamma_4\gamma_5 = (\mu_5\gamma_4\gamma_5)^\dagger, \tag{3.64}$$

and then the fermionic determinant is real in the presence of a chiral chemical potential. This allows us to directly measure the CME current at finite $\mu_5$ and $B$ in lattice QCD simulations. This was done in Refs. [116, 117] employing Wilson fermions, both in the quenched approximation and QCD at larger than physical quark masses. The results are shown in Fig. 3.3. In the left figure, we can see a clear non-zero signal for the CME

---

[3.10] More precisely, the derived WI and AWI fulfill the so-called Wess-Zumino consistency condition [139].

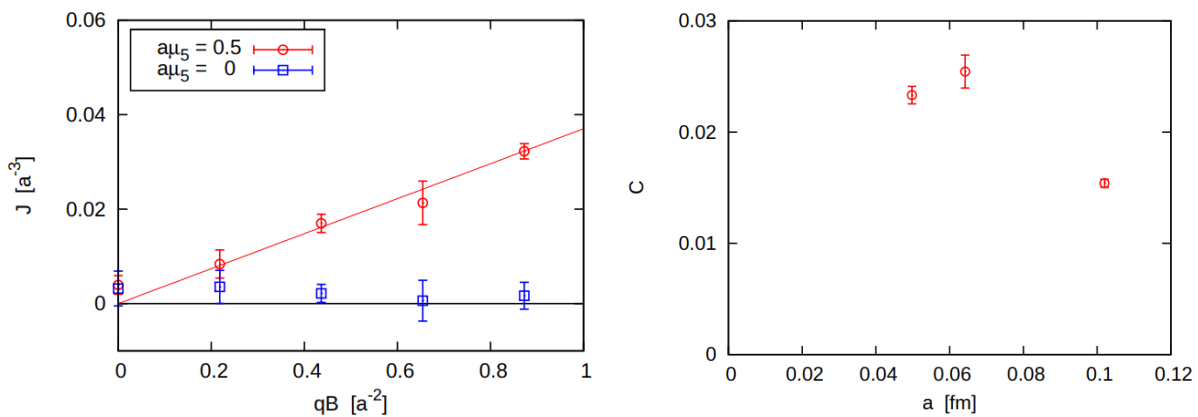

FIGURE 3.3: (Left): CME current at finite and zero $\mu_5$ as a function of the magnetic field. The slope of the linear fit of the finite $\mu_5$ points, rescaled by the appropriate factors, corresponds to $C_{\mathrm{CME}}$. The vector current used is the local version. (Right): Value of the $C_{\mathrm{CME}}$ as a function of the lattice spacing. The continuum limit was estimated to be in the range $[0.02, 0.03]$. Figures taken from Ref. [117].

current at non-zero $\mu_5$ and $B$. In the right figure, estimations for $C_{\mathrm{CME}}$ are shown in the quenched approximation as a function of the lattice spacing $a$, yielding approximately $C_{\mathrm{CME}} \approx 0.025 \approx 1/(4\pi^2) \neq 0$. This result is not only non-zero, but also not $1/(2\pi^2)$, a result that calls for clarification. We will study this in detail in Ch. 5, however we can briefly discuss the reason behind this surprising result. The vector current considered in Refs. [116, 117] is not the conserved Wilson fermion vector current (2.123). Instead, a *local* vector current

$$\bar{\psi}(n)\gamma_\mu \psi(n)$$

was used, which does not fulfill the lattice Ward identity. Although commonly used in other situations, for example to study hadron spectroscopy or usual transport effects, one can wonder if this has an implication for the CME. Bloch's theorem very suggestively indicates that this non-conserved version of the current could exist in equilibrium. In Ch. 5, we will show that this result is indeed just an artifact of the use of this current.

### 3.3.3 Chiral chemical potential

Although we have presented the chiral chemical potential in analogy with a usual quark chemical potential, it is important to clarify that $\mu_5$ is not a chemical potential in the usual thermodynamic sense, since the axial charge is not conserved, both due to the explicit breaking of chiral symmetry at finite mass and the U(1)$_A$ anomaly. Instead of a chemical potential, it is merely a parameter that controls the axial imbalance. It is possible to question whether the non-conserved nature of the axial charge allows $\mu_5$ to control the axial imbalance. We can prove this, both analytically and on the lattice. We focus on the linear term of the Taylor expansion of the axial density $N_5$ in the chiral chemical potential $\mu_5$,

$$N_5(\mu_5) = \chi_5 \mu_5 + \mathcal{O}(\mu_5^3), \qquad (3.65)$$

where $\chi_5$ is the **axial susceptibility**, which measures the linear response on the density to $\mu_5$. It is defined as

$$\chi_5^b = \frac{T}{V} \left.\frac{\partial^2 \log \mathcal{Z}}{\partial \mu_5^2}\right|_{\mu_5=0}. \qquad (3.66)$$

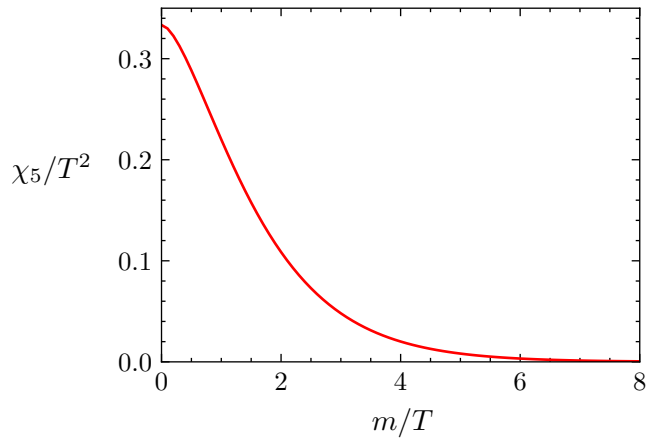

FIGURE 3.4: Analytic result (3.68) for the axial susceptibility $\chi_5$ for non-interacting fermions as a function of $m/T$. Figure from Ref. [138].

This is more convenient for a comparison with the lattice formulation, since once again this is an observable defined at $\mu_5 = 0$. For a system of free fermions, it can be calculated analytically using the PV regularization. The superscript $b$ indicates that this quantity is the bare one since $\log \mathcal{Z}$ contains a divergence in $\mu_5^2$ [142], which is a direct result of the non-conserved nature of the axial density. This divergence is temperature-independent, so we define the renormalized version of the axial susceptibility as

$$\frac{\chi_5}{T^2} = \frac{1}{T^2}\left[\chi_5^b(T) - \chi_5^b(0)\right]. \qquad (3.67)$$

For free fermions, the renormalized quantity takes then the form

$$\frac{\chi_5}{T^2} = \frac{4}{\pi^2} \int_0^\infty dp \, \frac{p^2}{\left(1+e^{\sqrt{(m/T)^2+p^2}}\right)\sqrt{(m/T)^2+p^2}} \xrightarrow{m/T \to 0} \frac{1}{3}. \qquad (3.68)$$

We refer to Ref. [142] for further details on the calculation. Its behavior as a function of $m/T$ is shown in Fig. 3.4. This proves that $\mu_5$ parameterizes the chiral imbalance for free fermions. We want to emphasize that, given these results, the arguments we presented before for the absence of CME in equilibrium are completely unrelated to a possible lack of chiral imbalance. In Ch. 5, we will show results for $\chi_5$ on the lattice, both for free fermions and in QCD, to prove that our lattice setup also possesses an axial imbalance.

We will close this section with a disclaimer. Our main focus of interest is the CME in global thermal equilibrium, which arises from the presence of a static background magnetic field and a static chiral chemical potential. This is an idealized setup, aimed at understanding the fundamental nature of this effect in QCD. We will not try to answer questions that are relevant in the experimental detection of the CME, such as how the global equilibrium is reached, or if the magnetic field in HICs lives long enough to produce the CME. We also want to emphasize that the absence of CME in equilibrium is not in contradiction with its experimental detection. For example, as we mentioned above, the CME can be indirectly detected in Weyl semimetals by measuring the Ohmic conductivity, which is an out-of-equilibrium process. This does not mean our results cannot be useful for the phenomenological and experimental high-energy physics community, especially the results we will present in Ch. 6, since the information that equilibrium QCD encodes can help build more realistic models to understand the properties of the CME in nature.

## 3.4 Chiral Separation Effect

Now we turn to the second anomalous transport phenomenon we will study in this thesis: the Chiral Separation Effect (CSE). As a reminder, it accounts for the generation of an axial current in the presence of finite density, parameterized by a quark chemical potential $\mu$, and a background magnetic field $B$. There are many approaches used to study the CSE theoretically, for example chiral kinetic theory [143], effective models of QCD [144], holography [145, 146], classical-statistical real-time simulations [119, 147], or the field correlator method [148]. There are also lattice studies, both with free fermions [118, 149] and with interactions [150, 151].

We can naively understand the CSE using the same type of argument as for the CME, which is depicted in Fig. 3.5. In this case, we consider a system of $u$ and $\bar{u}$ quarks with a given helicity. At finite *quark* density, the system has more particles than antiparticles. Thus the magnetic field separates now $u$ quarks with different chirality, hence generating an axial current.

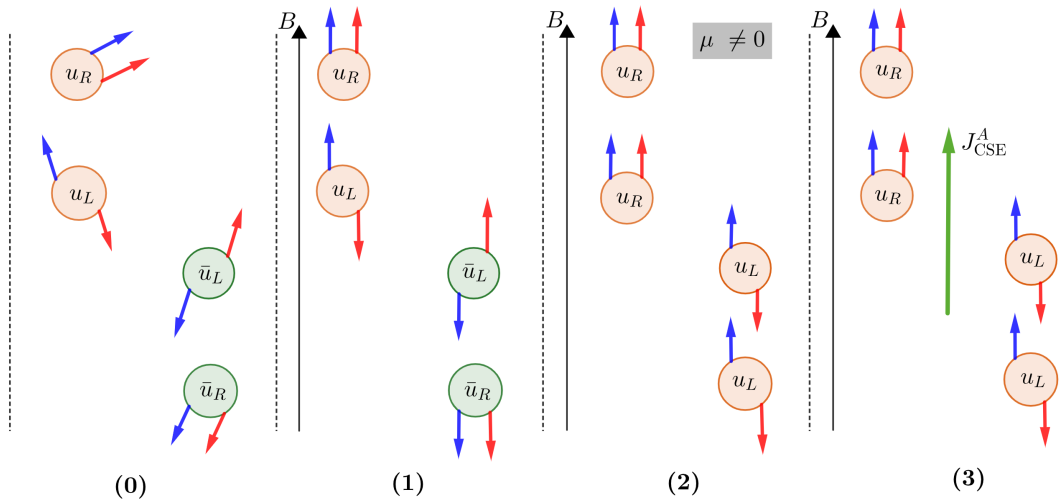

FIGURE 3.5: Illustration of the Chiral Separation Effect, now considering a system of $u$ and $\bar{u}$ quarks. The quark chemical potential $\mu > 0$ creates an imbalance of particles, which combine with the magnetic field to create the CSE.

### 3.4.1 CSE with free fermions

The CSE conductivity can be defined as

$$J_{35} = \sigma_{\text{CSE}}(\mu)\, eB_3 + \mathcal{O}(B^3)\,. \tag{3.69}$$

Analogously as for the CME, the conductivity can then be expanded in a Taylor series in the chemical potential,

$$J_{35} = C_{\text{CSE}}\, \mu\, eB_3 + \mathcal{O}(\mu^3)\,. \tag{3.70}$$

In this case, this expansion is required to avoid the sign problem in lattice QCD simulations at non-zero chemical potential. We can define $C_{\text{CSE}}$ as

$$\left.\frac{\partial J_{35}}{\partial \mu}\right|_{\mu=0} = C_{\text{CSE}}\, eB_3\,. \tag{3.71}$$

As opposed to the CME, the Chiral Separation Effect is not forbidden in thermal equilibrium by Bloch's theorem, since the axial current is not conserved. Therefore, we can concentrate solely on the equilibrium setup. In addition, we can show that the CSE is free of the ambiguities associated with the axial anomaly and the CME. For a system of non-interaction fermions, $C_{\text{CSE}}$ can be calculated using very similar diagrams as in the CME case, which we show in Fig. 3.6. In Ref. [152], we showed using diagram (B), that

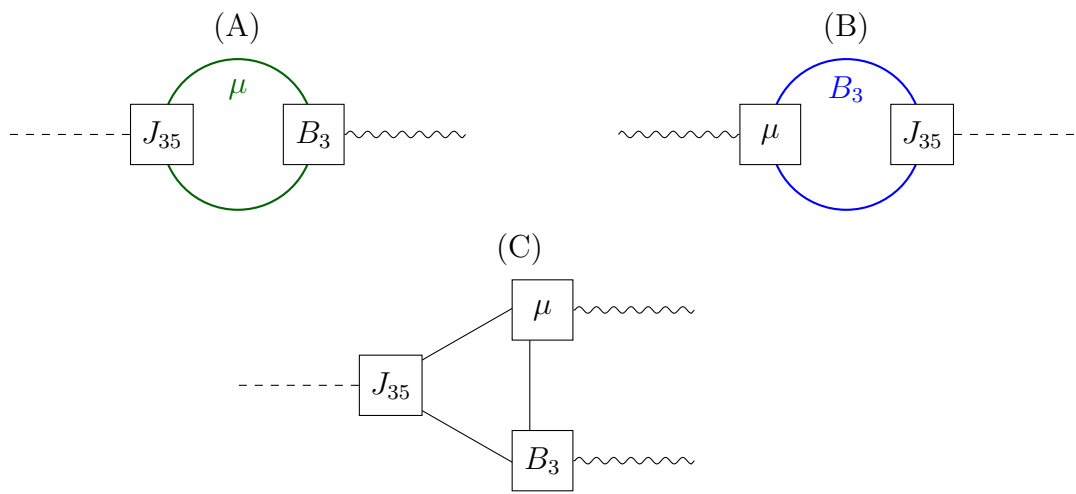

FIGURE 3.6: Diagrams that can be used to obtain $C_{\text{CSE}}$ in 1-loop perturbation theory (free fermions).

the CSE coefficient is well-defined even without regularization, and reads

$$C_{\text{CSE}}^{\text{free}} = \frac{1}{2\pi^2} \int_0^\infty \mathrm{d}p \left[1 + \cosh\left(\sqrt{p^2 + (m/T)^2}\right)\right]^{-1}. \tag{3.72}$$

This is a well-known result in the literature [12, 13], which we plot in Fig. 3.7. The conductivity starts from $C_{\text{CSE}}(m/T = 0) = 1/(2\pi^2)$ until reaching zero in the limit $m/T \to \infty$. These mass corrections and the fact that it does not require regularization, indicate that the CSE is not directly fixed by the anomaly, except for the special point $m/T = 0$, where the value $1/(2\pi^2)$ is recovered. In Ref. [148], it is shown explicitly how for the CSE, the conductivity is only fixed by the anomaly when the Dirac operator anticommutes with $\gamma_5$, i.e. when $m = 0$.

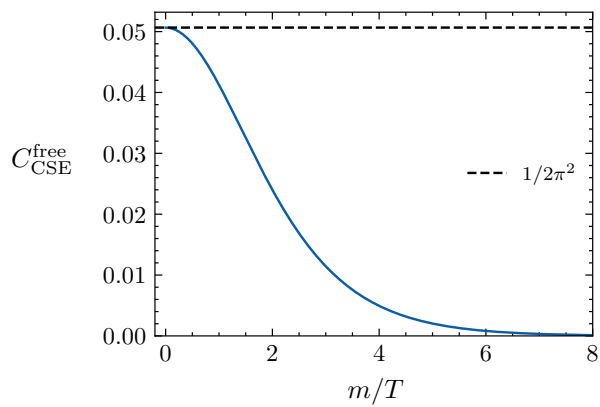

FIGURE 3.7: Analytical behavior of $C_{\text{CSE}}$ for a system of interacting fermions from Eq. (3.72), as a function of $m/T$. Figure from Ref. [152].

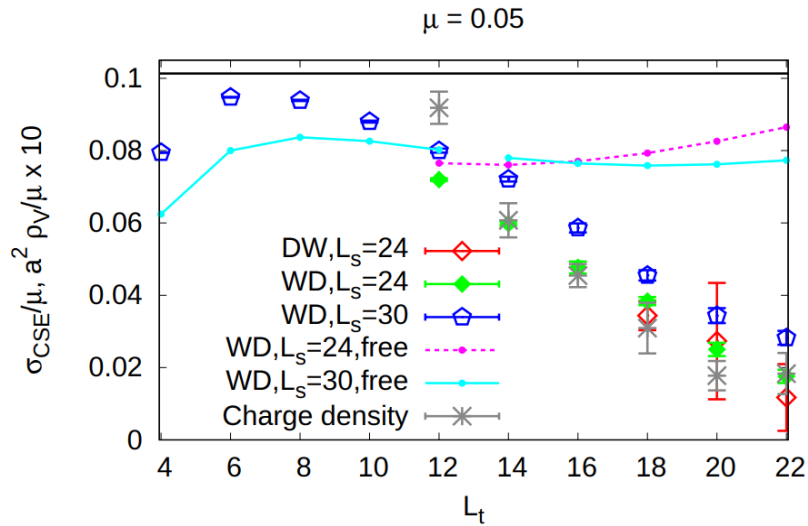

FIGURE 3.8: Results for $\sigma_{\text{CSE}}/\mu$ as a function of temperature in two-color QCD, at fixed lattice spacing. The results suggest a suppression of the CSE at low temperatures. Figure taken from Ref. [151].

### 3.4.2 CSE in QCD

In the presence of strong interactions, less is known about $C_{\text{CSE}}$. The main obstacle is that lattice simulations face the already mentioned sign problem at finite quark chemical potential. Two lattice works have employed approaches to circumvent this issue and study CSE with interactions. In Ref. [150], the authors used quenched simulations, which do not suffer from a sign problem since $\det M = \text{const.}$, to study $C_{\text{CSE}}$ with massless overlap fermions in QCD at two different temperatures, one above and one below the crossover temperature. The study found no significant corrections to the massless free case result $C_{\text{CSE}}^{\text{free}}(m/T = 0) = 1/(2\pi^2)$.

In Ref. [151], dynamical simulations were used but in two-color QCD, i.e. using the gauge group SU(2), with Wilson and Domain-Wall fermions with a heavier than physical pion mass. Using the group SU(2) instead of SU(3) has an important advantage, since the associated Dirac operator does not introduce a sign problem at finite $\mu$. This is due to an extra symmetry of the Dirac operator that guarantees that its eigenvalues always come in conjugate pairs at finite chemical potential, see e.g. Ref. [153]. This allows us to study not only $C_{\text{CSE}}$, but the whole conductivity $\sigma_{\text{CSE}}$. The results for $\sigma_{\text{CSE}}(\mu)/\mu$ at small $\mu$, which should be comparable to $C_{\text{CSE}}$ in this regime, are shown in Fig. 3.8 for a fixed lattice spacing at different values of $N_t$ (temperature dependence). These results point towards a suppression of CSE at low temperatures, which is qualitatively equivalent to the situation for free fermions, where at high values $m/T$ the coefficient $C_{\text{CSE}}$ approaches zero.

In the next chapter, we will present our results of the study of $C_{\text{CSE}}$ in dynamical QCD at the physical point [152]. We will discuss the temperature dependence of this coefficient, which is the first determination of the conductivity of an anomalous transport coefficient in physical QCD.

# Chapter 4

# Chiral Separation Effect

We start the discussion of the results with the Chiral Separation Effect. We will present the target observables and the lattice setup in detail, since the next two chapters will also rely on them. We will continue by showing results in two test cases: free fermions and quenched QCD, both with staggered and Wilson fermions. This will allow us to crosscheck our setup, before studying the CSE in dynamical QCD. To conclude, we present the main result of the chapter: the conductivity $C_{\mathrm{CSE}}$ in dynamical QCD at the physical point using staggered fermions. This chapter is based on Ref. [152].

## 4.1 CSE conductivity and chemical potentials

In Ch. 3, we discussed in detail the Chiral Separation Effect, focusing on the case of a single quark flavor. However, we aim to study the CSE in QCD, which includes more than one flavor. In particular, we will consider QCD with $2+1$ flavors, with degenerate $u$ and $d$ quarks and the heavier $s$ quark. In this case, we need to specify the type of chemical potential in flavor space. Let us consider the singlet axial current

$$J_{\nu 5}^i = \sum_f c_f^i \frac{T}{V} \int \mathrm{d}^4 x\, \bar{\psi}_f(x) \gamma_\nu \gamma_5 \psi_f(x)\,, \qquad (4.1)$$

with different quantum numbers $i = \mathcal{B}, \mathcal{Q}$ or $\mathcal{L}$ in flavor space,

$$c_f^\mathcal{B} = \frac{1}{3}\,, \qquad c_f^\mathcal{Q} = \frac{q_f}{e}\,, \qquad c_f^\mathcal{L} = \begin{cases} 1/3\,, & f = u, d\,, \\ 0\,, & \text{otherwise}\,, \end{cases} \qquad (4.2)$$

where $f = u, d, s$ labels the quark flavors and $q_f$ are the corresponding electric charges $q_u/2 = -q_d = -q_s = e/3$. These choices correspond to the current coupling to baryon number $\mathcal{B}$, electric charge $\mathcal{Q}$ or the so-called light quark baryon number $\mathcal{L}$ [154]. The latter has been used to perform lattice simulations of QCD with conditions similar to heavy-ion collision setups.

Analogously, the quark density can also be parameterized either by a baryon chemical potential $\mu_\mathcal{B}$, an electric charge chemical potential $\mu_\mathcal{Q}$ or a light baryonic chemical potential $\mu_\mathcal{L}$, which couples to the corresponding vector currents,

$$J_\nu^i = \sum_f c_f^i \frac{T}{V} \int d^4x \, \bar{\psi}_f(x) \gamma_\nu \psi_f(x) \,. \tag{4.3}$$

For the individual quark flavors, the chemical potentials are $\mu_f = \mu_\mathcal{B}/3$ in the baryonic case and $\mu_f = \mu_\mathcal{Q} q_f/e$ in the charge case, while they are set as $\mu_u = \mu_d = \mu_\mathcal{L}/3$, $\mu_f = 0$ ($f \neq u, d$) for the light quark baryon chemical potential.

In the study of the CSE, the most common choice is the quantum number $\mathcal{Q}$ for the axial current and $\mathcal{B}$ for the chemical potential. For this combination, considering a homogeneous magnetic field pointing in the third spatial direction, we can define the CSE conductivity coefficient as

$$\langle J_{35}^\mathcal{Q} \rangle = C_\text{CSE} \, \mu_\mathcal{B} \, eB + \mathcal{O}(\mu_\mathcal{B}^3, B^3) \,, \tag{4.4}$$

where we have used the simplified notation $C_\text{CSE} \equiv C_\text{CSE}^{\mathcal{QB}}$. Similarly, we define $C_\text{CSE}^{ij}$ for the different combinations of quantum numbers, where the first superscript $i \in \{\mathcal{B}, \mathcal{Q}, \mathcal{L}\}$ corresponds to the current and the second one $j \in \{\mathcal{B}, \mathcal{Q}, \mathcal{L}\}$ to the chemical potential.

As we already argued in Ch. 2, Monte Carlo simulations suffer from a sign problem at $\mu_f \neq 0$, which also extends to the cases $\mu_\mathcal{B} \neq 0$, $\mu_\mathcal{Q} \neq 0$ or $\mu_\mathcal{L} \neq 0$. This is the reason why we focus on the coefficients of Taylor expansion in the chemical potential, in particular $C_\text{CSE}$, since those can be computed at vanishing chemical potential. Following Eq. (3.71), we can obtain the CSE coefficient by taking a derivative with respect to $\mu_\mathcal{B}$ and then setting $\mu_\mathcal{B}$ to zero, yielding

$$\left.\frac{\partial \langle J_{35}^\mathcal{Q} \rangle}{\partial \mu_\mathcal{B}}\right|_{\mu_\mathcal{B}=0} = C_\text{CSE} \, eB \,. \tag{4.5}$$

This operator is to be evaluated at finite magnetic field, and $C_\text{CSE}$ can be extracted through a numerical derivative in $B$, i.e. the slope of a linear fit of the observable with respect to the magnetic field. We will come back to the specifics of this procedure below.

For each choice of the flavor structures, there is an overall normalization constant involving the quark baryon numbers or quark charges and the number of colors $N_c = 3$. By dividing this factor out in the conductivities, results for different combinations can be qualitatively compared with each other, as well as with the results for non-interacting fermions. These overall factors read

$$C_{\text{dof}} \equiv C_{\text{dof}}^{\mathcal{QB}} = C_{\text{dof}}^{\mathcal{BQ}} = \frac{N_c}{3} \sum_f \left(\frac{q_f}{e}\right)^2, \qquad C_{\text{dof}}^{\mathcal{QQ}} = N_c \sum_f \left(\frac{q_f}{e}\right)^3,$$

$$C_{\text{dof}}^{\mathcal{BB}} = \frac{N_c}{9} \sum_f \frac{q_f}{e}, \qquad C_{\text{dof}}^{\mathcal{QL}} = \frac{N_c}{3} \sum_{f=u,d} \left(\frac{q_f}{e}\right)^2, \qquad C_{\text{dof}}^{\mathcal{BL}} = \frac{N_c}{9} \sum_{f=u,d} \frac{q_f}{e}. \tag{4.6}$$

We emphasize that the different choices of flavor quantum numbers are not completely equivalent after rescaling out these factors, since for non-degenerate massive fermions each flavor contributes differently. In addition, the disconnected parts also differ for $\mathcal{B}$, $\mathcal{Q}$ or $\mathcal{L}$ in the presence of interactions. We also note that $C_{\text{dof}}^{\mathcal{BB}} = 0$ for the three lightest quarks, implying that the 1-loop result (3.72) vanishes for this combination. This will be relevant in the analysis of the different $C_{\text{CSE}}^{ij}$ in QCD in Sec. 4.3.3.

Finally, we can relate the observable in Euclidean space with the Minkowski one like

$$\text{Re} \left. \frac{\partial \langle J_{35}^{\mathcal{Q}} \rangle}{\partial \mu_{\mathcal{B}}} \right|_{\mu_{\mathcal{B}}=0}^{\text{M}} = -\text{Im} \left. \frac{\partial \langle J_{35}^{\mathcal{Q}} \rangle}{\partial \mu_{\mathcal{B}}} \right|_{\mu_{\mathcal{B}}=0}^{\text{E}}, \tag{4.7}$$

where we have used the relation between the Euclidean and Minkowski Dirac matrices given in Eq. (2.9). For the rest of the chapter, we consider the imaginary part of the Euclidean two-point function calculated on the lattice. The real part of the Euclidean observable was checked to be consistent with zero. Having discussed the available choices for the chemical potential basis, we discussed the precise form of the observable (4.5) in the two discretizations we have used, staggered and Wilson fermions.

## 4.2 Lattice setup

### 4.2.1 Staggered fermions

We start by discussing the setup with rooted staggered quarks, presented in Sec. 2.6.2, which we use to study the CSE in full dynamical QCD with up, down and strange quark flavors. The partition function for this theory is given by

$$\mathcal{Z} = \int \mathcal{D}U \exp[-S_g] \prod_f [\det M_f(U, q_f, m_f)]^{1/4}, \tag{4.8}$$

with $m_f$ the quark masses for each flavor $f = u, d, s$. The gauge action $S_g$ is the tree-level Symanzik action given by Eq. (2.46), and $M_f$ is the massive staggered Dirac operator with twice stout-smeared links, using the parameters discussed at the end of Sec. (2.6.2). The quark masses are tuned to the physical values, using the LCP obtained in Ref. [27]. The simulations are performed at a finite background magnetic field, taken to be homogeneous and pointing in the third spatial direction, which modifies the Dirac operator as we introduced in Sec. 2.8.

Regarding the chemical potential, although simulations are performed at $\mu_f = 0$, we still need to specify how this parameter enters the Dirac operator, to compute the correct Taylor coefficients. We choose an exponential introduction as in Eq. (2.161), to avoid the appearance of divergences in the theory associated with lattice artifacts. This introduction also affects the vector and axial currents, which we discussed in Sec. 2.6.2, whose generalized form we now present with a more compact notation. The conserved vector current and the anomalous anomalous axial current are the bilinears,

$$J_\nu^f(n) = \bar\chi_f(n)\Gamma_\nu^f(n,m)\chi_f(m), \qquad J_{\nu 5}^f(n) = \bar\chi_f(n)\Gamma_{\nu 5}^f(n,m)\chi_f(m), \tag{4.9}$$

involving the operators,

$$\Gamma_\nu^f(n,m) = \frac{\eta_\nu(n)}{2}\Bigg[ U_\nu(n) u_{f\nu}(n)\, e^{a\mu_f \delta_{\nu 4}} \delta_{n+\hat\nu, m} \tag{4.10}$$

$$+ U_\nu^\dagger(n-\hat\nu) u_{f\nu}^*(n-\hat\nu)\, e^{-a\mu_f \delta_{\nu 4}} \delta_{n-\hat\nu, m} \Bigg],$$

$$\Gamma_{\nu 5}^f = \frac{1}{3!} \sum_{\rho,\alpha,\beta} \epsilon_{\nu\rho\alpha\beta}\, \Gamma_\rho^f \Gamma_\alpha^f \Gamma_\beta^f, \tag{4.11}$$

with the convention $\epsilon_{1234} = +1$. Notice that the operators now depend on the magnetic field and the chemical potential, which is natural since they arise from variations of the action. In the case of the conserved vector current, it is straightforward to see that this is just the generalization of Eq. (2.151).

The anomalous axial current (4.10) is a 3-link operator that has been extensively studied in the literature, see for example Refs. [155–157]. Note that $\Gamma_{35}^f$ involves hoppings in the temporal direction and thus depends on $\mu_f$, which will turn out to be a crucial point in the calculation of $C_{\text{CSE}}$ (and $C_{\text{CME}}$ in the next two chapters). This operator also includes the average of different paths connecting the beginning and end points of the bilinear, as we anticipated in Sec 2.5.2. This is the reason behind the combinatorial factors and the order permutation given by the Levi-Civita symbol in Eq. (4.11).

However, an attentive reader would have realized that this is not the operator presented in Eq. (2.152), which is a 5-link operator as derived in the original study of the anomaly in the naive and staggered fermion formulations [34]. In the new notation, we can introduce this operator as follows. We define the $\gamma_5$ equivalent in the staggered formalism [155] as

$$\Gamma_5^f = \frac{1}{4!} \sum_{\nu,\rho,\alpha,\beta} \epsilon_{\nu\rho\alpha\beta}\, \Gamma_\nu^f \Gamma_\rho^f \Gamma_\alpha^f \Gamma_\beta^f \,. \tag{4.12}$$

This is precisely the operator appearing in the pseudoscalar density (2.154), now averaging over different paths. Then the 5-link anomalous axial current reads

$$\Gamma_{\nu 5}^{\text{5-link},f} = \Gamma_\nu^f \Gamma_5^f \,. \tag{4.13}$$

Notice that $\Gamma_{\nu 5}$ and $\Gamma_{\nu 5}^{\text{5-link}}$ are not the same operator, as opposed to for example $-\gamma_1\gamma_2\gamma_3$ and $\gamma_4\gamma_5$ in the continuum, since these objects do not form a Clifford algebra at finite lattice spacing. However, we will show that both lattice operators have the same continuum limit for the case of $C_{\text{CSE}}$.

Another interesting case we can study has to do with the introduction of the U(1) links in both the vector and axial currents. We can check the effect of not including them by considering the operators $\Gamma_\nu$ and $\Gamma_{\nu 5}$ without the U(1) links, i.e. at $B = 0$. We define

$$\begin{aligned}\Gamma_\nu^{B=0,f} &= \Gamma_\nu^f(B=0)\,, \\ \Gamma_{\nu 5}^{B=0,f} &= \Gamma_{\nu 5}^f(B=0)\,.\end{aligned} \tag{4.14}$$

Again we will show that this does not affect the continuum limit of $C_{\text{CSE}}$ for non-interacting fermions. The explicit form of the observable for free fermions in these two cases is discussed in App. B.

With these definitions, we obtain the expectation value of the axial current,

$$\langle J_{35}^\mathcal{Q} \rangle = \frac{T}{V}\frac{1}{4} \sum_f \frac{q_f}{e} \left\langle \text{Tr}\left(\Gamma_{35}^f M_f^{-1}\right) \right\rangle \,, \tag{4.15}$$

where the factor 1/4 results from rooting. Next, we can calculate the derivative (4.5) to obtain $C_{\text{CSE}}$. This yields the usual disconnected and connected terms, together with an additional *tadpole term* arising due to the derivative of $\Gamma_{35}^f$ with respect to $\mu_{\mathcal{B}}$,

$$C_{\text{CSE}}\, eB = \frac{\partial \langle J_{35}^{\mathcal{Q}} \rangle}{\partial \mu_{\mathcal{B}}}\bigg|_{\mu_{\mathcal{B}}=0} = \frac{T}{V}\Bigg[\frac{1}{16}\sum_{f,f'} c_f^{\mathcal{Q}} c_{f'}^{\mathcal{B}} \left\langle \text{Tr}\left(\Gamma_{35}^f M_f^{-1}\right)\text{Tr}\left(\Gamma_4^{f'} M_{f'}^{-1}\right)\right\rangle \\ -\frac{1}{4}\sum_f c_f^{\mathcal{Q}} c_f^{\mathcal{B}} \left\langle \text{Tr}\left(\Gamma_{35}^f M_f^{-1}\Gamma_4^f M_f^{-1}\right)\right\rangle \\ +\frac{1}{4}\sum_f c_f^{\mathcal{Q}} c_f^{\mathcal{B}} \left\langle \text{Tr}\left(\frac{\partial \Gamma_{35}^f}{\partial \mu_f} M_f^{-1}\right)\right\rangle \Bigg]. \qquad (4.16)$$

The tadpole term (the last line in Eq. (4.16)) appears due to the exponential form of introducing the chemical potential, analogously to the calculation of quark number susceptibilities for staggered quarks, see e.g. Ref. [158]. With a linear introduction of the chemical potential, which leads to additional divergences, this tadpole term would not appear. Hence the tadpole term guarantees gauge invariance in the staggered formulation, which is of vital importance in studying anomalous transport phenomena, as we discussed in the previous chapter. We also mention that in deriving (4.16) we exploited the $\mathcal{C}$-symmetry of the system at $\mu_{\mathcal{B}} = 0$, i.e. that the expectation value of the baryon density and that of the axial current vanish at $\mu_{\mathcal{B}} = 0$ (the equivalent of the term $\langle \text{Tr}\, M^{-1}\rangle^2$ in (2.56)). We will refer back to this point in Sec. 4.3.2.

The expectation values in Eq. (4.16) are to be evaluated at $\mu_{\mathcal{B}} = 0$ but nonzero magnetic field $eB$. The conductivities $C_{\text{CSE}}^{ij}$ for the other flavor quantum numbers can be calculated analogously, and differ from Eq. (4.16) by factors of $c_f^i c_{f'}^j$ in the disconnected and $c_f^i c_f^j$ in the connected and tadpole terms under the flavor sums.

### 4.2.2 Wilson fermions

Next we describe our setup involving Wilson quarks, a discretization that we only consider for free fermions and in the quenched approximation. In this case, the currents are constructed as the bilinears,

$$J_\nu^f(n) = \bar{\psi}_f(n)\Gamma_\nu^f(n,m)\psi_f(m), \qquad J_{\nu 5}^f(n) = \bar{\psi}_f(n)\Gamma_{\nu 5}^f(n,m)\psi_f(m), \qquad (4.17)$$

with the generalized version of the conserved vector current (2.123) and the anomalous axial current (2.126),

$$\Gamma^f_\nu(n,m) = \frac{1}{2}\Big[(\gamma_\nu - r)U_\nu(n)u_\nu(n)\,e^{a\mu_f\delta_{\nu 4}}\delta_{m,n+\hat{\nu}} \tag{4.18}$$
$$+ (\gamma_\nu + r)U^\dagger_\nu(n-\hat{\nu})u^*_\nu(n-\hat{\nu})\,e^{-a\mu_f\delta_{\nu 4}}\delta_{m,n-\hat{\nu}}\Big],$$

$$\Gamma^f_{\nu 5}(n,m) = \frac{1}{2}\Big[\gamma_\nu\gamma_5 U_\nu(n)u_\nu(n)\,e^{a\mu_f\delta_{\nu 4}}\delta_{m,n+\hat{\nu}} \tag{4.19}$$
$$+ \gamma_\nu\gamma_5 U^\dagger_\nu(n-\hat{\nu})u^*_\nu(n-\hat{\nu})\,e^{-a\mu_f\delta_{\nu 4}}\delta_{m,n-\hat{\nu}}\Big].$$

Notice that the chemical potential also couples to the Wilson term in Eq. (4.18), as dictated by the conservation equation of the vector current. Using the above definitions, the expectation value of the axial current reads,

$$\langle J^{\mathcal{Q}}_{35}\rangle = \frac{T}{V}\sum_f \frac{q_f}{e}\left\langle \mathrm{Tr}\left(\Gamma^f_{35}M_f^{-1}\right)\right\rangle. \tag{4.20}$$

As opposed to the staggered formulation, now $\Gamma^f_{35}$ only involves hoppings in the 3 direction, and it is independent of the chemical potential. Thus the observable does not contain a tadpole term,

$$C_{\mathrm{CSE}}\,eB = \left.\frac{\partial\langle J^{\mathcal{Q}}_{35}\rangle}{\partial\mu_{\mathcal{B}}}\right|_{\mu_{\mathcal{B}}=0} = \frac{T}{V}\Bigg[\sum_{f,f'} c^{\mathcal{Q}}_f c^{\mathcal{B}}_{f'}\left\langle \mathrm{Tr}\left(\Gamma^f_{35}M_f^{-1}\right)\mathrm{Tr}\left(\Gamma^{f'}_4 M_{f'}^{-1}\right)\right\rangle$$
$$-\sum_f c^{\mathcal{Q}}_f c^{\mathcal{B}}_f \left\langle \mathrm{Tr}\left(\Gamma^f_{35}M_f^{-1}\Gamma^f_4 M_f^{-1}\right)\right\rangle\Bigg]. \tag{4.21}$$

Again the conductivities for the other flavor quantum numbers follow similarly.

Finally, we can also consider a *local* vector current

$$J^{f,\mathrm{loc}}_\nu(n) = \bar{\psi}_f(n)\gamma_\nu\psi_f(n) \tag{4.22}$$

instead. Even though it is not conserved, it has the same quantum numbers as $J^f_\nu$ and is often employed in Wilson fermion simulations, as we already mentioned above. More importantly, this current was also used to study the CME in quenched and dynamical QCD [117], obtaining a non-zero result. Although we will analyze directly the effect of considering this local operator in the next chapter, we can already use the CSE as a case study for this current. We can introduce a chemical potential $\mu^{\mathrm{loc}}_{\mathcal{B}}$ that couples to these local currents in the action. The analogue of (4.21) now reads

$$C_{\text{CSE}}^{\mathcal{QB},\text{loc}} eB = \left.\frac{\partial \langle J_{35}^{\mathcal{Q}} \rangle}{\partial \mu_{\mathcal{B}}^{\text{loc}}}\right|_{\mu_{\mathcal{B}}^{\text{loc}}=0} = \frac{T}{V}\Bigg[\sum_{f,f'} c_f^{\mathcal{Q}} c_{f'}^{\mathcal{B}} \left\langle \text{Tr}\left(\Gamma_{35}^f M_f^{-1}\right)\text{Tr}\left(\gamma_4 M_{f'}^{-1}\right)\right\rangle \\ - \sum_f c_f^{\mathcal{Q}} c_f^{\mathcal{B}} \left\langle \text{Tr}\left(\Gamma_{35}^f M_f^{-1} \gamma_4 M_f^{-1}\right)\right\rangle \Bigg]. \tag{4.23}$$

Notice that this way of introducing the chemical potential corresponds to the linear version, and hence it is expected to lead to divergences (see the discussion in Sec. 2.7). We will test the impact of these ultraviolet divergences on the observable (4.23) below.

## 4.3 Results

Having discussed the form of the observables, we present now the obtained results. To estimate the traces appearing in Eqs. (4.16), (4.21) and (4.23), we use the standard noisy estimator technique, described in App. A. We considered $N_V = 100$ Gaussian noise vectors, which were found to be sufficient to reliably calculate the observables. For free staggered fermions, we developed another method to calculate the required traces that does not rely on stochastic methods, only using the exact eigensystem instead. We explain the details of this calculation in App. B.

To determine $C_{\text{CSE}}$, we calculated the current derivative for different values of the magnetic field, and obtained the coefficient from its slope with respect to $eB$, by fitting the data with a usual $\chi^2$ minimization method.[4.1] We considered a linear fit function with only one free parameter (no intercept). The error analysis for each simulation is performed using the jackknife method with 10 bins (see App. A), which is also used to propagate the error to the fit. This is the *statistical* error of $C_{\text{CSE}}$. Furthermore, we repeat the fit, successively eliminating data points at the largest value of $eB$, until we are left with only one point. The largest difference in the slopes between the original fit and the different repetitions is what we consider the *systematical* error of $C_{\text{CSE}}$. Both errors are added in quadrature, to yield the total value displayed as the error of $C_{\text{CSE}}$.

---

[4.1] We note that for the quenched simulations, the observable calculated for a particular set of parameters at a given magnetic field, is correlated with the value at a different magnetic field, since it is measured on the same configurations. This can be taken into account in the fitting method, performing a *correlated* fit. However, we found this correlation to not have a strong impact on the calculation of $C_{\text{CSE}}$. We will come back to this point in the next chapter, since it will turn out to have a stronger impact in the calculation of $C_{\text{CME}}$.

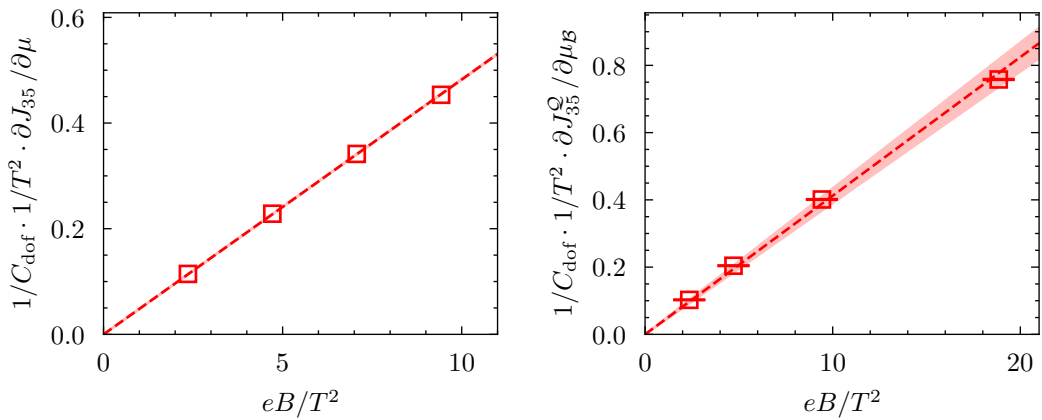

FIGURE 4.1: Derivative of the axial current with respect to the chemical potential for staggered fermions, given by Eq. (4.16), as a function of the magnetic field for non-interacting fermions in a $24^3 \times 6$ lattice (left) and in full QCD with $2+1$ flavors and physical quark masses at $T = 305$ MeV (right). The dashed line shows the result of the linear fit, while the band is the estimation of the total error. Notice that the points for free fermions do not have an error bar, since they are not obtained stochastically. Figure from Ref. [152].

In Fig. 4.1 we show an example of the obtained results and the fits, both for free fermions and full QCD with staggered quarks. In both situations we obtain a linear result as a function of the magnetic field, as expected for weak enough fields. A linear fit yields the value of $C_{\text{CSE}}$, which we analyze now in detail for different scenarios.

### 4.3.1 Free quarks

Since the free case is analytically solvable, see Eq. (3.72), we can use it as a check of the correctness of our setup. We consider a single-color fermion with charge $q$ and mass $m$. This implies that the flavor quantum numbers are irrelevant and it is sufficient to treat a single chemical potential $\mu$ and axial current $J_{35}$. The overall proportionality constant is thus $C_{\text{dof}} = (q/e)^2$, which is used to normalize the results.

In Fig. 4.2 we show the results for free staggered fermions at different $m/T$ values. For the largest one, we also show results with Wilson fermions. To compare to the analytic formula, we need to consider both the continuum limit ($a \to 0$) and the thermodynamic limit ($L \to \infty$). This can be achieved using dimensionless combinations, in particular the continuum limit can be reached by increasing $N_s$ and $N_t$, while keeping the aspect ratio $LT = N_s/N_t$ and $m/T$ constant. In this situation, a larger aspect ratio $LT$ corresponds to a larger volume, i.e. going towards the thermodynamic limit. The different panels of Fig. 4.2 correspond to the continuum extrapolations for different aspect ratios, at a given value of $m/T$. These results indicate that finite-size effects are more sizeable when $m/T$ is small, as expected. In particular, $LT = 4$ is already found to agree with the infinite volume limit for all masses.

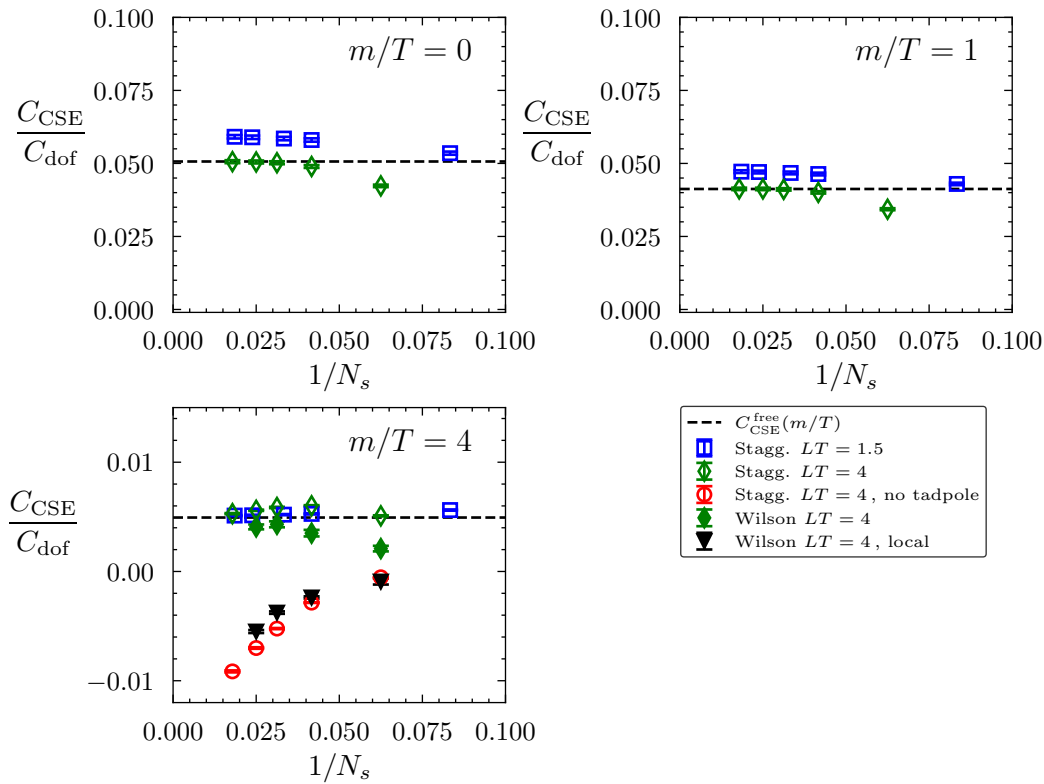

FIGURE 4.2: CSE conductivity coefficient for free fermions at different values of $m/T$ using the staggered discretization. For $m/T = 4$, Wilson fermions results are also shown, as well as non-conserved versions of the vector currents. The dashed line corresponds to the analytical value of $C_{\text{CSE}}$ for different values of $m/T$, given by Eq. (3.72). Figure from Ref. [152].

From these results, we learn that $C_{\text{CSE}}$ approaches the value given by Eq. (3.72) in the continuum limit, if the volume is large enough. In the case of Wilson fermions, we also see that the setup that gives the correct continuum limit is to consider a conserved vector current and the anomalous axial current. If we deviate from this and use the local vector current (4.22) instead, the analytical result is no longer recovered, even showing a possibly divergent behavior in the continuum limit. Something very similar is observed for staggered fermions if we exclude the tadpole term of (4.16). This emphasizes the importance of a detailed analysis of the observables on the lattice, since operators that naively have the same continuum limit do not necessarily yield the same result.

A further check we can perform for free staggered fermions is the different functional forms of the staggered operators, as discussed in Eqs. (4.13), (4.14). In Fig. 4.3, we show the continuum limit of $C_{\text{CSE}}$ for these three definitions. We see that all of the combinations yield the correct continuum limit, but with different lattice artifacts. From these results, we learn that the use of the 1-link operator $\Gamma_\nu$ for the vector current and 3-links operators $\Gamma_{\nu 5}$ for the axial one, both with the U(1) links included in the operators, i.e. Eqs. (4.10), (4.11), provides the fastest scaling towards the continuum limit. Therefore we will use this particular definition for the rest of the calculation.

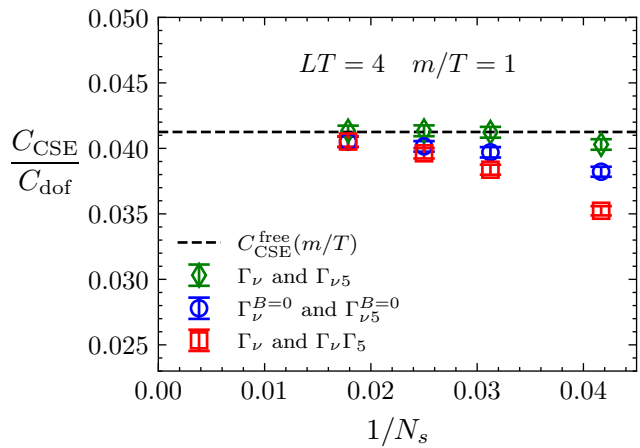

FIGURE 4.3: Continuum limit for the different operators in the staggered formulation. We show here the results for a particular $m/T = 1$, although the continuum limit agreement of all the possibilities also holds for other values.

### 4.3.2 Quenched theory

Before moving on to full QCD, there is an intermediate step where one can analyze the impact of gluons on the CSE: the quenched approximation. Furthermore, the quenched theory reveals an interesting feature of the CSE related to the presence of an exact center symmetry.

As we discussed in Sec. 2.7, in the absence of dynamical fermions the system undergoes a first-order phase transition from confined to deconfined matter at around $T_c^q \approx 270$ MeV [64], for which the Polyakov loop (2.165) acts as the order parameter. This is related to the spontaneous breaking of the $Z_3$ symmetry at high temperatures. The key observation is that in the broken phase, the Polyakov loop sectors can be mimicked by an *imaginary* baryon chemical potential $i\mu_{\mathcal{B}}/T \equiv \mu_I/T = 0, \pm 2\pi/3$. This has a direct effect on the CSE, since an imaginary chemical potential modifies non-trivially the conductivity. This can be shown analytically for free fermions by generalizing the calculation of $C_{\text{CSE}}$ at finite imaginary chemical potential, whose results are shown in Fig. 4.4 (see Ref. [152] for a more detailed discussion on the calculation). We see that choosing the value $\mu_I/T = \pm 2\pi/3$ heavily modifies the functional form of $C_{\text{CSE}}$, even reaching negative values. Therefore, we can expect that the imaginary sectors of the Polyakov loops have an effect on the CSE conductivity at high temperatures.

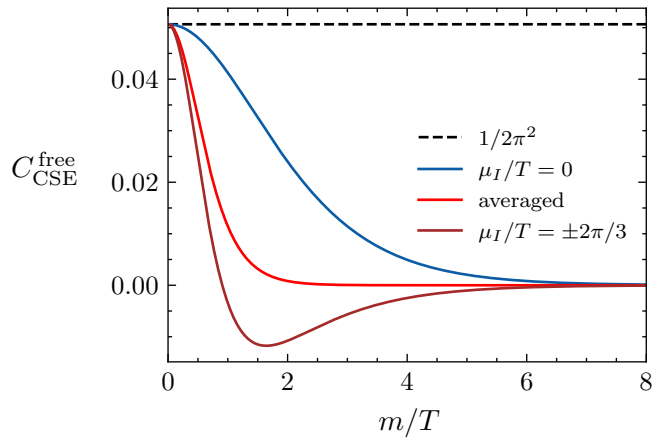

FIGURE 4.4: Value of $C_{\text{CSE}}$ as a function of $m/T$ for different values of the imaginary chemical potential $\mu_I$. Here *averaged* refers to take the average over $\mu_I = 0, \pm 2\pi/3$. Figure from Ref. [152].

In Fig. 4.5, we present an example of this effect at a temperature of $T \approx 400$ MeV. In the left panel, we observe clearly the three different Polyakov loop sectors, while in the right panel we show the contributions of the individual sectors to the observable. We note that, at $\mu_I \neq 0$, the expectation values of the baryon density and the axial current are nonzero, giving rise to an additional term $\propto \left\langle \text{Tr}(\Gamma^f_{35} M_f^{-1}) \right\rangle \left\langle \text{Tr}(\Gamma^{f'}_4 M_{f'}^{-1}) \right\rangle$ in Eq. (4.16). This term is included for the comparison in Fig. 4.5. The fluctuations of the current derivative are observed to be enhanced drastically in the imaginary sectors. However, this is not a physical effect. In full QCD, the quarks explicitly break the $Z_3$ symmetry and, consequently, the theory is always in the real Polyakov loop sector, corresponding to a vanishing imaginary baryon chemical potential. This indicates that the relevant QCD contribution to the CSE originates from configurations with real Polyakov loops. Therefore, for the quenched analysis of $C_{\text{CSE}}$ below, we rotate all our gauge configurations to this sector by the appropriate center transformation.

Using this rotation to the real Polyakov loop sector, we present the results for $C_{\text{CSE}}$ in the quenched approximation in Fig. 4.6. We use configurations generated with the Wilson gauge action (2.44) at $\beta = 5.845, 5.9, 6.0, 6.2, 6.26$ and $6.47$, and use both staggered and Wilson fermions in the operator. Some of these gauge configurations were already employed in Refs. [38, 159]. For the staggered discretization, the quark masses were tuned to a pion mass of $m_\pi \approx 415$ MeV, while for Wilson fermions we use $m_\pi \approx 710$ MeV. The latter choice is motivated by the CME study [117].

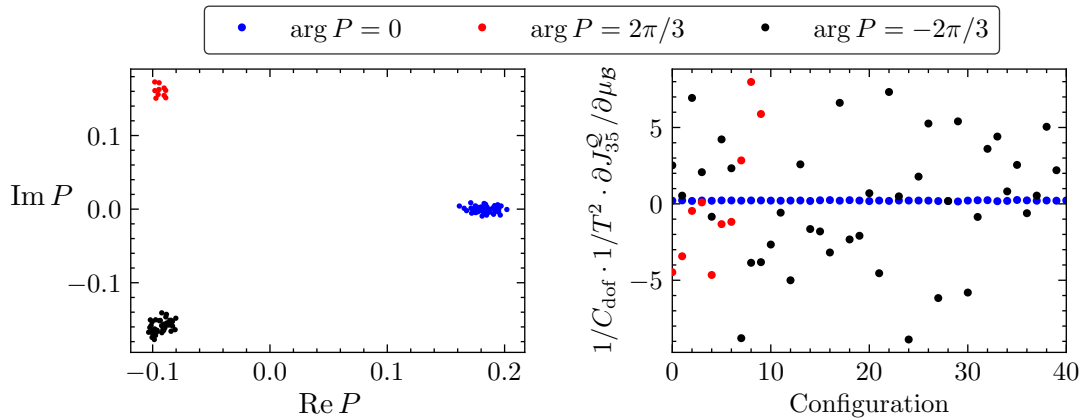

FIGURE 4.5: Results for the Polyakov loop (left) and the derivative of the current with staggered fermions (right) in an ensemble of 100 configurations on a $32^3 \times 8$ lattice at $T \approx 400$ MeV and at a magnetic field of $eB = 0.74$ GeV$^2$. The three sectors are clearly visible at this temperature, where the center symmetry is spontaneously broken, and the contribution of configurations with imaginary Polyakov loops is observed to be very different as compared to those in the real sector. Figure from Ref. [152].

Although we only present results at a few temperatures, this already enables us to distinguish two differentiated regimes: in the confined phase the CSE is severely suppressed, reaching zero at $T \approx 0$, while at temperatures well above $T_c^q$, the result approaches the massless free fermion value $C_{\text{CSE}} = 1/(2\pi^2)$. The suppression at low temperatures is in agreement with the two-color QCD study [151]. One may also compare this with the findings of the quenched study [150], where no corrections due to QCD interactions were found. However, in that case the quenched configurations were measured with a massless overlap Dirac operator, making a direct comparison to our setup complicated. Moreover, we emphasize that according to our results, the transition between the regime where CSE is suppressed and the one where the massless case is approached, appears to occur in the vicinity of $T_c^q \approx 270$ MeV. All these features hinted at by the quenched results will be confirmed by the full QCD simulations, which we present next.

### 4.3.3 QCD at physical quark masses

Finally, we present the main result of this chapter: the conductivity coefficient $C_{\text{CSE}}$ in full QCD, using $N_f = 2+1$ flavors of staggered fermions at physical quark masses. We emphasize that this is the first fully non-perturbative result for $C_{\text{CSE}}$ at the physical point. The measurements were partly performed on an already existing ensemble of configurations for different magnetic fields [67, 160]. In Fig. 4.7, we present the dependence of the conductivity on the temperature using several finite-temperature lattice ensembles $24^3 \times 6$, $24^3 \times 8$, $28^3 \times 10$, $36^3 \times 12$ as well as two zero-temperature ensembles $24^3 \times 32$ and $32^3 \times 48$. The temperature has a very prominent effect on $C_{\text{CSE}}$, as we have already seen in the quenched results. At temperatures below QCD crossover temperature $T_c \approx 150$ MeV, the conductivity is severely suppressed, reaching zero at $T \approx 100$ MeV. This suppression of the CSE at low temperatures is again consistent with a previous study in two-color QCD [151] and our quenched results. Around $T_c$, $C_{\text{CSE}}$ experiences a very sharp increase, approaching the value corresponding to free massless quarks at high temperatures, as expected due to asymptotic freedom.

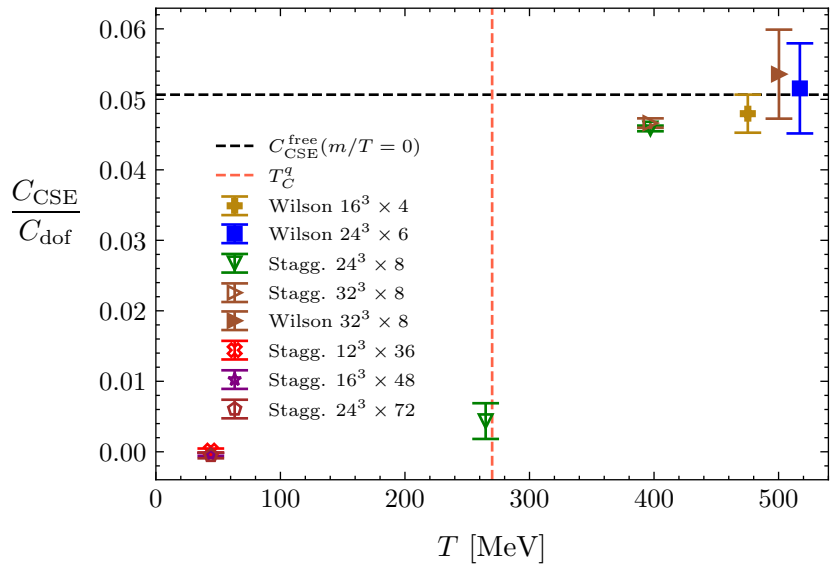

FIGURE 4.6: Results for $C_{\text{CSE}}$ with Wilson and staggered fermions in the quenched approximation, measured on configurations generated using the plaquette gauge action (2.44), where the configurations were rotated to the real Polyakov loop sector. The pion mass is set to $M_\pi \approx 415$ MeV for staggered quarks and $M_\pi \approx 710$ MeV for Wilson quarks. The quenched critical temperature $T_c^q$ is indicated by the dashed vertical line. Figure from Ref. [152].

In the low-temperature region, we can understand this suppression by considering a simple model to describe the system: a non-interacting gas of hadronic degrees of freedom. This accounts to calculate $C_{\text{CSE}}$ using Eq. (3.72), summing over the different hadronic resonances, whose masses can be found in Ref. [76]. For this particular observable, only electrically charged hadrons contribute that couple to chirality (i.e. the Dirac indices of $\gamma_5$).[4.2] Thus, in our model we only consider a gas of proton $p$ and $\Sigma^\pm, \Xi^-$ baryons. The

---

[4.2] We do not consider the impact of Wess-Zumino-Witten type terms [161] in our model.

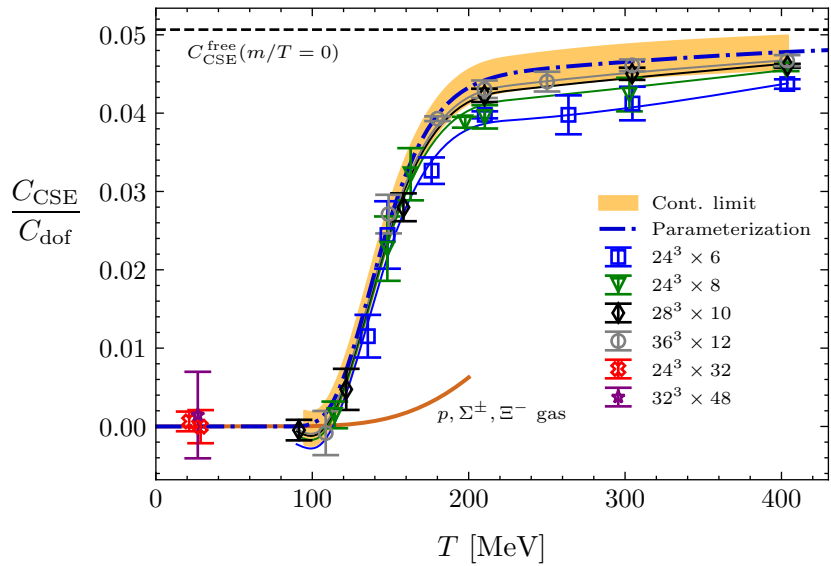

FIGURE 4.7: $C_{\text{CSE}}$ for a broad range of temperatures in full QCD with $2+1$ flavors of staggered fermions. The continuum limit estimation is indicated by the orange band. The result of our low-energy model, involving $p$ and $\Sigma^{\pm}, \Xi^{-}$ baryons, is shown by the continuous line at low $T$, while the parameterization for $C_{\text{CSE}}$ in the whole range of temperatures is displayed as a dashed-dotted line. Figure from Ref. [152].

contribution of heavier charged spin-1/2 baryons was found to be negligible in the relevant temperature range, and we do not include spin-3/2 baryons either. The so-constructed model, which is shown in Fig. 4.7, also features a strong suppression of $C_{\text{CSE}}$ in the confined regime and is found to agree with the lattice results for $T \lesssim 120$ MeV. It is also in qualitative agreement with other approaches used to studying the CSE [143, 148].

The continuum limit is performed using the lattice data in the range of temperatures 90 MeV $\lesssim T \lesssim 400$ MeV. To perform a reliable extrapolation to the continuum,[4.3] we use a spline-fit procedure combined with the continuum limit. For this, we consider a spline fit in $T$ of all the lattice data, with the coefficients depending on the lattice spacing $a$. The best fitting surface in the $a-T$ plane is found by minimizing $\chi^2/\text{dof}$. The details of the spline fit can be found in [164]. The statistical error of this procedure is again calculated using the jackknife method, while we estimated the systematic error by repeating the spline fit removing the coarsest lattice ($N_t = 6$). The maximum difference between the continuum limits obtained with these two data sets is taken as the systematic error, and the total error estimation is calculated by adding the statistical and systematic error

---

[4.3] We note that the flavor-singlet axial vector current entering $C_{\text{CSE}}$ is subject to multiplicative renormalization. The corresponding (perturbative) renormalization constant approaches unity in the continuum limit, both for Wilson and for staggered fermions [162, 163]. Although it might be used to reduce discretization errors, we performed the continuum limit without including multiplicative renormalization constants. We will come back to this finite renormalization factor in the next chapter.

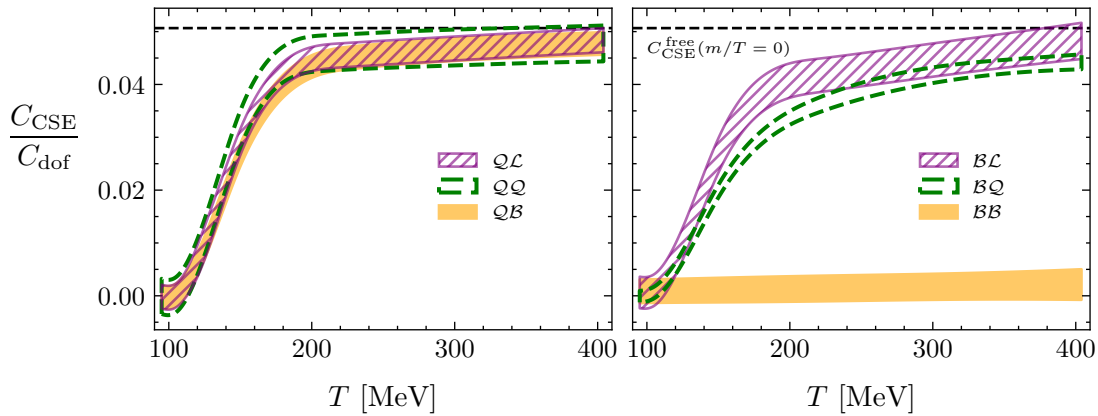

FIGURE 4.8: Continuum limits of $C_{\text{CSE}}$ for different flavor quantum numbers for the axial current and the chemical potential. We present six possible combinations, the phenomenologically relevant cases, involving an electrically charged axial current (left panel) and a baryon axial current (right panel). Figure from Ref. [152].

in quadrature. We also provided a parameterization of $C_{\text{CSE}}$ in the whole range of temperatures, which can be used for comparisons of our result with model calculations. The parameterization smoothly connects the continuum extrapolation of the lattice results with the perturbative limit and the non-interacting $p, \Sigma^{\pm}, \Xi^{-}$ gas model, see Ref. [152] for the precise functional form.

In summary, we conclude that the CSE conductivity is an observable very sensitive to the finite temperature QCD crossover, and it may even be used to define its characteristic temperature. In particular, the inflection point of the parameterization is found at $T_c = 136(1)$ MeV. As an alternative definition, we also consider the temperature where the conductivity assumes half its asymptotic value $1/(2\pi^2)$, yielding $T_c = 149(4)$ MeV.

These results correspond to the combination of flavor numbers $C_{\text{CSE}} = C_{\text{CSE}}^{\mathcal{QB}}$, i.e. the response of the electrically charged axial current to the baryon chemical potential. Next, we compare results for other combinations of the flavor quantum numbers $\mathcal{Q}, \mathcal{B}$ and $\mathcal{L}$. Even after normalizing the conductivities by the appropriate proportionality factor $C_{\text{dof}}$ from Eq. (4.6), the results are not equivalent, since the different quark flavors interact with each other via gluons. Out of the nine possibilities, in Fig. 4.8 we present the continuum limit for six of these combinations, focusing on the most interesting ones from a phenomenological point of view.

All combinations (except the $\mathcal{BB}$ case) show very similar qualitative behaviors, being suppressed at low $T$ and approaching the massless free fermion result at high temperatures. However, it is possible to see differences in the intermediate temperature regime, indicating that the sensitivity to the crossover temperature gets modified in each case.

The $\mathcal{BB}$ case is special, because the three-flavor 1-loop perturbative result (free fermions) in this case vanishes[4.4] (see the discussion in Sec. 4.1). Our full QCD results for $C_{\text{CSE}}^{\mathcal{BB}}$ also approaches zero for high temperatures. Although at intermediate temperatures the perturbatively expected cancellation between the contributions of the three flavors does not completely hold, since higher order loop contributions are not negligible, we still find the result to be suppressed compared to the other combinations. In addition, the data for $C_{\text{CSE}}^{\mathcal{BB}}$ was found to be rather noisy due to the disconnected term in Eq. (4.16), particularly around the crossover temperature. This complicates a precise characterization of the temperature dependence in this case, hence we give a conservative estimation for the continuum limit of $C_{\text{CSE}}^{\mathcal{BB}}$, which lies within the orange band shown in the right panel of Fig. 4.8.

This concludes our study of the CSE in QCD. We now briefly summarize our findings:

- We have emphasized the importance of using a conserved vector current to study the CSE, as well as a careful analysis of the precise form of the observables.

- In quenched QCD, we have found a novel interplay between the center symmetry and the CSE.

- We have completely characterized $C_{\text{CSE}}$ in physical QCD in a broad range of temperatures, obtaining that the CSE is greatly suppressed at low temperatures, experimenting a very sharp increase in the vicinity of $T_c$, until reaching the free massless fermions value at high temperatures. This is the first determination of an anomalous transport conductivity in QCD at the physical point.

As we have mentioned in Ch. 3, the CSE is a challenging effect to detect experimentally. However, these results could shed light on the detection of other anomalous transport phenomena, the Chiral Magnetic Wave. Since our findings point towards a suppression of the CSE at low temperatures in QCD, at least in its equilibrium formulation, this may suggest that an experimental signal of the CMW could be enhanced at higher temperatures.

---

[4.4] Since $C_{\text{dof}}^{\mathcal{BB}} = 0$, in Fig. 4.8 we have only divided the $\mathcal{BB}$ data by $N_c$.

# Chapter 5

# Chiral Magnetic Effect

In this chapter, we present our study of the Chiral Magnetic Effect in QCD. The main objective of this chapter is the clarification of the nature of the CME in global thermal equilibrium, being of particular interest the comparison to the non-vanishing results obtained in lattice simulations with Wilson fermions [116, 117]. This chapter is based on Ref. [138].

## 5.1 CME conductivity and chemical potentials

First, we need to discuss the choice of the basis for the currents and chiral chemical potentials, analogously as we did for the CSE in the previous chapter. However, contrary to the CSE case, these different choices do not play a role for the CME. Therefore, we consider the most popular election for the quantum numbers in the CME literature, namely a charged vector current $J_\nu^\mathcal{Q}$ and a baryon chiral chemical potential $\mu_5^\mathcal{B}$, and we suppress these indexes for clarity.[5.1]

We consider the Taylor-expansion of the current expectation value, this time in the chiral chemical potential, evaluated at non-zero external magnetic fields but vanishing $\mu_5$. The $C_\text{CME}$ coefficient is defined as

$$\left.\frac{\partial \langle J_3 \rangle}{\partial \mu_5}\right|_{\mu_5=0} = C_\text{CME}\, eB\,, \tag{5.1}$$

---

[5.1] We also do not include the factor of $1/3$ in the baryonic chiral chemical potential (see Eq. (4.2)), since it is just an overall factor.

where again we have considered a magnetic field pointing in the third spatial direction. We want to emphasize that the use of this expansion is motivated by the technical challenges of implementing a chiral chemical potential in the staggered formulation, not due to a sign problem, like in the case of the CSE. For Wilson fermions, simulations at $\mu_5 \neq 0$ are free of any difficulty and have already been performed in other studies, see Refs. [116, 117, 165]. However, we will follow the same strategy in the analysis of $C_{\text{CME}}$ using Eq. (5.1) with Wilson fermions, as a crosscheck for the setup with staggered quarks. Since we are only considering the $\mathcal{QB}$ case, the overall proportionality factor simply reads

$$C_{\text{dof}} = N_c \sum_f \left(\frac{q_f}{e}\right)^2 . \tag{5.2}$$

It is also easy to check that again we need to calculate the imaginary part of the Euclidean correlator,

$$\left.\text{Re}\, \frac{\partial \langle J_3 \rangle}{\partial \mu_5}\right|^{\text{M}}_{\mu_5=0} = -\left.\text{Im}\, \frac{\partial \langle J_3 \rangle}{\partial \mu_5}\right|^{\text{E}}_{\mu_5=0} . \tag{5.3}$$

## 5.2 Lattice setup

The considered lattice setups are completely equivalent to the ones used in the former chapter. However, there are some subtleties regarding the introduction of $\mu_5$ and the precise form of the observables that we present now.

Before presenting the observables, we briefly discuss the issue of renormalization at finite $\mu_5$. The chiral chemical potential, like the quark mass, has a non-trivial multiplicative renormalization constant, and also induces additive divergences terms in $\log \mathcal{Z}$ as we mentioned in Sec. 3.3.3. The key difference is that, unlike the mass renormalization constant $Z_S^{-1}$, the relevant singlet axial renormalization constant $Z_A^{-1}$ approaches unity perturbatively in the continuum limit [162, 163], as we briefly mentioned in the previous chapter, see Fn. 4.3 . This can be traced back to the fact that $Z_A^{-1}$ arises from a two-loop diagram. Thus, we can expect that its use would only affect discretization errors in our lattice simulations and again in this chapter we do not include such a multiplicative renormalization for our observables. In addition, we also do not expect an impact of $\mu_5$ in our LCP, since such effects would in any case enter at $\mathcal{O}(\mu_5^2)$ due to $\mathcal{C}$-symmetry, and therefore do not impact our observable (5.1), which contains a first derivative with respect to $\mu_5$ evaluated at $\mu_5 = 0$.

## 5.2.1 Staggered fermions

Now we proceed to discuss the observables in the staggered formulation. The chiral chemical potential enters the definition of the vector and axial currents, analogously to the case of a quark chemical potential. However, the exact functional form is more elaborated. First, we choose an exponential introduction of $\mu_5$, as we did for $\mu$. We can then rewrite Eqs. (4.10), (4.11) in terms of the chiral chemical potential,

$$\Gamma^f_\nu(n,m) = \frac{\eta_\nu(n)}{2}\Big[U_\nu(n)u_{f\nu}(n)\,e^{h(\mu_5)}\delta_{n+\hat{\nu},m} \tag{5.4}$$
$$+ U_\nu^\dagger(n-\hat{\nu})u_{f\nu}^*(n-\hat{\nu})\,e^{-h(\mu_5)}\delta_{n-\hat{\nu},m}\Big],$$

$$\Gamma^f_{\nu 5} = \frac{1}{3!}\sum_{\rho,\alpha,\beta}\epsilon_{\nu\rho\alpha\beta}\,\Gamma^f_\rho\Gamma^f_\alpha\Gamma^f_\beta. \tag{5.5}$$

with $h(\mu_5)$ a general function of the chiral chemical potential. To linear order in $\mu_5$, the precise functional form can be found by imposing that

$$\left.\frac{\partial M^f_{\text{staggered}}}{\partial \mu_5}\right|_{\mu_5=0} = \Gamma^f_{45}. \tag{5.6}$$

This leads to

$$h(\mu_5) = a\mu_5\,\Sigma^f_\nu(\mu_5), \tag{5.7}$$

with the staggered representations of the spin operator,

$$\Sigma^f_\nu(\mu_5) = \frac{1}{3!}\epsilon_{\nu\rho\alpha 4}\Gamma^f_\rho\Gamma^f_\alpha(\mu_5). \tag{5.8}$$

Altogether, the Dirac operator in the presence of a chiral chemical potential in the staggered formalism can be written as (notice that $\Sigma^f_4 = 0$),

$$\slashed{D}^f_{\text{staggered}}(n,m) = \frac{1}{2a}\sum_\nu \eta_\nu(n)\Big[U_\nu(n)u_{f\nu}(n)\,e^{a\mu_5\Sigma^f_\nu(\mu_5)}\delta_{m,n+\hat{\nu}} \tag{5.9}$$
$$- U_\nu^\dagger(n-\hat{\nu})u_{f\nu}^*(n-\hat{\nu})\,e^{-a\mu_5\Sigma^f_\nu(\mu_5)}\delta_{m,n-\hat{\nu}}\Big].$$

Here we can see the difficulties of simulating at non-zero $\mu_5$ with staggered fermions. The Dirac operator (5.9) has a recursive dependence on $\mu_5$, since each spatial hopping involves the operator $\Sigma^f_i$ operator, which itself contains spatial hoppings. Hence we would need to truncate this dependence at a given order, to make numerical simulations feasible. Even in this case, the Dirac operator would still become a highly non-local operator involving the exponentials of the $\Sigma_i^f$ matrices. An alternative approach is to introduce the chiral chemical potential in a *linear* way, see Refs. [156, 157, 165]. In this formulation, one has to carefully analyze the observables, since there are potential extra divergences that can appear due to this way of introducing $\mu_5$. Both the exponential and the linear way lead to the same expressions for first derivatives like the CME current (5.1), however, they differ for higher derivatives like the axial susceptibility (3.66). We get back to the latter point below in Sec. 5.3.3.

The derivative (5.1), required to extract $C_{\mathrm{CME}}$, gives a very similar form to the one used to calculate the $C_{\mathrm{CSE}}$ (4.16). It contains disconnected and connected terms, as well as a tadpole term, since now $\Gamma_3^f$ depends on $\mu_5$,

$$C_{\mathrm{CME}}\, eB = \left.\frac{\partial \langle J_3 \rangle}{\partial \mu_5}\right|_{\mu_5=0} = \frac{T}{V}\left[\frac{1}{16}\sum_{f,f'}\frac{q_f}{e}\left\langle \mathrm{Tr}\left(\Gamma_{45}^{f'} M_{f'}^{-1}\right)\mathrm{Tr}\left(\Gamma_3^f M_f^{-1}\right)\right\rangle \right.$$
$$-\frac{1}{4}\sum_f \frac{q_f}{e}\left\langle \mathrm{Tr}\left(\Gamma_{45}^f M_f^{-1} \Gamma_3^f M_f^{-1}\right)\right\rangle \quad (5.10)$$
$$\left. +\frac{1}{4}\sum_f \frac{q_f}{e}\left\langle \mathrm{Tr}\left(\frac{\partial \Gamma_3^f}{\partial \mu_5} M_f^{-1}\right)\right\rangle \right].$$

Again these expectation values are to be evaluated at $\mu_5 = 0$ but nonzero magnetic field $eB$. It is also possible to consider different staggered operators using the equivalent of Eq. (4.14) and Eq. (4.13) at finite $\mu_5$. We will show below that these have no impact on the result for non-interacting fermions.

In Ch. 3, we also introduced the axial susceptibility $\chi_5$ (3.66), which measures the linear response of the chiral density to the introduction of $\mu_5$. It is an important cross-check that $\chi_5$ is non-zero in our system, since the absence of a chiral imbalance would automatically imply a vanishing CME. In the staggered formulation, this observable reads

$$\frac{\chi_5^b}{T^2} = \frac{1}{TV}\left[\frac{1}{16}\sum_{f,f'}\left\langle \mathrm{Tr}\left(\Gamma_{45}^f M_f^{-1}\right)\mathrm{Tr}\left(\Gamma_{45}^{f'} M_{f'}^{-1}\right)\right\rangle - \frac{1}{4}\sum_f \left\langle \mathrm{Tr}\left(\Gamma_{45}^f M_f^{-1} \Gamma_{45}^f M_f^{-1}\right)\right\rangle \right.$$
$$\left. +\frac{1}{4}\sum_f \left\langle \mathrm{Tr}\left(\frac{\partial^2 M_f}{\partial \mu_5^2} M_f^{-1}\right)\right\rangle\right], \quad (5.11)$$

where again a tadpole term is present. This observable is subject to additive renormalization (see Eq. (3.67)), which we will carry out by subtracting the $T=0$ observable. We come back to the specifics of this procedure in Sec. 5.3.

## 5.2.2 Wilson fermions

Next we describe our setup involving Wilson quarks. First, we consider how the chiral chemical potential enters the Wilson fermion formulation. Again considering an exponential introduction, we can write the hopping operator at finite $\mu_5$ as

$$H = \sum_n \sum_\nu \Big[(r - \gamma_\nu\, e^{a\mu_5 \delta_{\nu 4}})U_\nu(n)u_{f\nu}(n)\delta_{n+\hat\nu,m} \\ + (r + \gamma_\nu\, e^{-a\mu_5 \delta_{\nu 4}})U_\nu^\dagger(n-\hat\nu)u_{f\nu}^*(n-\hat\nu)\delta_{n-\hat\nu,m}\Big]. \quad (5.12)$$

Notice that the chiral chemical potential does not couple to the Wilson term, since the topological part of the axial anomaly for Wilson fermions arises from terms proportional to $r$, see Sec 2.6.1.

The conserved vector current (2.123) and the anomalous axial current (2.126) are trivial to generalize at finite $\mu_5$,

$$\Gamma_\nu^f(n,m) = \frac{1}{2}\Big[(\gamma_\nu e^{a\mu_5\delta_{\nu 4}} - r)U_\nu(n)u_{f\nu}(n)\,\delta_{m,n+\hat\nu} \\ + (\gamma_\nu e^{-a\mu_5\delta_{\nu 4}} + r)U_\nu^\dagger(n-\hat\nu)u_{f\nu}^*(n-\hat\nu)\,\delta_{m,n-\hat\nu}\Big], \quad (5.13)$$

$$\Gamma_{\nu 5}^f(n,m) = \frac{1}{2}\Big[\gamma_\nu\gamma_5 U_\nu(n)u_{f\nu}(n)\,e^{a\mu_5\delta_{\nu 4}}\delta_{m,n+\hat\nu} \\ + \gamma_\nu\gamma_5 U_\nu^\dagger(n-\hat\nu)u_{f\nu}^*(n-\hat\nu)\,e^{-a\mu_5\delta_{\nu 4}}\delta_{m,n-\hat\nu}\Big]. \quad (5.14)$$

Then the observable reads

$$C_{\text{CME}}\, eB = \left.\frac{\partial\langle J_3\rangle}{\partial \mu_5}\right|_{\mu_5=0} = \frac{T}{V}\Bigg[\sum_{f,f'}\frac{q_f}{e}\left\langle \text{Tr}\left(\Gamma_{45}^{f'}M_{f'}^{-1}\right)\text{Tr}\left(\Gamma_3^f M_f^{-1}\right)\right\rangle \\ - \sum_f \frac{q_f}{e}\left\langle \text{Tr}\left(\Gamma_{45}^f M_f^{-1}\Gamma_3^f M_f^{-1}\right)\right\rangle\Bigg], \quad (5.15)$$

involving only disconnected and connected terms, lacking a tadpole term as for the case of the CSE, since $\Gamma_3^f$ does not depend on $\mu_5$.

We can again consider the local vector current (4.22), which now is of central importance to compare to the study of the CME in quenched and dynamical QCD [116, 117].

The analogue of Eq. (5.15) reads

$$C_{\text{CME}}^{\text{loc}} eB = \left.\frac{\partial \langle J_3^{\text{loc}}\rangle}{\partial \mu_5}\right|_{\mu_5=0} = \frac{T}{V}\left[\sum_{f,f'}\frac{q_f}{e}\left\langle \text{Tr}\left(\Gamma_{45}^{f'}M_{f'}^{-1}\right)\text{Tr}\left(\gamma_3 M_f^{-1}\right)\right\rangle \right.$$
$$\left. - \sum_f \frac{q_f}{e}\left\langle \text{Tr}\left(\Gamma_{45}^f M_f^{-1}\gamma_3 M_f^{-1}\right)\right\rangle\right]. \tag{5.16}$$

Finally, the axial susceptibility in this case takes the form

$$\frac{\chi_5}{T^2} = \frac{1}{TV}\left[\sum_{f,f'}\left\langle \text{Tr}\left(\Gamma_{45}^f M_f^{-1}\right)\text{Tr}\left(\Gamma_{45}^{f'}M_{f'}^{-1}\right)\right\rangle - \sum_f \left\langle \text{Tr}\left(\Gamma_{45}^f M_f^{-1}\Gamma_{45}^f M_f^{-1}\right)\right\rangle\right.$$
$$\left. + \sum_f \left\langle \text{Tr}\left(\frac{\partial^2 M_f}{\partial \mu_5^2}M_f^{-1}\right)\right\rangle\right], \tag{5.17}$$

where now a tadpole term appears also in the Wilson formulation since $\Gamma_{45}^f$ does depend on $\mu_5$.

## 5.3 Results

For our numerical calculations, we follow the same approach as in Sec. 4.3, for the calculation of the traces (again for free staggered fermions we use the exact eigensystem, see App. B), the extraction of $C_{\text{CME}}$ and the calculation of the uncertainties. However, for the quenched data, we use a slightly different approach. To take the correlation into account, which we found to have an important impact on this observable, we perform correlated fits using the *covariance matrix* of the data in the $\chi^2$ minimization method. The statistical error of these fits can be again calculated with the jackknife method. However, we consider a less conservative estimation of the systematic error in this case. We perform a wide number of different fits by changing the number of points considered and including fits with a cubic term in the fitting function. We then construct a histogram of the so-obtained values of $C_{\text{CME}}$, weighted by the Akaike information criterion, see e.g. [166], and we take the standard deviation of this distribution as the systematic error of the fit. In Fig. 5.1 we present two examples of the behavior of the current derivative with $eB$ and the fits, both for free fermions and full QCD with staggered quarks. In these plots, we can see that a vanishing result is obtained with staggered fermions. For a complete understanding of the CME on the lattice with different discretizations and setups, we now present a detailed analysis of the obtained results for $C_{\text{CME}}$ in a wide range of situations.

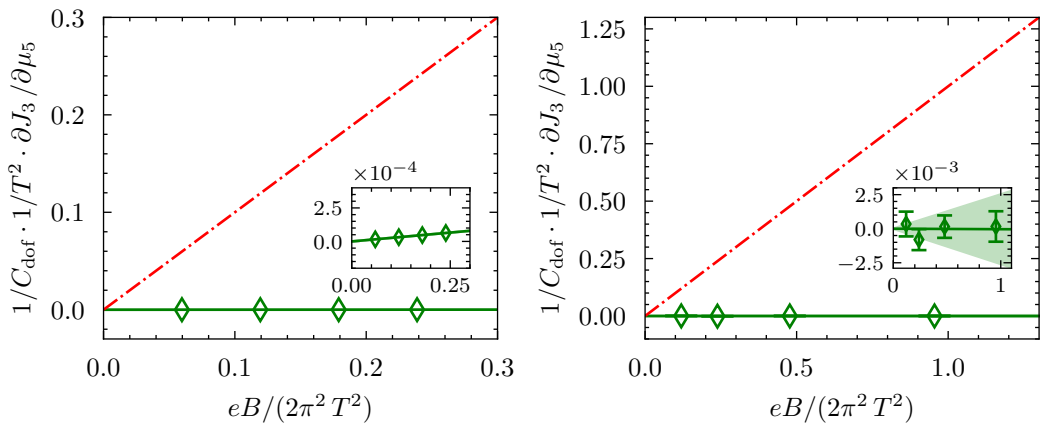

FIGURE 5.1: Derivative of the vector current with respect to the chiral chemical potential as a function of the magnetic field for a $24^3 \times 6$ lattice with staggered quarks in the free case with $m/T = 1$ (left) and in full QCD with $2 + 1$ flavors and physical quark masses at $T = 305$ MeV (right). The continuous green line represents the result of the fit, and the green band represents the total error estimation. For comparison, the dashed-dotted red line indicates a linear function with unit slope (notice the rescaling of the horizontal axis). For free fermions, the data points were obtained using exact diagonalization (see App. B) and thus have no errors. A nonzero slope is found, suppressed by four orders of magnitude, which is due to residual finite volume effects and lattice artifacts. In the interacting case, the error band is directly compatible with a vanishing CME. Figure from Ref. [138].

### 5.3.1 Free quarks

As we already learned in the previous chapter, a system of non-interacting fermions is the perfect test case for the lattice setup. Hence we start again by analyzing this case. In Fig. 5.2, we present the results for $C_{\text{CME}}$ at three different values of $m/T$ with staggered fermions, and for the highest $m/T$, we also include a comparison to Wilson fermions. For small volumes ($LT = 1.5$), the continuum extrapolation of the results shows a non-vanishing result for small values of $m/T = 0, 1$. As we already anticipated in the discussion of Bloch's theorem, this is allowed by this no-go theorem, since it requires the infinite volume limit to take effect. When we consider a bigger aspect ratio ($LT = 4$), the results are found to be compatible with zero for these values of $m/T$ at every lattice spacing.

In this system of free fermions, we already learn a very important lesson: the CME vanishes on a Euclidean lattice if a *conserved* vector current and an *anomalous* axial current are considered, both for staggered and Wilson fermions. For $m/T = 4$, we show the results of deviating from this correct setup. For staggered quarks, if the tadpole term is not considered, it leads to a deviation of the expected vanishing result in the continuum limit. The same occurs for Wilson fermions if a local vector current is used instead of a conserved one. The latter is exactly the setup used in Ref. [117], whose results for the quenched theory are marked by an orange band in the bottom left plot. This is already a very strong indication that the reason behind the non-zero result in that work is closely related to the use of a local vector current. This will be confirmed next, by performing a direct comparison in the quenched theory.

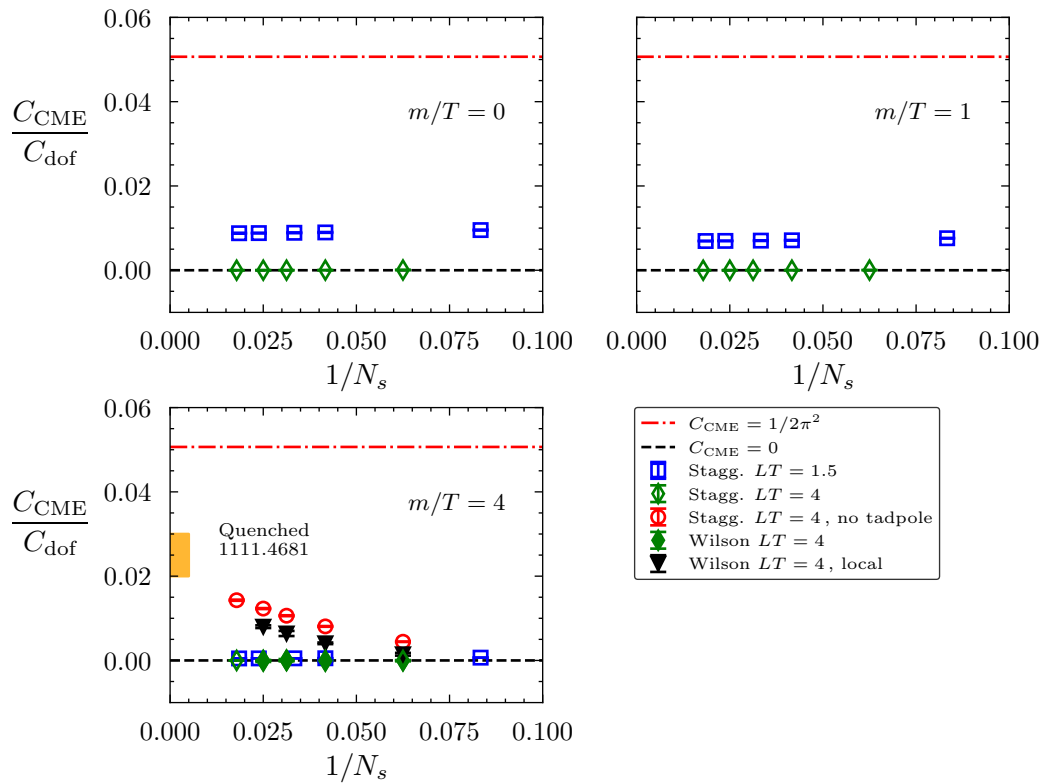

FIGURE 5.2: Continuum extrapolation of the CME conductivity for free fermions at different values of $m/T$ using the staggered discretization, as well as Wilson fermions for $m/T = 4$. The red dot-dashed line corresponds to $C_{\text{CME}} = 1/(2\pi^2)$, while the black dashed line to $C_{\text{CME}} = 0$. The quenched continuum limit estimation with Wilson fermions from Ref. [117] is depicted as an orange band for comparison. Figure from Ref. [138].

Analogously to the CSE case, we can study if modifying the staggered operators, i.e. considering the operator defined in Eq. (4.13) or Eq. (4.14), plays a role for the calculation of $C_{\text{CME}}$. In Fig. 5.3 we see that all the combinations are compatible with zero, with no difference in the lattice artifacts. Hence we use the same operators as in the previous chapter.

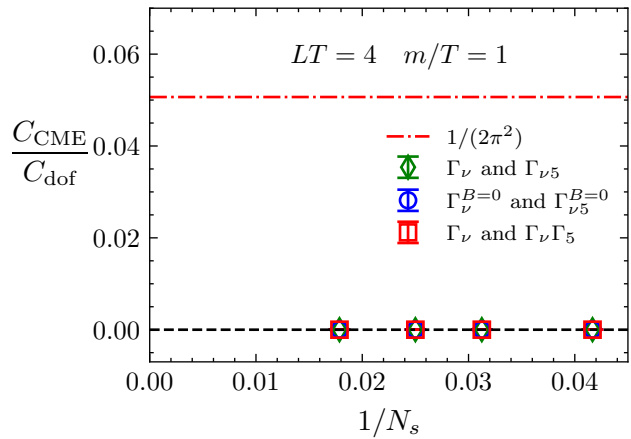

FIGURE 5.3: Continuum limit for the different operator in the staggered formulation, As opposed to the case of $C_{\text{CSE}}$ (see Fig. 4.3) All the combinations display the same behavior for $C_{\text{CME}}$.

Before continuing with the quenched theory, there is another important crosscheck that can be done. We can show that $\mu_5$ parameterizes the chiral imbalance both in the Wilson and staggered discretizations, by calculating the axial susceptibility $\chi_5$ and comparing to the analytical result (3.66). The additive renormalization of $\chi_5$, defined as the $T = 0$ subtraction at Eq. (3.67), can be translated for the free fermions case, where there are no physical scales, as

$$\frac{\chi_5}{T^2} = \frac{1}{T^2}\left[\chi_5^b(N_s^3 \times N_t, am) - \chi_5^b(N_s^4, am)\right]. \tag{5.18}$$

In Fig. 5.4, we present results for $\chi_5$ for staggered and Wilson fermions with $m/T = 3$, where the required traces have been estimated with Gaussian random vectors in the two cases. For both discretizations, the continuum extrapolation agrees with the analytic calculation (3.68). This clearly shows that the chiral density would be non-zero at finite $\mu_5$ in our setup, illustrating that the vanishing result for CME arises from a non-trivial cancellation, and not from the absence of chiral imbalance.

### 5.3.2 Quenched theory

As the next step towards the full QCD theory, we present our results in the quenched setup. This theory not only allows us to have a first look at how interactions affect the CME, but also enables a direct comparison to the non-vanishing results in Ref. [117].

In Fig. 5.5, we present the calculation of $C_{\text{CME}}$ in the quenched approximation, following the same setup as in Sec. 4.3.2. We now use configurations at $\beta = 5.845, 5.9, 6.0, 6.25, 6.26$ and 6.47, and again use both staggered and Wilson fermions for the observable. The correlator is calculated using a conserved vector current and an anomalous axial current for both discretizations, and the result clearly indicates a vanishing CME. We note that at zero temperature, the results show a $2\sigma$ deviation from zero, which we interpret as finite volume effects.

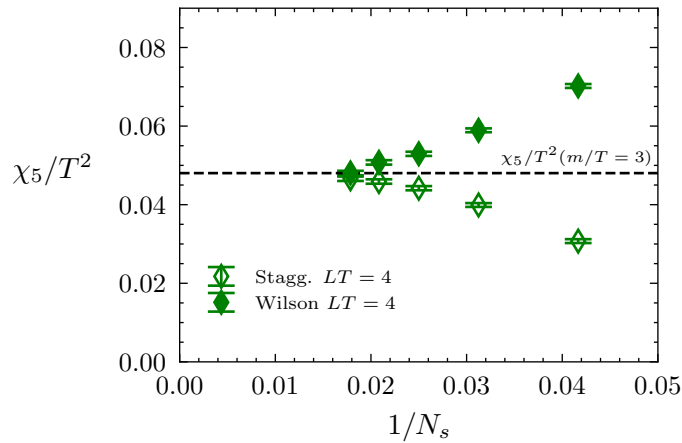

FIGURE 5.4: Axial susceptibility for free fermions with $m/T = 3$, using both the staggered and the Wilson discretizations. The black dashed line represents the value from the analytical expression (3.68) at this particular value of $m/T$. This agreement is also reproduced for other values of $m/T$. Figure from Ref. [138].

As mentioned above, we can compare the quenched results from [117] by simulating directly with the same parameters, but with a different approach, since the results in that work were obtained by simulating at finite chiral chemical potential. In Fig. 5.6, we present the results for Wilson fermions in the quenched theory, with a conserved vector current and with a local one. Using the conserved current yields a result consistent with a vanishing $C_\text{CME}$ for every lattice spacing, while the results with a local vector current are non-zero and in agreement with the results of Ref. [117]. We can now conclude that the conservation of the vector current is crucial to reproduce the expected physical picture of a vanishing CME in equilibrium, and the non-zero results for $C_\text{CME}$ in Ref. [117] are just an artifact of using a non-conserved vector current.

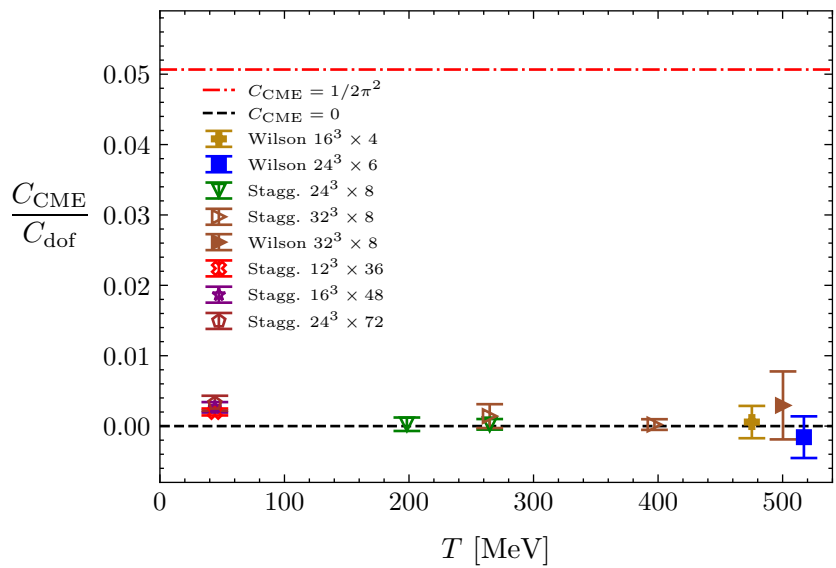

FIGURE 5.5: Results for the CME conductivity in the quenched theory, both with Wilson and staggered fermions. The conductivity is consistent with zero in all the cases (within $2\sigma$) even without taking the continuum limit. Figure from Ref. [138].

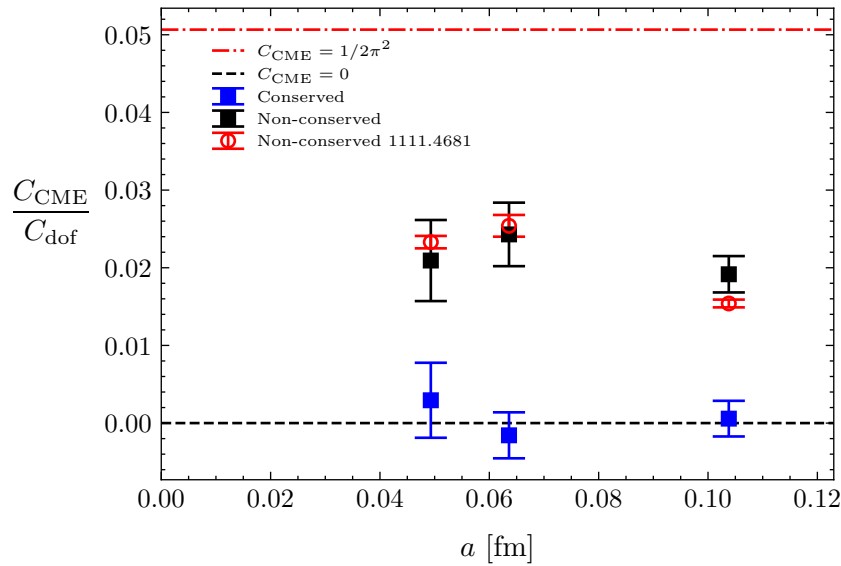

FIGURE 5.6: Direct comparison of the study in Ref. [117] (open red circles), with our setup with a conserved vector current (filled blue squares) and a non-conserved one (filled black squares). The results with non-conserved currents agree with each other within errors and deviate from the correct value. The result with a conserved vector current is consistent with a vanishing $C_\text{CME}$. Figure from Ref. [138].

Finally, we note that in the quenched ensembles, the Polyakov loop backgrounds do not affect the CME conductivity appreciably, contrary to the case of the CSE in Sec. 4.3.2. This is because the complex Polyakov loop backgrounds effectively amount to an imaginary baryon chemical potential, not affecting the chiral density.

### 5.3.3 QCD at physical quark masses

In this last section, we present our results for $C_\text{CME}$ in full QCD using $2+1$ flavors of staggered fermions with physical quark masses. The measurements were performed in the same ensembles used in Sec. 4.3.3. The CME coefficient vanishes for every considered temperature within errors. For completeness, we also perform the continuum limit for temperatures from 100 MeV to 400 MeV, considering the same spline-fit procedure as for $C_{\text{CSE}}$. The obtained continuum extrapolation is again compatible with zero for all temperatures.

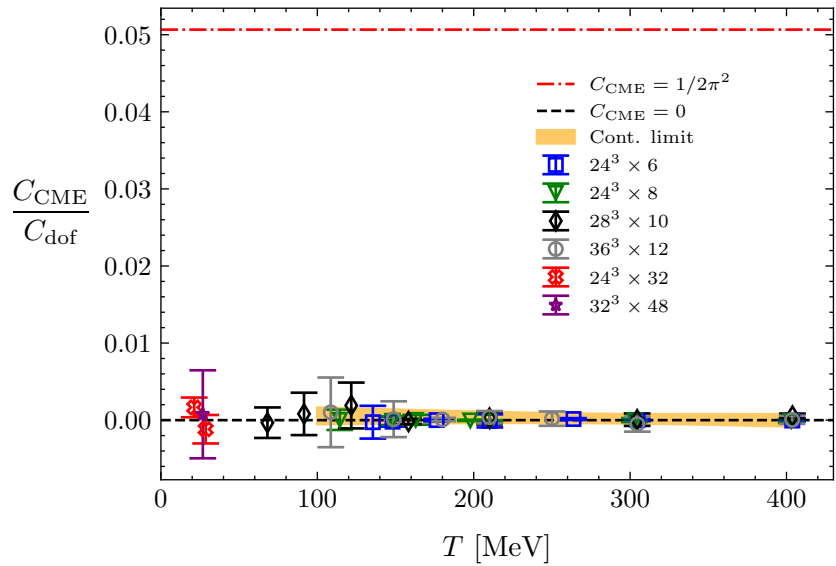

FIGURE 5.7: Full dynamic QCD results with $2+1$ flavors of staggered fermions at the physical point in a wide range of temperatures. The continuum limit for points at $T > 100$ MeV is shown as an orange band. Figure from Ref. [138].

We mention that the use of different combinations of flavor quantum numbers $\mathcal{B}, \mathcal{Q}, \mathcal{L}$ for the current and $\mu_5$ does not change our conclusion, since we checked that $C_{\text{CME}}$ vanishes for any combination. This is expected, since the general argument implying that CME vanishes in equilibrium, applies in all the different setups. Our result clearly demonstrates that the CME vanishes in equilibrium in full QCD. In addition, it shows that the implications of Bloch's theorem also apply to finite temperature QCD.

Finally, in Fig. 5.8, we show the results for the chiral susceptibility $\chi_5$ in full QCD for different temperatures. The renormalized observable is now defined via the $T=0$ subtraction at the same lattice spacing $a$,

$$\frac{\chi_5}{T^2} = \frac{1}{T^2}\left[\chi_5^b(T,a) - \chi_5^b(0,a)\right]. \tag{5.19}$$

Since the renormalization[5.2] of the observable requires this subtraction, approaching the continuum limit is computationally more challenging than for $C_{\text{CME}}$. For this reason,

---

[5.2] We emphasize again that we have not included an axial vector multiplicative renormalization factor in our observables, since it approaches unity in the continuum limit for our discretizations, see the discussion in Sec. 5.2.

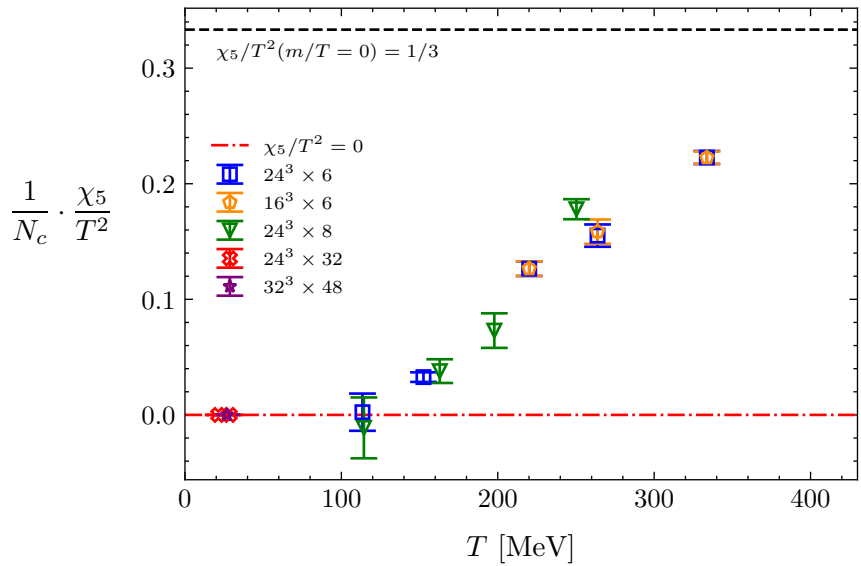

FIGURE 5.8: Axial susceptibility for $2+1$ flavors of staggered quarks in full QCD at the physical point. The results are normalized by the number of colors $N_c = 3$ to directly compare to the free fermions result (3.68). We observe $\chi_5$ to slowly grow above the crossover temperature, approaching the analytic result for massless non-interacting fermions (indicated by the black dashed line). Figure from Ref. [138].

we only present results for $N_t = 6, 8$ lattices with different volumes. To reduce cutoff effects, we follow a similar improvement program for the observable as in Ref. [27]. We multiply our results by improvement coefficients, which can be calculated for very high temperatures (free quarks with $m/T \approx 0$) for the different values of $N_t$. Using Eq. (3.66), we can consider the analytic value $\chi_5(m/T = 0)/T^2 = 1/3$, and divide it by the infinite volume limit of the lattice value at a given $N_t$ for massless fermions $\chi_5^{\text{free}}(m/T = 0, N_t)$. This defines the improvement coefficients,

$$c^{\text{impr.}}(N_t) = \frac{1/3}{\chi_5^{\text{free}}(m/T = 0, N_t)}, \qquad (5.20)$$

which gives an estimate of the lattice artifacts. Therefore, multiplying by these coefficients helps to diminish the discretization effects, bringing the QCD data closer to the continuum limit. In particular, we find these coefficients to be $c^{\text{impr.}}(N_t = 6) = 1.2505$ and $c^{\text{impr.}}(N_t = 8) = 1.1196$. We observe that the susceptibility starts growing around the crossover temperature $T_c$, and it slowly approaches the analytic result for massless free fermions. The asymptotic value of $1/3$ appears to be reached at temperatures higher than 1 GeV, a behavior also observed in the QCD pressure [27], and in contrast with the very sharp increase after $T_c$ observed for $C_{\text{CSE}}$ in Fig. 4.7. These results show that our lattice setup would experience a chiral imbalance at $\mu_5 \neq 0$ at high temperatures, confirming that the absence of the CME is a highly non-trivial result.

We now summarize the results of this chapter:

- We found that the use of a conserved vector current is crucial to reproduce the absence of CME in equilibrium, which can be understood as a consequence of Bloch's theorem.

- The non-vanishing results in dynamical and quenched QCD with Wilson fermions from Refs. [116, 117] are an artifact of using a local (non-conserved) vector current.

- In physical QCD, the CME does not exist in global thermal equilibrium, when the correct lattice setup is used.

- This vanishing result for the CME is completely unrelated to a possible lack of chiral imbalance, which we have shown to be non-zero at high temperatures in QCD.

## Chapter 6

# Localized Chiral Magnetic Effect

For the last chapter of the thesis, we present a generalization of the study in the previous chapter. Following the discussion of Bloch's theorem in Ch. 3, a *local* CME signal is allowed by the theorem, as long as the current integrates to zero in the whole volume. Hence we pose the following question: is it possible to obtain a non-zero local signal for the CME in equilibrium? To achieve this, we require a new ingredient in our study: a non-homogeneous magnetic field. This chapter is based on Ref. [142].

## 6.1 CME and non-uniform magnetic fields

In heavy-ion collision experiments, strong magnetic fields are believed to be created in the early stages of the collision, estimated to reach values of the order of $\sqrt{eB} = 0.1$ GeV [86]. A key observation is that these magnetic fields would not be spatially *homogeneous*, but they would be localized in the center of the collision, decaying with the distance [167].

Since the CME is actively sought for in HIC experiments, the previous observation suggests studying the effect of non-homogeneous magnetic fields on the CME. This is precisely the question we will try to answer in this chapter, by considering a system with a background inhomogeneous magnetic field in global thermal equilibrium. The aim is to study if the CME acquires a non-zero *local* signal due to these non-uniform magnetic fields in QCD, and we can generalize the lattice setup presented in the last two chapters to study these questions.

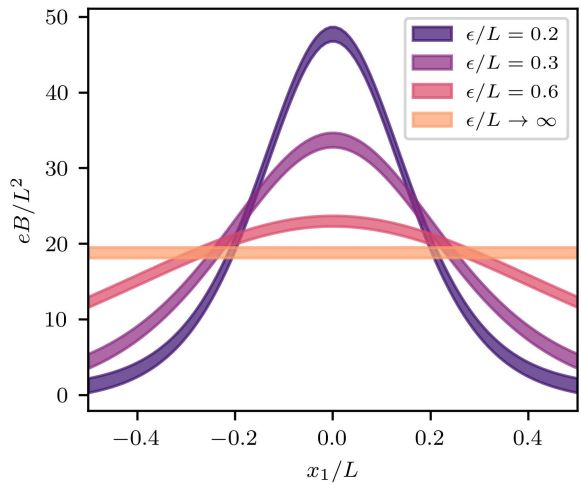

FIGURE 6.1: Magnetic field profile (6.1) for different values of the parameter $\epsilon$, that controls the width. In the limit $\epsilon \to \infty$, the homogeneous magnetic field profile is recovered.

For the implementation on the lattice, we consider a simplified functional form of the magnetic field. We choose a $x_1$-dependent profile of the form[6.1]

$$\vec{B}(x_1) = B \cosh^{-2}(x_1/\epsilon)\, \hat{e}_3\,, \tag{6.1}$$

motivated by its analytical properties [168]. The parameter $\epsilon$ sets the width of the field profile, as we show in Fig. 6.1. Notice that the limit $\epsilon \to \infty$ corresponds to the homogeneous $B$ case. The profile (6.1) has already been used in QCD models [169], as well as on the lattice to study thermodynamics [170, 171]. We proceed now to generalize the discussion of the CME observables to the local case.

First, we consider again a continuum charged vector current and a baryonic axial current, but this time averaged over a space-time slice,

$$J_\nu(x_1) = \frac{T}{L^2} \sum_f \frac{q_f}{e} \int d^4x'\, \bar{\psi}_f(x')\gamma_\nu \psi_f(x')\delta(x_1 - x'_1)\,, \tag{6.2}$$

$$J_{\nu 5}(x_1) = \frac{T}{L^2} \sum_f \int d^4x'\, \bar{\psi}_f(x')\gamma_\nu \gamma_5 \psi_f(x')\delta(x_1 - x'_1)\,. \tag{6.3}$$

Similarly to the previous chapter, we will not discuss further generalizations of this combination to other flavor quantum numbers.

---

[6.1] In this chapter we will follow a different convention where the coordinates go from $-L/2 \leq x < L/2$ in a finite volume. This facilitates the visualization of local lattice results.

We define the most general form of the CME current in the presence of magnetic fields and chiral chemical potentials, where both depend on the $x_1$ coordinate. To linear order in $\mu_5$ and $B$, the current parallel to the magnetic field reads

$$\langle J_3(x_1)\rangle = \int \mathrm{d}x_1' \mathrm{d}x_1'' \, \chi_{\mathrm{CME}}(x_1 - x_1'; x_1 - x_1'') eB(x_1')\mu_5(x_1'') \,, \tag{6.4}$$

where $\chi_{\mathrm{CME}}$ is a general form factor. In our simulations, which are performed at $\mu_5 = 0$ following Ch. 5, this quantity can be accessed via the first derivative of the current with respect to an inhomogeneous $\mu_5(x_1'')$,

$$H(x_1, x_1'') \equiv \left.\frac{\delta \langle J_3(x_1)\rangle}{\delta \mu_5(x_1'')}\right|_{\mu_5=0} = \int \mathrm{d}x_1' \, \chi_{\mathrm{CME}}(x_1 - x_1'; x_1 - x_1'') \, eB(x_1') \,, \tag{6.5}$$

which describes the electromagnetic current generated at $x_1$ due to a weak chiral imbalance present at $x_1''$. We note that $H(x_1, x_1'')$ depends on both spatial arguments separately and not on $|x_1' - x_1''|$, since the inhomogeneous magnetic field breaks translational invariance.

The response to a homogeneous chiral imbalance follows from replacing a local $\mu_5(x_1'')$ by a global, homogeneous $\mu_5$ in Eq. (6.4), resulting in the integral of $\chi_{\mathrm{CME}}$ over its second variable, which yields a local version of the $C_{\mathrm{CME}}$,

$$C_{\mathrm{CME}}(x_1) = \int \mathrm{d}x_1'' \, \chi_{\mathrm{CME}}(x_1; x_1'') \,, \tag{6.6}$$

which we will also refer to as the CME coefficient. For a homogeneous $\mu_5$ setup, Eq. (6.5) simplifies to

$$G(x_1) \equiv \left.\frac{\partial \langle J_3(x_1)\rangle}{\partial \mu_5}\right|_{\mu_5=0} = \int \mathrm{d}x_1' \, C_{\mathrm{CME}}(x_1 - x_1') \, eB(x_1') \,, \tag{6.7}$$

which can also be constructed directly from Eq. (6.5) as $G(x_1) = \int \mathrm{d}x_1'' H(x_1, x_1'')$. Equivalently, Eq. (6.7) in Fourier space reads

$$\widetilde{G}(q_1) = \widetilde{C}_{\mathrm{CME}}(q_1) \, e\widetilde{B}(q_1) \,. \tag{6.8}$$

In the case of a homogeneous magnetic field, Eq. (6.7) trivially reduces to the global effect, parameterized by a single coefficient $C_{\mathrm{CME}}$. This coincides with the zero momentum limit of Eq. (6.8), $\widetilde{C}_{\mathrm{CME}}(q_1 = 0)$, which we showed to be zero in QCD in the previous chapter. As we will show below, this picture is not changed by an inhomogeneous field. However, while Bloch's theorem forbids global currents to flow in equilibrium, it allows the appearance of local currents, and hence non-vanishing expectation values of $G(x_1)$ and $H(x_1, x_1')$. For the results in QCD, our main discussion in Sec. 6.4.1 revolves around $G(x_1)$, while the more general behavior of $H(x_1, x_1')$ is discussed in Sec. 6.4.2.

## 6.2 Lattice setup

We consider the same lattice setup for staggered fermions in QCD at the physical point as in Chs. 4 and 5, with the exception of the inhomogeneous magnetic field. For our specific profile (6.1), the quantization condition takes the form [170],

$$eB = \frac{3\pi N_b}{\epsilon L \tanh(L/2\epsilon)}, \quad \text{where } N_b \in \mathbb{Z}, \qquad (6.9)$$

with the following U(1) links,

$$\begin{aligned}
u_1(x_1, x_2) &= \begin{cases} \exp\bigl[-2iqB\epsilon(x_2 + L_2/2)\tanh\bigl(\tfrac{L_1}{2\epsilon}\bigr)\bigr] & \text{if } x_1 = L_1/2 - a\,, \\ 1 & \text{otherwise}\,, \end{cases} \\
u_2(x_1, x_2) &= \exp\biggl\{ iqB\epsilon a \biggl[\tanh\Bigl(\frac{x_1}{\epsilon}\Bigr) + \tanh\Bigl(\frac{L_1}{2\epsilon}\Bigr)\biggr] \biggr\}\,, \\
u_3(x_1, x_2) &= 1\,, \\
u_4(x_1, x_2) &= 1\,,
\end{aligned} \qquad (6.10)$$

where $L_1 = aN_s$ and $x_1 = an_1$.

The precise form of $H$ can be obtained from Eq. (5.10), by defining the projectors $\mathcal{P}_{x_1}$ on a slice of the lattice,

$$\begin{aligned}
H(x_1, x_1') = \frac{T}{4L^2} \sum_f \biggl[ &\frac{1}{4} \sum_{f'} \frac{q_f}{e} \Bigl\langle \mathrm{Tr}\Bigl[\mathcal{P}_{x_1} \Gamma_3^f M_f^{-1}\Bigr] \mathrm{Tr}\Bigl[\mathcal{P}_{x_1'} \Gamma_{45}^{f'} M_{f'}^{-1}\Bigr] \Bigr\rangle \\
&- \frac{q_f}{e} \Bigl\langle \mathrm{Tr}\Bigl[\mathcal{P}_{x_1} \Gamma_3^f M_f^{-1} \mathcal{P}_{x_1'} \Gamma_{45}^f M_f^{-1}\Bigr] \Bigr\rangle \\
&+ \frac{q_f}{e} \Bigl\langle \mathrm{Tr}\Bigl[\mathcal{P}_{x_1} \frac{\delta \Gamma_3^f}{\delta \mu_5(x_1')} M_f^{-1}\Bigr] \Bigr\rangle \biggr]\,.
\end{aligned} \qquad (6.11)$$

Summing over the $x_1'$ coordinate (i.e. the coordinate corresponding to the $\mu_5$ insertion) in (6.11) yields the correlator $G(x_1)$, in accordance with Eq. (6.7). The operator defined in Eq. (6.11) includes the conserved vector currrent (5.4) and the anomalous axial current (5.5), what is of vital importance for the study of the CME as we learned in the previous chapter.

## 6.3 Results for non-interacting fermions

The setup we have described in the previous sections is analytically solvable for a system of free fermions, which again yields an opportunity to test our lattice setup in the non-interacting case. In particular, the general behavior of the CME for an arbitrary magnetic field is required, which can be obtained through the calculation of the triangle diagram (C) in Fig. 3.2. In order to circumvent any regularization issue, we consider again the PV regularization, see Ref. [142] for further details on the calculation. The result for the momentum dependence of the CME coefficient $\widetilde{C}_{\mathrm{CME}}(q)$, defined by Eq. (6.8), is given by

$$\widetilde{C}_{\mathrm{CME}}(q_1) = -\frac{1}{2\pi^2 q_1} \int_0^\infty \mathrm{d}k\, k \left[ \frac{m^2(1/2 - n_F(E_k))}{E_k^3} - \frac{k^2}{E_k^2} n'_F(E_k) \right] \log \frac{(2k-q_1)^2}{(2k+q_1)^2} \\ -\frac{1}{2\pi^2}, \quad (6.12)$$

where $n_F(E_k)$ is the Fermi-Dirac distribution and $n'_F(E_k)$ is its derivative with respect to the energy $E_k = \sqrt{k^2 + m^2}$. We emphasize that the last term in this expression is a contribution coming solely from the regulator. By Fourier transforming $\widetilde{C}_{\mathrm{CME}}(q_1)$, we can obtain the observable $G(x_1)$ for a general magnetic field, which we choose to be the profile introduced in Eq. (6.1).

For the lattice calculation, we use the exact eigensystem of the staggered Dirac operator as in the previous results with free staggered fermions, to not rely on stochastic estimators, see App. B for a discussion on the required modifications of the formalism. Now we are ready to present the results. In Fig. 6.2, we show an example of the continuum extrapolation of the CME correlator for $m/T = 1$, $N_b = 2$ and aspect ratio $LT = 4$. As in Secs. 4.3.1 and 5.3.1, this choice was found to be sufficiently close to the thermodynamic limit so that finite volume effects are negligible. The continuum limit shows a local signal of the CME, which follows non-trivially the magnetic field profile, and agrees with the analytical calculation. The spatial structure features a current flowing in opposite directions in the center ($x_1 \approx 0$) and towards the edges ($|x_1| \gtrsim \epsilon$). This causes the $x_1$-integral of $G(x_1)$ to vanish, or in other words, $\widetilde{G}(q_1 = 0) = 0$, confirming that no global CME current flows in equilibrium, in agreement with the discussion in th previous chapter.

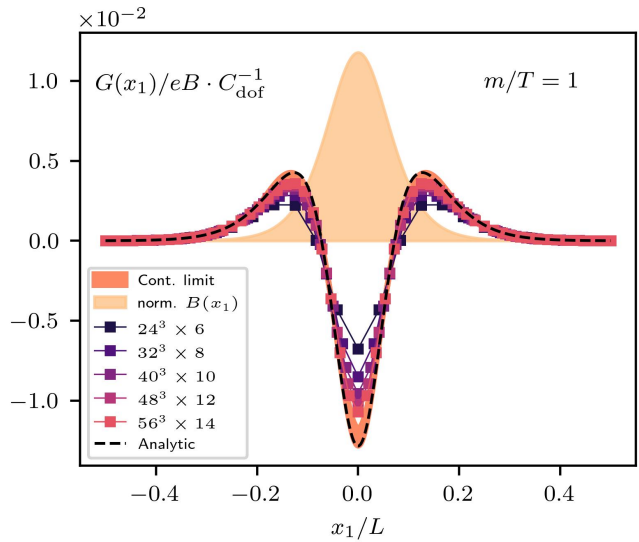

FIGURE 6.2: Lattice data and continuum extrapolation of the CME correlator with $eB/T^2 = 14.14$ and $\epsilon T = 1/3$, normalized by the magnetic field, for free fermions. The analytical result is recovered in the continuum limit, confirming the validity of our setup. For comparison, the shaded area depicts the magnetic field profile (6.1) in an arbitrary normalization. Figure from Ref. [142].

Therefore, this result does not contradict Bloch's theorem, and it can exist in thermal equilibrium. We refer to this phenomenon as **localized Chiral Magnetic Effect**.

Next, we discuss how the CME correlator is affected by the properties of the magnetic field profile. In the top panel of Fig. 6.3, we show the impact of changing the magnitude of the magnetic field by increasing $N_b$. We observe that the middle point ($x_1 = 0$) scales linearly for weak $B$, a behavior that is found to also persist in QCD, as we will discuss in Sec. 6.4.1. In the bottom panel, we show the change due to the width of the profile. By increasing $\epsilon$, the shape of the CME correlator flattens, until a homogeneous $B$ field is reached for $\epsilon \to \infty$. In this limit, we see that the correlator vanishes identically for every $x_1$, which corresponds to the case analyzed in Ch. 5.

The vanishing of the global CME can also be understood in momentum space, where the role of the regulator is more transparent. In Fig. 6.4, we show the Fourier transform of the CME coefficient, as defined in Eq. (6.8). As we have discussed in Sec. 3.3.1 and can be seen in Eq. (6.8), the Pauli-Villars regulator fields contribute with a constant term $-1/(2\pi^2)$, which is the value approached in the infinite momentum limit of Fig. 6.4. Due to this contribution, the total CME coefficient vanishes at zero momentum, in agreement with the results obtained for free fermions in the homogeneous case, see Sec. 5.3.1. We note that the higher momentum components are increasingly more difficult to extract on the lattice, since large momenta are tougher to resolve at finite lattice spacing. We perform the continuum limit by fitting the data including lattice artifacts up to quartic order in $a$ and excluding the coarsest lattice. The systematic error was estimated by performing similar fits, considering $\mathcal{O}(a^2)$ lattice artifacts and including all data points.

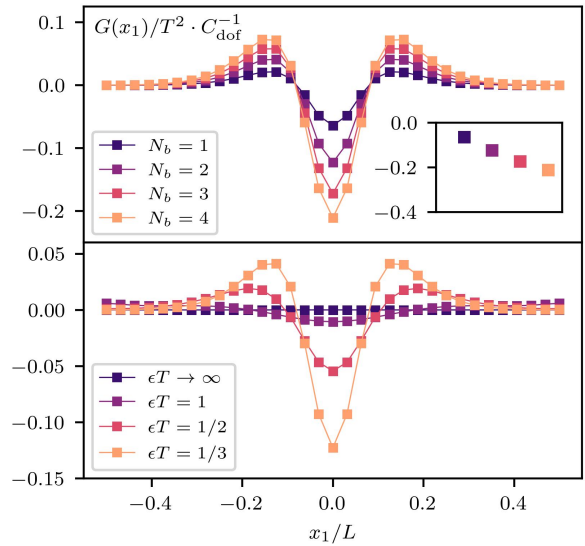

FIGURE 6.3: Top panel: CME correlator for free fermions as a function of $x_1/L$ for different values of $N_b$. The results correspond to a $32^3 \times 8$ lattice with $m/T = 1$ and $\epsilon T = 1/3$. The inset shows the magnetic field dependence of the central point. Bottom plot: $\epsilon$ dependence of the CME correlator, calculated on a $32^3 \times 8$ lattice with $m/T = 1$ and $N_b = 2$. Notice that the current vanishes in the limit of homogeneous magnetic fields, i.e. $\epsilon \to \infty$. Figure from Ref. [142].

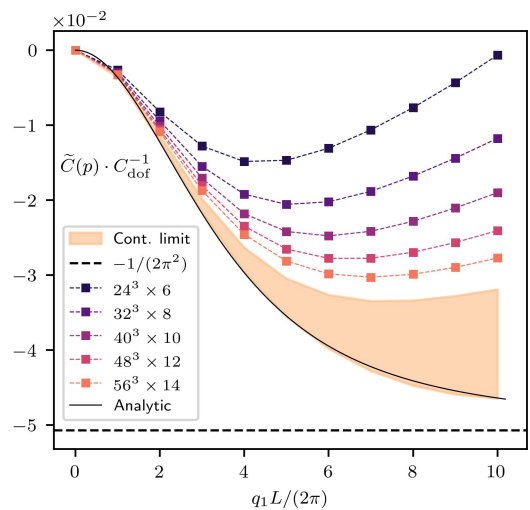

FIGURE 6.4: Fourier transform of the CME coefficient as a function of the momentum for different lattice sizes. The solid line represents the analytical result. In the infinite-momentum limit, the result converges to $-1/(2\pi^2)$, which arises purely from the Pauli-Villars regulator fields in the analytic calculation, see Eq. (6.12). Figure from Ref. [142].

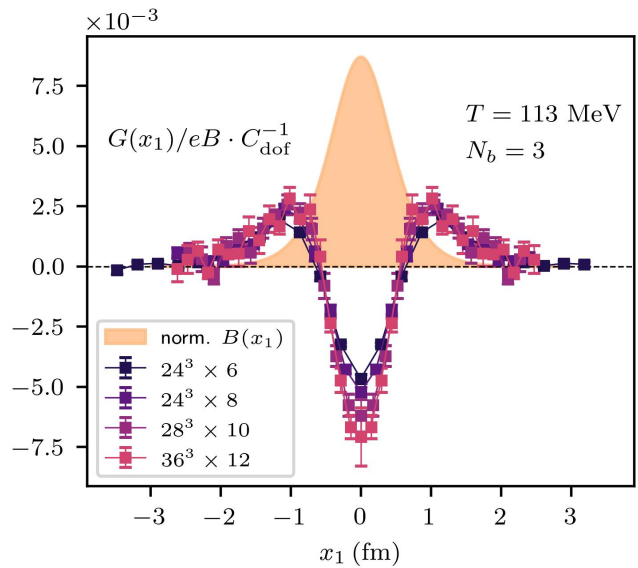

FIGURE 6.5: Lattice data for the CME correlator in QCD as a function of $x_1$ at $T = 113$ MeV and $N_b = 3$ for different lattice spacings. The connecting lines serve to guide the eye. For comparison, the shaded area depicts the magnetic field profile from Eq. (6.1) in an arbitrary normalization. Figure from Ref. [142].

The error determined in this way is shown by the orange band in Fig. 6.4, which overlaps with the analytical result.

We have found, both analytically and on the lattice, that a system of free fermions with an inhomogeneous magnetic field background develops a localized CME. Now we turn on the gluon interactions, to study if this interesting phenomenon is present in physical QCD.

## 6.4 Results in QCD

### 6.4.1 Homogeneous chiral chemical potential

We first consider the results for the $x_1$-dependence of the CME correlator (6.7) for a homogeneous $\mu_5$. In the QCD simulations, we choose the parameter $\epsilon \sim 0.6$ fm, motivated by HIC phenomenology [167].

In Fig. 6.5, we show $G(x_1)$ as a function of $x_1$ for different lattice spacings and a weak magnetic field.[6.2] The figure shows that in this background, the CME correlator develops

---
[6.2] Notice that, since the volumes of the lattices are not exactly equal (except for the $24^3 \times 8$ and $36^3 \times 12$ lattices), the precise value of $eB$ changes slightly for each of the considered lattices.

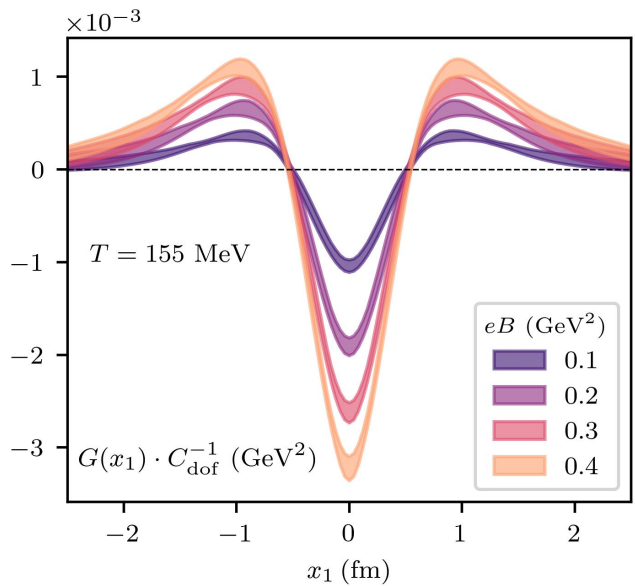

FIGURE 6.6: Continuum limit of the CME correlator in QCD as a function of $x_1$ at $T = 155$ MeV. The bands show this correlator at three different magnetic field strengths: $eB = 0.1, 0.2, 0.3$ and $0.4$ GeV$^2$. Figure from Ref. [142].

a non-trivial spatial structure, very similar to the behavior found for non-interacting fermions. In addition, using the results obtained on four different lattice spacings and different weak magnetic fields, we can carry out the continuum extrapolation of $G(x_1)$. Due to the higher complexity of this procedure compared to the two previous chapters, where only spline fits in the $T$ direction were required, we consider a different approach. In particular, we employ a multi-dimensional spline fit in $x_1$, $a$ and $eB$ with node points generated via Monte Carlo methods, see Refs. [66, 172] for a detailed discussion on the implementation. In Fig. 6.6, we show the obtained continuum limit of the $x_1$-dependent CME correlator for various values of $eB$ at the crossover temperature $T_c = 155$ MeV. We see that the localized CME signal is also present in the continuum limit, and it maintains its distinct features. This is a remarkable result, which shows that is possible to develop a non-zero local signal of the CME in QCD when an inhomogenous magnetic field is considered. We note that the integral of the continuum extrapolated results for $G(x_1)$ was checked to be consistent with zero, agreeing with Bloch's theorem and with the findings of the absence of the *global* CME in equilibrium QCD we presented in Ch. 5.

Now we turn to analyze the dependence of the observable with the magnetic field. In Fig. 6.7, we show the CME correlator at the center ($x_1 = 0$) and near the edge ($x_1 = 0.9$ fm) as a function of the magnetic field. An interesting property that can be observed is that the $B$-dependence is practically independent of the temperature in the studied range, maintaining a linear behavior at low magnetic fields below, at and above the QCD crossover temperature. Based on the slope of the $x_1 = 0$ curve in Fig. 6.7, we can quantify the strength of this effect by comparing it to the expected out-of-equilibrium CME conductivity for free massless fermions $C_{\text{CME}} = 1/(2\pi^2)$ (see the discussion in Sec. 3.3). We find the slope to be $-1/(2\pi^2) \times 0.19(2)$, hence suppressed by an order of magnitude with respect to the expected out-of-equilibrium conductivity. However, we want to emphasize that the precise value of this slope depends on the considered magnetic field profile, so the comparison above has to be understood just as a qualitative one.

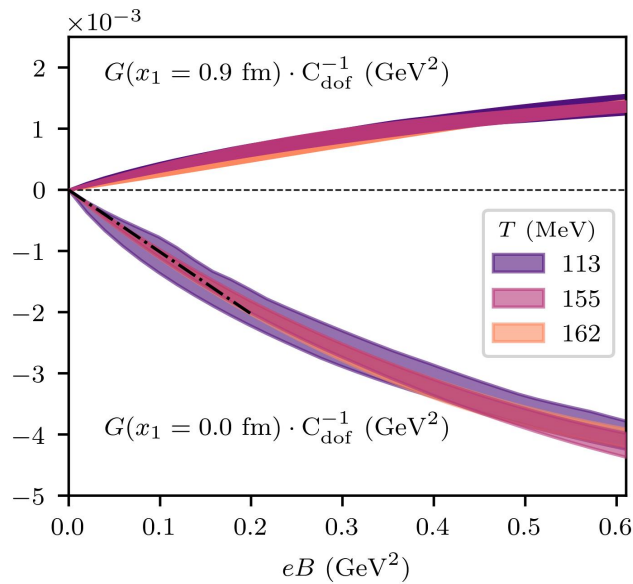

FIGURE 6.7: CME correlator in the continuum limit at $x_1 = 0$ fm and $x_1 = 0.9$ fm as a function of $eB$ for temperatures below, at, and above the crossover. The weak-field behavior of the correlator at the center is indicated by the dot-dashed line. Figure from Ref. [142].

### 6.4.2 Beyond homogeneous chiral chemical potentials

To finish this chapter, we can consider the most general situation possible, where the chiral chemical potential is also inhomogeneous in the $x_1$-direction. The induced current at point $x_1$ is now given by the correlator $H(x_1, x_1')$ from Eq. (6.5). This generalized correlator can be used to obtain a clearer picture of the role of $\mu_5$ in this effect. In particular, a possible interesting use of this operator is to convolute it with a physically motivated parameterization of the axial imbalance, yielding a more realistic picture of the localized CME.

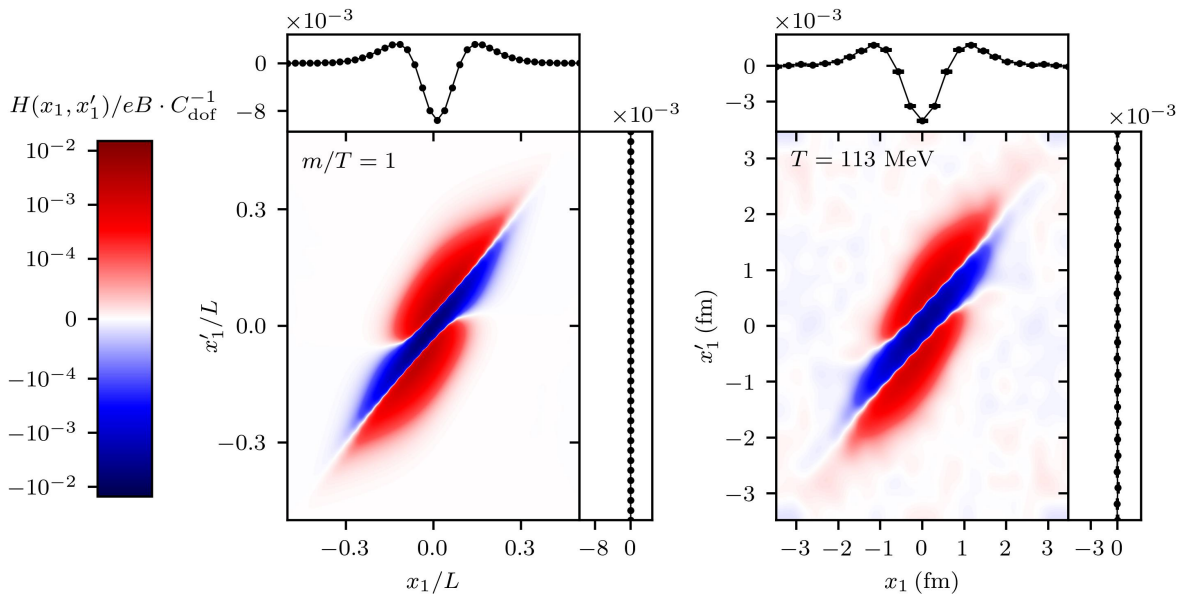

FIGURE 6.8: Left panel: heat plot of the correlator $H(x_1, x_1')$ normalized by $eB$ in the free case in the $x_1 - x_1'$ plane for $m/T = 1$ on a $40^3 \times 10$ lattice. The color scheme is shown on a logarithmic scale for the absolute value of the observable from 0.015 to $10^{-4}$, and a linear scale from the latter to 0. Red (blue) colors indicate the sign of the correlator. The projections on the top and right axes correspond to integrating over $x_1'$ and $x_1$, respectively. Right panel: the connected part of the same observable in QCD at $T = 113$ MeV on a $24^3 \times 6$ lattice. To highlight the main features of the heat plot, the data was smoothed via a bicubic interpolation. Figure from Ref. [142].

In Fig. 6.8, we show the result for $H(x_1, x_1')$ for a system of free fermions and full QCD at a temperature below the crossover temperature ($T = 113$ MeV). In the interacting case, we have neglected the disconnected contributions, the first term in Eq. (6.11), which were found to only enhance the statistical noise. The heat maps reveal more details about the local CME currents generated at different coordinates and their interplay. Let us consider the integral over one of the two coordinates of $H(x_1, x_1')$. The sum over $x_1'$ corresponds to integrating out the spatial dependence of $J_{45}$ (projection on the top axis of Fig. 6.8), which leads to the homogeneous $\mu_5$ effect and agrees with the profiles that we computed before, as expected. On the other hand, summing over the $x_1$ coordinate corresponds to the zero-momentum component of $J_3$ (projection on the right axis in Fig. 6.8), which vanishes due to Bloch's theorem.

Now we can summarize this chapter:

- We have presented a novel localized signal for the Chiral Magnetic Effect in global equilibrium. This effect is also present in physical QCD.

- The finding of this effect is not in contradiction with the results presented in Ch. 5, since the localized CME signal vanishes when integrating over the whole volume.

The existence of this local effect could have interesting consequences for the phenomenological modeling of the CME and its experimental search in heavy-ion collision experiments.

# Chapter 7

# Conclusions

## 7.1 Summary

In this thesis, we have presented a complete study of the Chiral Magnetic Effect (CME) and the Chiral Separation Effect (CSE) in QCD using lattice simulations. After introducing and motivating the study of anomalous transport phenomena, we have discussed the lattice gauge theory formulation of QCD in Ch. 2, focusing on the more relevant aspects for this thesis. We then continued with the theoretical foundation of anomalous transport in Ch. 3, in particular for CME and CSE. We explained and reviewed the arguments supporting that the Chiral Magnetic Effect does not exist in global thermal equilibrium, emphasizing that several results in the literature are in contradiction with this fact, in particular lattice QCD simulations with Wilson fermions [116, 117]. Furthermore, we showed the known results for the free CSE conductivity coefficient $C_{\text{CSE}}^{\text{free}}$ and how lattice simulations in two-color QCD point towards the CSE being suppressed at low temperatures [151].

We then moved to present the results achieved during this thesis. We began with the results for the Chiral Separation Effect in QCD in Ch. 4. We started by crosschecking our setup with free fermion simulations, where the result is known. After confirming the validity of our setup, we continued to quenched simulations with Wilson fermions, finding that the suppression of CSE observed in the SU(2) case in Ref. [151] is severe, yielding a vanishing result at $T \approx 0$. In this quenched setup, we studied a novel interplay between the Polyakov loop sectors at high temperature and $C_{\text{CSE}}$, where the sectors with non-zero imaginary part contribute to the effect as an imaginary chemical potential. Finally, we presented our results with staggered fermions in QCD at physical quark masses. We calculated $C_{\text{CSE}}$ in a wide range of temperatures, from $T \approx 0$ to 400 MeV, and performed a continuum extrapolation of the results. We found the CSE to be severely suppressed at $T \approx 0$, actually being zero within errors, and experiencing a sharp increase around the pseudocritical temperature of QCD $T_c = 155$ MeV. At high temperatures, the conductivity approaches the free fermion result $C_{\text{CSE}}^{\text{free}}(m/T = 0) = 1/(2\pi^2)$ as expected due to asymptotic freedom. This complete characterization of $C_{\text{CSE}}$ in QCD addresses the first research question Q1) we posed in Sec. 1.5. This is the first determination of the conductivity of an anomalous transport effect in physical QCD, and shows the power of lattice simulations to guide the theoretical understanding of these phenomena in the presence of strong interactions. Our results could play a role in the experimental search of the Chiral Magnetic Wave (CMW), since the temperature dependence of the CSE, at least in its equilibrium version, indicates that the CMW could be more prominent at high temperatures.

In Ch. 5, we presented our study of the Chiral Magnetic Effect in thermal equilibrium. We discussed the impact of the conserved versus non-conserved versions of the lattice vector current in this study of the CME, showing that for free fermions only the non-conserved version yields a non-zero CME. We then compared the same setup with quenched Wilson fermions from Refs. [116, 117], where a local (non-conserved) current was used and a non-vanishing $C_{\text{CME}}$ was obtained. We have seen that this non-zero result is an artifact of the use of the local vector current, and that the conserved lattice current yields a result compatible with zero for every lattice spacing studied. Finally, we showed that in QCD at the physical point with staggered fermions, a vanishing CME is found for all temperatures, confirming that the arguments for the absence of the CME in thermal equilibrium also apply in QCD. These studies address the second research question Q2), clarifying the nature of the CME in equilibrium QCD. We want to emphasize again that this absence of CME in equilibrium does not forbid its experimental detection, both in heavy-ion collision experiments and Weyl semi-metals, since these detections are also sensitive to out-of-equilibrium effects.

Finally in Ch. 6, we analyzed the effect that inhomogeneous magnetic fields, which are present in heavy-ion collisions, have on the Chiral Magnetic Effect. By observing that Bloch's theorem only forbids the presence of a volume-averaged CME, we observed a novel phenomenon: a localized CME current that develops in the direction of the magnetic field. Using continuum extrapolated results with staggered fermions, we show that this localized CME signal exists in physical QCD, respecting Bloch's theorem since it vanishes when the volume average is taken. This shows that even in thermal equilibrium, it is possible to observe a CME current if the magnetic field is inhomogeneous, which could have implications for the phenomenology of the CME. The findings of this chapter address and clarify the last research question Q3).

## 7.2 Outlook

Most of the discussion in this thesis has been directed at the global equilibrium aspects of the CSE and the CME. This is due to the inherent nature of lattice QCD simulations. However, it is possible to use Euclidean formalism to obtain information about the out-of-equilibrium properties of anomalous transport phenomena, by calculating the spectral function, which encodes this out-of-equilibrium information from the Euclidean correlator, see Ref. [173] for an introduction. Unfortunately, this is not easy: the spectral function can only be reconstructed by solving an ill-posed inverse problem, known as spectral reconstruction. Although many techniques have been developed to diminish the arising issues, it is an infamously difficult problem in the lattice community, as well as in many other sciences.

However, the out-of-equilibrium properties of anomalous transport phenomena, the CME in particular, are of great theoretical and phenomenological interest. The out-of-equilibrium CME can be probed indirectly through the calculation of the Ohmic conductivity in QCD at finite magnetic field,[7.1] which has been performed using lattice QCD combined with spectral reconstructions methods [126, 127]. However, a more direct and complete study of the out-of-equilibrium CME conductivity in QCD is missing, see Refs. [120, 174, 175], for some preliminary works in this direction. This motivates us to study this problem, using the methods and knowledge acquired in this thesis, which have led to a conference paper with preliminary results [176]. The calculation of the out-of-equilibrium CME conductivity would be of interest both for the theoretical and experimental high-energy physics community and it would help to clarify the nature of this effect in the presence of strong interactions.

## 7.3 Disclosure of author responsibility

With the appearance of new technologies allowing for fast and accessible communication, science has become more than ever a collaborative effort. Since one of the goals of a

---

[7.1] This way of studying the CME is very similar to idea behind its detection in Weyl-semimetals [105].

PhD thesis is to perform original and independent research, it is important to clarify the role of the author, in this case myself, in the publications whose results are presented in the dissertation. This thesis is based on two peer-reviewed papers [138, 152], one paper currently under revision [142] and a conference proceedings [177]. In Refs. [138, 152, 177], I:

- Derived and implemented the required formulas to calculate the CME and the CSE with non-interacting staggered fermions using the exact eigensystem, as described in App. B. I also performed the required simulations to compare with the analytical results to crosscheck the numerical setup.

- Implemented and measured the observables in the quenched approximation with Wilson fermions in an existing lattice code, as well as the generation of configurations when pre-existing ensembles were not available. In addition, I developed a CUDA C version of the inverter in the code to speed up the measurements.

- Implemented and measured the observables in full QCD with staggered fermions in an existing lattice code, as well as the generation of configurations when required.

- Performed the whole numerical analysis of the results, including the statistical analysis and the continuum limits.

In Ref. [142], I:

- Adapted the calculation for CME with non-interacting staggered fermions using the exact eigensystem to the case of a local observable, performing also the simulations and the analysis of the non-interacting fermionic theory, comparing again with the analytical result.

In all the aforementioned publications, I also took part in the writing of the manuscripts and presented the results at several international conferences.

I acknowledge financial support from the DFG (Collaborative Research Center CRC-TR 211 "Strong-interaction matter under extreme conditions" - project number 315477589 - TRR 211) and the Helmholtz Graduate School for Hadron and Ion Research (HGS-HIRe for FAIR). The numerical results presented in this thesis were obtained in the GPU Cluster at Bielefeld University.

# Appendix A

# Numerical methods in a nutshell

In this appendix, we will present the basics of the numerical methods which are usually employed in lattice gauge theory. When this formalism wants to be used in real calculations, it is obvious that analytical methods are unfeasible in many situations, given the number of degrees of freedom. For example, let us consider the calculation of a one-point function

$$\langle \bar{\psi}\psi \rangle = \frac{T}{V} \int \mathcal{D}U \, e^{-S_g} \det M \, \text{Tr} \, M^{-1} \, . \tag{A.1}$$

For definiteness, we consider a lattice with $N_s = N_t = 24$. For naive fermions, the matrix $M$ has a dimension $N_s^3 \times N_t \times 4 \times N_c = 3,981,312$. By looking at this number, it is clear that numerical methods are of central importance in lattice QCD. For example, in Eq. (A.1) one has to consider how to efficiently calculate the trace, determinant and inverse of a matrix of such a huge dimension. In addition, the integration under the Haar measure yields a very high-dimensional integral, of the same order of magnitude as the dimension of $M$, which is challenging to calculate even numerically.

We will not cover in detail all the numerical aspects required for lattice simulations, and we will only discuss briefly some basic aspects that are connected to the main text. We refer to lattice QCD textbooks like Refs. [29–31] for a more detailed introduction. In particular, this appendix follows closely Ref. [31].

## A.1 Monte Carlo methods

Let us consider a simple one dimensional integral of function $f(x)$ in an interval $(a,b)$ with respect to a probability distribution $\rho(x)$, i.e. its expectation value

$$\langle f \rangle = \frac{\int_a^b \mathrm{d}x\, \rho(x) f(x)}{\int_a^b \mathrm{d}x\, \rho(x)}\,. \tag{A.2}$$

Consider a sample of $N$ points $\{x_i\}_{i=1}^N$ where $x_n$ is ditributed according to $\rho(x)$. Then it can be shown that

$$\langle f \rangle = \sum_{i=1}^N f(x_i) + \mathcal{O}\!\left(\frac{1}{\sqrt{N}}\right). \tag{A.3}$$

In the limit of infinite samples, the total expectation value is recovered. This integration method is known as **importance sampling Monte Carlo**.

On the lattice, the statistical distribution is given by the fermionic determinant and $e^{-S_g}$, known as the Boltzmann weight. Therefore, we can make use of Monte Carlo methods to perform the path integral. The normalized probability density is given by

$$\rho[U] = \frac{1}{\mathcal{Z}}\, e^{-S_G} \det M\,. \tag{A.4}$$

We will denote the total probability associated with this density as $P_\rho[U]$. The observables are now evaluated in a set of links distributed according to this distribution. Each of the elements $\mathcal{U}_i$ of the sets $\{\mathcal{U}_i\}_{i=1}^N$ are known as a **configuration**. Then the expectation value of an observable $O$ is given by

$$\langle O \rangle = \sum_{i=1}^N O[\mathcal{U}_i] + \sigma_E\,, \tag{A.5}$$

where $\sigma_E$ is the error due to the truncation of the sum. At first glance, it would seem like this is simply the standard deviation, however, this is not entirely correct as we will discuss below. The important concept is that it can be shown that $\sigma_E \sim \mathcal{O}(1/\sqrt{N})$, so it is always possible to decrease the error by the generation of more configurations.

But the problem is, how to generate these configurations in the first place? A possibility consists of using a stochastic process called **Markov proccesses**. The defining characteristic of these processes is that they have *no memory*. This means that the next step in the chain depends only on the current configuration, and it is blind to the preceding part of the chain. Formally, a Markov process is characterized by a transition probability $P(\mathcal{U}_n = \mathcal{U}_j | \mathcal{U}_{n-1} = \mathcal{U}_i) \equiv T_{ij}$ of hopping from the configuration $\mathcal{U}_i$ to the configuration $\mathcal{U}_j$, which clearly has to obey

$$0 \leq T_{ij} \leq 1, \tag{A.6}$$

$$\sum_j T_{ij} = 1. \tag{A.7}$$

The no-memory property is reflected in the **balance condition** with respect to the target probability $P_\rho$,

$$\sum_i T_{ij} P_\rho[\mathcal{U}_i] = \sum_j T_{ji} P_\rho[\mathcal{U}_j]. \tag{A.8}$$

This expression implies that the total probability of ending in the configuration $\mathcal{U}_j$ is equal to the total probability of hopping out of $\mathcal{U}_j$, i.e. there are no sources or sinks of probabilities. This equation can also be shown to imply that the target probability $P_\rho$ is a *fixed point* of the Markov Chain. This means that the distribution will eventually be reached upon iterative applications of $T$, guaranteeing the convergence of the Monte Carlo integration. A particular solution of the balance equation is imposing the condition term by term

$$T_{ij} P_\rho[\mathcal{U}_i] = T_{ji} P_\rho[\mathcal{U}_j]. \tag{A.9}$$

This is known as **detailed balance condition**. Another important property that Markov processes have to fulfill is **ergodicity**, which means that every possible configuration in the space can be reached in a finite amount of steps.

There are many different possible ways of implementing these conditions, that correspond to different Monte Carlo algorithms. For completeness, we will briefly discuss one of the better-known ones, the **Metropolis algorithm**:

Step 1: Choose a candidate configuration $\mathcal{U}_j$ from the current configuration $\mathcal{U}_i$ according to an *a priori* selection probability $T_{ij}^O$. We consider this probability to be symmetric $T_{ij}^O = T_{ji}^O$, although this is not required.

Step 2: Accept the candidate configuration $\mathcal{U}_j$ in the Markov chain configuration with a probability

$$T_{ij}^A = \min\left(1, \frac{P_\rho[\mathcal{U}_j]}{P_\rho[\mathcal{U}_i]}\right) \tag{A.10}$$

If the change is rejected, the current configuration is included again in the chain.

Step 3: Repeat from Step 1.

The total probability $T = T^O T^A$ can be shown to fulfill the detailed balance condition (A.9). This algorithm has a very transparent interpretation of the acceptance probability $T^A$. The new configuration is always accepted if it *decreases* the quantum action $P_\rho$, i.e. if $P_\rho[\mathcal{U}_j]/P_\rho[\mathcal{U}_i] > 1$, and it accepts configurations that increase it with a certain probability, but penalizing large fluctuations. This is a reflection of the principle of least action in quantum field theory.

The Metropolis algorithm is a good and convenient choice to simulate theories without a fermionic determinant, like quenched QCD. However, in simulations with dynamical fermions, the Metropolis algorithm is usually substituted for more elaborated algorithms. One of the most popular choices for simulations in QCD is the **Hybrid Monte Carlo algorithm** [178], which has been used for the generation of configurations in this thesis. We will not cover this algorithm in further detail and we refer to Refs. [30, 31] for a detailed discussion.

After discussing Markov chains, we are ready to understand one subtlety regarding the estimation of the statistical error of the observables. Since the configurations are coming from the Markov process where each step depends on the one before, the value of the observable in one configuration is correlated with its value in another configuration coming from the same chain. This correlation is unphysical, it is just an artifact of the Markov chain and thus should be avoided. An easy strategy consists of considering only configurations separated by a given distance in the Markov chain. However, the autocorrelation is still present and our statistical analysis should reflect it. This is the reason why the standard deviation is not a good estimator for the statistical error of the observable and one has to rely on other methods. In our case, we have considered the well-known **Jackknife method**, see Ref. [29] for a detailed discussion.

## A.2   Inversions and traces

Given the high dimensionality of $M$, the calculation of its inverse is a challenging task. An efficient method to perform this operation is to consider a slightly modified problem, the application of the inverse matrix to a vector $b$, which can be further modified to

$$M^{-1}b = x \quad \Leftrightarrow \quad Mx = b \,, \tag{A.11}$$

Since $M$ and $b$ are known, we have transformed the inversion problem into solving a simple system of linear equations for $x$. A common method to solve this problem in lattice simulations is the **conjugate gradient algorithm** [179].

Given that we can only calculate $M^{-1}$ applied to a vector, how can we calculate $\text{tr}[M^{-1}]$? In principle, one could obtain the eigenvector of the $M$ and calculate the trace exactly. However, this is also a very costly method, which is the reason why it is usually more convenient to use stochastic techniques to calculate the trace, in particular, we will discuss now the method of **noisy estimators**. We consider a set of uncorrelated complex random numbers $\xi_i$, distributed according to a distribution with unit variance, for example a Gaussian distribution or a $Z_N$ distribution. We can then built vectors $\xi^k$ with components $\xi_i^k$. Asymptotically, the distribution of the $\xi$-s guarantees that

$$\sum_{k=1}^{\infty} \xi_i^k \, (\xi_j^k)^* = \delta_{ij} \,. \tag{A.12}$$

Using this property, the trace of the matrix $M^{-1}$ can be calculated as

$$\text{tr}\left[M^{-1}\right] = \sum_{k=1}^{\infty} (\xi^k)^\dagger \, M^{-1} \, \xi^k \,. \tag{A.13}$$

For obvious reasons, in numerical simulations it is not feasible to use an infinite number of vectors. We can truncate the sum, averaging over $N_V$ vectors,

$$\text{tr}\left[M^{-1}\right] \approx \sum_{k=1}^{N_V} (\xi^k)^\dagger \, M^{-1} \, \xi^k \,. \tag{A.14}$$

The precise value of $N_V$ needs to be tuned according to the desired accuracy for each observable.

The noisy estimators method combined with the conjugate gradient algorithm can also be used to calculate the fermionic determinant $\det M$, which is required for Monte Carlo integration as we discussed above. We will not discuss this here, see for example Ref. [31], but it shows that the calculation of the inverse of the Dirac operator is of central importance in lattice simulations, and reducing the required number of inversions correlates directly with lowering the computational cost of lattice QCD.

To close this appendix, we would like to emphasize the importance of High Performance Computing (HPC) in lattice simulations. The development of novel algorithms needs to be accompanied by the means to simulate sizeable lattices in the relevant parameter space, all in a sensible amount of computer time. Therefore, the success of lattice QCD cannot be understood without the sophistication of hardware and software in the last 20 years. A crucial part of lattice simulations is the parallelization of calculations, which achieves tremendous speed-ups. This can be done using sets of coordinated Central Processing Units (CPUs) or more recently, through Graphical Processing Units (GPUs). The proliferation and refinement of these tools in the last years have given lattice QCD simulations a needed boost to achieve calculations in full QCD at physical and lower-than-physical quark masses, a milestone that was completely out of reach in the early days of lattice QCD. As a consequence, HPC programming skills have become a very useful asset in the toolbox of lattice practitioners.

# Appendix B

# Free staggered fermions

In this appendix, we illustrate how to calculate Eqs. (4.5), (5.1), and (6.11) for free staggered quarks using explicitly the eigenvalues and eigenvectors of the Dirac operator. Therefore, here we turn off the gluon fields and consider one color component only single quark flavor with mass $m$, charge $e$, one quark/chiral chemical potential $\mu/\mu_5$ and one conserved vector/anomalous axial current $J_3/J_{35}$. Below we work on an $N_s^3 \times N_t$ lattice and write everything in lattice units, setting $a = 1$. In the non-interacting theory, the disconnected term vanishes, so we only need to calculate the connected part and the tadpole term in the expressions.

## B.1 Chiral Separation Effect

Let us consider first Eq. (4.5) in this simplified setup,

$$\left.\frac{\partial J_{35}}{\partial \mu}\right|_{\mu=0} = \frac{1}{N_s^3 N_t}\left[-\frac{1}{4}\text{Tr}\left(\Gamma_4 M^{-1}\Gamma_{35}M^{-1}\right) + \frac{1}{4}\text{Tr}\left(\frac{\partial \Gamma_{35}}{\partial \mu}M^{-1}\right)\right], \quad \text{(B.1)}$$

where the expectation value indicating the fermion path integral was omitted for brevity. Here we consider the operators $\Gamma_\nu$ and $\Gamma_{\nu5}$ defined in Eq. (4.10), (4.11).[B.1] Furthermore, for convenience we redefine the staggered phases[B.2] as

$$\eta_1(n) = 1, \quad \eta_2(n) = (-1)^{n_1}, \quad \eta_3(n) = (-1)^{n_1+n_2+n_4}, \quad \eta_4(n) = (-1)^{n_1+n_2}. \quad \text{(B.2)}$$

---

[B.1] An analogous calculation holds for the other cases considered in Fig. 4.3, we give the corresponding expressions at the end of the section.

[B.2] Notice that this corresponds to choose $T(n) = \gamma_1^{n_1}\gamma_2^{n_2}\gamma_4^{n_4}\gamma_3^{n_3}$ for the staggered transformation. This change will be beneficial to analyze the case of CME in the next two sections.

The massless staggered Dirac operator $\slashed{D} = \sum_\nu \eta_\nu D_\nu$, defined by Eq. (2.137), is antihermitian, $\slashed{D}^\dagger = -\slashed{D}$, therefore its eigenvalues are purely imaginary. Moreover, due to staggered chiral symmetry, $\{\slashed{D}, \eta_5\} = 0$, the eigenvalues come in complex conjugate pairs,

$$\slashed{D}\Psi_i = \pm i\lambda_i \Psi_i \,. \tag{B.3}$$

In addition, we will need the analogous eigensystem for the massive operator $MM^\dagger = (\slashed{D}+m)(\slashed{D}+m)^\dagger = \slashed{D}\slashed{D}^\dagger + m^2$, so

$$MM^\dagger \Psi_i = (\lambda_i^2 + m^2)\Psi_i \,. \tag{B.4}$$

Notice that each eigenvalue is doubly degenerate due to the conjugate pairs in Eq. (B.3). Using this eigensystem as a basis, the traces in Eq. (B.1) can be written as

$$\left.\frac{\partial J_{35}}{\partial \mu}\right|_{\mu=0} = \frac{1}{N_s^3 N_t} \left[ -\frac{1}{4} \sum_{i,j} \frac{1}{(\lambda_i^2+m^2)(\lambda_j^2+m^2)} \Psi_i^\dagger \Gamma_4 M^\dagger \Psi_j \Psi_j^\dagger \Gamma_{35} M^\dagger \Psi_i \right. \\ \left. + \frac{1}{4} \sum_i \frac{1}{\lambda_i^2+m^2} \Psi_i^\dagger \frac{\partial \Gamma_{35}}{\partial \mu} M^\dagger \Psi_i \right], \tag{B.5}$$

where we inserted a complete set of eigenstates $\sum_j \Psi_j \Psi_j^\dagger = \mathbb{1}$ in the first term and used $M^{-1} = M^\dagger (MM^\dagger)^{-1}$.

Next, we will exploit the separability of the problem and reduce Eq. (B.4) to one two-dimensional and two one-dimensional eigenproblems. This property allows us to determine the complete spectrum on much larger lattices than in the full four-dimensional case. To this end, we write the free Dirac operator as $\slashed{D} = \slashed{D}_{12} + \slashed{D}_3 + \slashed{D}_4$ with $\slashed{D}_{12} = \slashed{D}_1 + \slashed{D}_2$ and the explicit form of the Dirac operator

$$\slashed{D}_\nu(n,m) = \frac{\eta_\nu(n)}{2}\left[u_\nu(n)\,e^{\mu\delta_{\nu 4}}\,\delta_{n+\hat\nu,m} - u_\nu^*(n-\hat\nu)\,e^{-\mu\delta_{\nu 4}}\,\delta_{n-\hat\nu,m}\right] = \eta_\nu(n) D_\nu(n,m)\,. \tag{B.6}$$

Similarly, the staggered Dirac matrices (4.10) simplify to

$$\Gamma_\nu(n,m) = \frac{\eta_\nu(n)}{2}\left[u_\nu(n)\,e^{\mu\delta_{\nu 4}}\,\delta_{n+\hat\nu,m} + u_\nu^*(n-\hat\nu)\,e^{-\mu\delta_{\nu 4}}\,\delta_{n-\hat\nu,m}\right] \equiv \eta_\nu(n) S_\nu(n,m)\,. \tag{B.7}$$

Notice that the operators $S_\nu$ satisfy the property

$$[S_1, S_3] = [S_1, S_4] = [S_2, S_3] = [S_2, S_4] = [S_3, S_4] = 0\,, \tag{B.8}$$

because the U(1) links only enter in $S_1$ and $S_2$, see Eq. (2.179). Moreover,

$$\left.\frac{\partial S_4}{\partial \mu}\right|_{\mu=0} = D_4 \,, \tag{B.9}$$

due to the way that the chemical potential enters the temporal hoppings.

Now we consider the squared massive Dirac operator $MM^\dagger = MM_{12}^\dagger + MM_3^\dagger + MM_4^\dagger$, since it separates into three terms that act in the respective subspaces and only depend on the corresponding coordinates, thus commuting with each other. Therefore, the eigenvectors in (B.4) factorize,

$$\Psi_{\{i_{12},i_3,i_4\}}(n_1, n_2, n_3, n_4) = \rho_{i_{12}}(n_1, n_2) \, \phi_{i_3}(n_3) \, \xi_{i_4}(n_4) \,, \tag{B.10}$$

with $0 \le i_{12} < N_s^2$, $0 \le i_3 < N_s$ and $0 \le i_4 < N_t$.[B.3] Below we use a shorthand notation and simply write $\Psi_i = \rho_i \phi_i \xi_i$. Notice that the eigenvalues and eigenvectors in the 3, 4 directions are calculated at $B = 0$.

We proceed by expanding the operators appearing in the matrix elements in Eq. (B.5) into separate contributions that depend only on $n_{1,2}$, $n_3$ or $n_4$,

$$\Gamma_4 = (-1)^{n_1+n_2} S_4 \,, \tag{B.11}$$

$$M^\dagger = -\slashed{D}_{12} - (-1)^{n_1+n_2+n_4} D_3 - (-1)^{n_1+n_2} D_4 + m \,. \tag{B.12}$$

Concerning the operator $\Gamma_{35}$, it can be shown with some algebra and Eq. (B.8) that the combinatorial factors appearing in Eq. (4.11) cancel with permutations of the operators once the staggeredd phases $\eta_\nu$ are factorized, yielding,

$$\Gamma_{35} = S_{12} S_4 (-1)^{n_2} \,, \tag{B.13}$$

with $S_{12} = \{S_1, S_2\}/2$. Finally, for the tadpole term we need the derivative of $\Gamma_{35}$ with respect to the chemical potential. Since the latter only appears in $S_4$, we obtain

$$\frac{\partial \Gamma_{35}}{\partial \mu} = S_{12} D_4 (-1)^{n_2} \,. \tag{B.14}$$

where we have used Eq. (B.9).

---

[B.3] Note that the same separation does not hold for the eigensystem (B.3), since for example $[\slashed{D}_{12}, \slashed{D}_3] \ne 0$ due to the staggered phases. This is because $[MM^\dagger, \eta_5] = 0$ but $[\slashed{D}, \eta_5] \ne 0$.

Considering all these properties, the necessary products appearing in Eq. (B.5) are

$$\begin{aligned}
\Gamma_4 M^\dagger = & -[(-1)^{n_1+n_2} \slashed{D}_{12}]_{12} \cdot [\mathbb{1}]_3 \cdot [S_4]_4 \\
& -[\mathbb{1}]_{12} \cdot [D_3]_3 \cdot [S_4(-1)^{n_4}]_4 \\
& -[\mathbb{1}]_{12} \cdot [\mathbb{1}]_3 \cdot [S_4 D_4]_4 \\
& +m[(-1)^{n_1+n_2}]_{12} \cdot [\mathbb{1}]_3 \cdot [S_4]_4 \,,
\end{aligned} \quad (B.15)$$

where $[.]_{12}$, $[.]_3$ and $[.]_4$ denotes operators that only act in the corresponding spaces and only depend on the respective coordinates. Similarly, we obtain

$$\begin{aligned}
\Gamma_{35} M^\dagger = & -[S_{12}(-1)^{n_2} \slashed{D}_{12}]_{12} \cdot [\mathbb{1}]_3 \cdot [S_4]_4 \\
& -[S_{12}(-1)^{n_1}]_{12} \cdot [D_3]_3 \cdot [S_4(-1)^{n_4}]_4 \\
& -[S_{12}(-1)^{n_1}]_{12} \cdot [\mathbb{1}]_3 \cdot [S_4 D_4]_4 \\
& +m[S_{12}(-1)^{n_2}]_{12} \cdot [\mathbb{1}]_3 \cdot [S_4]_4 \,,
\end{aligned} \quad (B.16)$$

and

$$\begin{aligned}
\frac{\partial \Gamma_{35}}{\partial \mu} M^\dagger = & -[S_{12}(-1)^{n_2} \slashed{D}_{12}]_{12} \cdot [\mathbb{1}]_3 \cdot [D_4]_4 \\
& -[S_{12}(-1)^{n_1}]_{12} \cdot [D_3]_3 \cdot [D_4(-1)^{n_4}]_4 \\
& -[S_{12}(-1)^{n_1}]_{12} \cdot [\mathbb{1}]_3 \cdot [D_4 D_4]_4 \\
& +m[S_{12}(-1)^{n_2}]_{12} \cdot [\mathbb{1}]_3 \cdot [D_4]_4 \,.
\end{aligned} \quad (B.17)$$

Finally, we can define

$$\begin{aligned}
A_{ij} &\equiv \rho_i^\dagger (-1)^{n_1+n_2} \slashed{D}_{12} \rho_j \,, & B_{ij} &\equiv \rho_i^\dagger (-1)^{n_1+n_2} \rho_j \,, \\
C_{ij} &\equiv \rho_i^\dagger S_{12}(-1)^{n_2} \slashed{D}_{12} \rho_j \,, & E_{ij} &\equiv \rho_i^\dagger S_{12}(-1)^{n_1} \rho_j \,, \\
F_{ij} &\equiv \rho_i^\dagger S_{12}(-1)^{n_2} \rho_j \,, & J_{ij} &\equiv \phi_i^\dagger D_3 \phi_j \,, \\
G_{ij} &\equiv \xi_i^\dagger S_4 \xi_j \,, & H_{ij} &\equiv \xi_i^\dagger S_4 D_4 \xi_j \,, \\
I_{ij} &\equiv \xi_i^\dagger S_4 (-1)^{n_4} \xi_j \,, & K_{ij} &\equiv \xi_i^\dagger D_4 \xi_j \,, \\
L_{ij} &\equiv \xi_i^\dagger D_4 D_4 \xi_j \,, & N_{ij} &\equiv \xi_i^\dagger D_4 (-1)^{n_4} \xi_j \,,
\end{aligned} \quad (B.18)$$

so that Eq. (B.5) becomes

$$\left.\frac{\partial J_{35}}{\partial \mu}\right|_{\mu=0} = -\frac{1}{N_s^3 N_t}\frac{1}{4}\sum_{i,j}\frac{1}{(\lambda_i^2+m^2)(\lambda_j^2+m^2)}$$
$$\left\{[A_{ij}C_{ji}]_{12}\cdot[\delta_{ij}]_3\cdot[G_{ij}G_{ji}]_4 + [A_{ij}E_{ji}]_{12}\cdot[\delta_{ij}]_3\cdot[G_{ij}H_{ji}]_4\right.$$
$$+ [A_{ij}E_{ji}]_{12}\cdot[\delta_{ij}J_{ji}]_3\cdot[G_{ij}I_{ji}]_4 - m[A_{ij}F_{ji}]_{12}\cdot[\delta_{ij}]_3\cdot[G_{ij}G_{ji}]_4$$
$$+ [\delta_{ij}C_{ji}]_{12}\cdot[\delta_{ij}]_3\cdot[H_{ij}G_{ji}]_4 + [\delta_{ij}E_{ji}]_{12}\cdot[\delta_{ij}]_3\cdot[H_{ij}H_{ji}]_4$$
$$+ [\delta_{ij}E_{ji}]_{12}\cdot[\delta_{ij}J_{ji}]_3\cdot[H_{ij}I_{ji}]_4 - m[\delta_{ij}F_{ji}]_{12}\cdot[\delta_{ij}]_3\cdot[H_{ij}G_{ji}]_4$$
$$+ [\delta_{ij}C_{ji}]_{12}\cdot[J_{ij}\delta_{ji}]_3\cdot[I_{ij}G_{ji}]_4 + [\delta_{ij}E_{ji}]_{12}\cdot[J_{ij}\delta_{ji}]_3\cdot[I_{ij}H_{ji}]_4$$
$$+ [\delta_{ij}E_{ji}]_{12}\cdot[J_{ij}J_{ji}]_3\cdot[I_{ij}I_{ji}]_4 - m[\delta_{ij}F_{ji}]_{12}\cdot[J_{ij}\delta_{ji}]_3\cdot[I_{ij}G_{ji}]_4$$
$$- m[B_{ij}C_{ji}]_{12}\cdot[\delta_{ij}]_3\cdot[G_{ij}G_{ji}]_4 - m[B_{ij}E_{ji}]_{12}\cdot[\delta_{ij}]_3\cdot[G_{ij}H_{ji}]_4$$
$$\left.- m[B_{ij}E_{ji}]_{12}\cdot[\delta_{ij}J_{ji}]_3\cdot[G_{ij}I_{ji}]_4 + m^2[B_{ij}F_{ji}]_{12}\cdot[\delta_{ij}]_3\cdot[G_{ij}G_{ji}]_4\right\}$$
$$- \frac{1}{N_s^3 N_t}\frac{1}{4}\sum_i \frac{1}{\lambda_i^2+m^2}$$
$$\left\{[C_{ii}]_{12}\cdot[\delta_{ii}]_3\cdot[K_{ii}]_4 + [E_{ii}]_{12}\cdot[J_{ii}]_3\cdot[N_{ii}]_4\right.$$
$$\left.+ [E_{ii}]_{12}\cdot[\delta_{ii}]_3\cdot[L_{ii}]_4 - m[F_{ii}]_{12}\cdot[\delta_{ii}]_3\cdot[K_{ii}]_4\right\}. \quad \text{(B.19)}$$

Notice that the appearance of the Kronecker deltas is a consequence of the orthonormality of the eigenvectors.

We employed the LAPACK library to separately solve the eigenvalue problems. The eigensystem in the 12 plane gives the Hofstadter's butterfly [180], the spectrum of a well-known solid-state physics model. This problem is also relevant in lattice QCD, as pointed out in Ref. [181] and its generalization to the presence of gluonic interactions [182, 183] and inhomogeneous magnetic fields [170]. The coefficient $C_{\text{CSE}}$ can be obtained using the techniques described in Sec. 4.3.

Similarly, we can obtain the expressions for the observables considered in Fig. 4.3. In the case of $\Gamma_\nu^{B=0}, \Gamma_{\nu 5}^{B=0}$ defined in Eq. (4.14), it is straightforward to see that the expression is equivalent to Eq. (B.19) with the substitution $S_{12} = S_1 S_2$, since the operators $S_1$ and $S_2$ commute when the U(1) links are not included.

The case using $\Gamma_\nu \Gamma_5$ as the (5-link) axial current, defined in Eq. (4.13), requires more work. The observable takes the form

$$\left.\frac{\partial J_{35}^{\text{5-link}}}{\partial \mu}\right|_{\mu=0} = \frac{1}{N_s^3 N_t}\left[-\frac{1}{4}\text{Tr}\left(\Gamma_4 M^{-1}\Gamma_3\Gamma_5 M^{-1}\right) + \frac{1}{4}\text{Tr}\left(\frac{\partial\,(\Gamma_3\Gamma_5)}{\partial \mu} M^{-1}\right)\right]. \quad \text{(B.20)}$$

Similarly as above, we can factor out the staggered phases of the operators, obtaining

$$\Gamma_5 = S_1 S_2 S_3 S_4 (-1)^{n_1+n_4}, \tag{B.21}$$

and

$$\Gamma_3 \Gamma_5 = -S_3 S_1 S_2 S_3 S_4 (-1)^{n_2}. \tag{B.22}$$

The tadpole term then yields

$$\frac{\partial (\Gamma_3 \Gamma_5)}{\partial \mu} = -S_3 S_1 S_2 S_3 D_4 (-1)^{n_2}. \tag{B.23}$$

With this we can construct the explicit formula for Eq. (B.20), which compared to Eq. (B.19) only gets modified in the matrix elements in the third spatial direction and by a global minus sign,

$$
\begin{aligned}
\left.\frac{\partial J_{35}^{\text{5-link}}}{\partial \mu}\right|_{\mu=0} = &+\frac{1}{N_s^3 N_t}\frac{1}{4}\sum_{i,j}\frac{1}{(\lambda_i^2+m^2)(\lambda_j^2+m^2)} \\
&\Big\{[A_{ij}C_{ji}]_{12}\cdot[\delta_{ij}O_{ji}]_3\cdot[G_{ij}G_{ji}]_4 + [A_{ij}E_{ji}]_{12}\cdot[\delta_{ij}O_{ji}]_3\cdot[G_{ij}H_{ji}]_4 \\
&+ [A_{ij}E_{ji}]_{12}\cdot[\delta_{ij}P_{ji}]_3\cdot[G_{ij}I_{ji}]_4 - m[A_{ij}F_{ji}]_{12}\cdot[\delta_{ij}O_{ji}]_3\cdot[G_{ij}G_{ji}]_4 \\
&+ [\delta_{ij}C_{ji}]_{12}\cdot[\delta_{ij}O_{ji}]_3\cdot[H_{ij}G_{ji}]_4 + [\delta_{ij}E_{ji}]_{12}\cdot[\delta_{ij}O_{ji}]_3\cdot[H_{ij}H_{ji}]_4 \\
&+ [\delta_{ij}E_{ji}]_{12}\cdot[\delta_{ij}P_{ji}]_3\cdot[H_{ij}I_{ji}]_4 - m[\delta_{ij}F_{ji}]_{12}\cdot[\delta_{ij}O_{ji}]_3\cdot[H_{ij}G_{ji}]_4 \\
&+ [\delta_{ij}C_{ji}]_{12}\cdot[J_{ij}O_{ji}]_3\cdot[I_{ij}G_{ji}]_4 + [\delta_{ij}E_{ji}]_{12}\cdot[J_{ij}O_{ji}]_3\cdot[I_{ij}H_{ji}]_4 \\
&+ [\delta_{ij}E_{ji}]_{12}\cdot[J_{ij}P_{ji}]_3\cdot[I_{ij}I_{ji}]_4 - m[\delta_{ij}F_{ji}]_{12}\cdot[J_{ij}O_{ij}]_3\cdot[I_{ij}G_{ji}]_4 \\
&- m[B_{ij}C_{ji}]_{12}\cdot[\delta_{ij}O_{ji}]_3\cdot[G_{ij}G_{ji}]_4 - m[B_{ij}E_{ji}]_{12}\cdot[\delta_{ij}O_{ji}]_3\cdot[G_{ij}H_{ji}]_4 \\
&- m[B_{ij}E_{ji}]_{12}\cdot[\delta_{ij}P_{ji}]_3\cdot[G_{ij}I_{ji}]_4 + m^2[B_{ij}F_{ji}]_{12}\cdot[\delta_{ij}O_{ji}]_3\cdot[G_{ij}G_{ji}]_4\Big\} \\
&+ \frac{1}{N_s^3 N_t}\frac{1}{4}\sum_i \frac{1}{\lambda_i^2+m^2} \\
&\Big\{[C_{ii}]_{12}\cdot[O_{ii}]_3\cdot[K_{ii}]_4 + [E_{ii}]_{12}\cdot[P_{ii}]_3\cdot[N_{ii}]_4 \\
&+ [E_{ii}]_{12}\cdot[O_{ii}]_3\cdot[L_{ii}]_4 - m[F_{ii}]_{12}\cdot[O_{ii}]_3\cdot[K_{ii}]_4\Big\},
\end{aligned}
\tag{B.24}
$$

with

$$\begin{aligned}
A_{ij} &\equiv \rho_i^\dagger (-1)^{n_1+n_2} \slashed{D}_{12} \rho_j\,, & B_{ij} &\equiv \rho_i^\dagger (-1)^{n_1+n_2} \rho_j\,, \\
C_{ij} &\equiv \rho_i^\dagger S_{12}(-1)^{n_2} \slashed{D}_{12} \rho_j\,, & E_{ij} &\equiv \rho_i^\dagger S_{12}(-1)^{n_1} \rho_j\,, \\
F_{ij} &\equiv \rho_i^\dagger S_{12}(-1)^{n_2} \rho_j\,, & J_{ij} &\equiv \phi_i^\dagger D_3 \phi_j\,, \\
O_{ij} &\equiv \phi_i^\dagger S_3 S_3 \phi_j\,, & P_{ij} &\equiv \phi_i^\dagger S_3 S_3 D_3 \phi_j\,, \\
G_{ij} &\equiv \xi_i^\dagger S_4 \xi_j\,, & H_{ij} &\equiv \xi_i^\dagger S_4 D_4 \xi_j\,, \\
I_{ij} &\equiv \xi_i^\dagger S_4 (-1)^{n_4} \xi_j\,, & K_{ij} &\equiv \xi_i^\dagger D_4 \xi_j\,, \\
L_{ij} &\equiv \xi_i^\dagger D_4 D_4 \xi_j\,, & N_{ij} &\equiv \xi_i^\dagger D_4 (-1)^{n_4} \xi_j\,.
\end{aligned} \tag{B.25}$$

## B.2 Chiral Magnetic Effect

In the case of the CME, we need to calculate Eq. (5.1),

$$\left. \frac{\partial J_3}{\partial \mu_5} \right|_{\mu_5=0} = \frac{1}{N_s^3 N_t} \left[ -\frac{1}{4} \text{Tr}\left(\Gamma_3 M^{-1} \Gamma_{45} M^{-1}\right) + \frac{1}{4} \text{Tr}\left(\frac{\partial \Gamma_3}{\partial \mu_5} M^{-1}\right) \right]. \tag{B.26}$$

The derivation is very similar as in the previous section, but instead of repeating the required steps, we are going to exploit a symmetry of the two-point function. If we now consider the usual staggered phases

$$\eta_1(n) = 1, \quad \eta_2(n) = (-1)^{n_1}, \quad \eta_3(n) = (-1)^{n_1+n_2}, \quad \eta_4(n) = (-1)^{n_1+n_2+n_3}\,, \tag{B.27}$$

then it is easy to realize that the calculations for the CME and the CSE are analogous under the exchange of the coordinates $3 \leftrightarrow 4$., i.e. the magnetic field independent ones, and a global minus sign,

$$\left. \frac{\partial J_3}{\partial \mu_5} \right|_{\mu_5=0} = -\left. \frac{\partial J_{35}}{\partial \mu} \right|_{\mu_5=0} (n_3 \leftrightarrow n_4)\,. \tag{B.28}$$

For completeness, we give the explicit expression

$$\begin{aligned}
\frac{\partial J_3}{\partial \mu_5}\bigg|_{\mu_5=0} = &+\frac{1}{N_s^3 N_t}\frac{1}{4}\sum_{i,j}\frac{1}{(\lambda_i^2+m^2)(\lambda_j^2+m^2)}\\
&\Big\{[A_{ij}C_{ji}]_{12}\cdot[G_{ij}G_{ji}]_3\cdot[\delta_{ij}]_4 + [A_{ij}E_{ji}]_{12}\cdot[G_{ij}H_{ji}]_3\cdot[\delta_{ij}]_4\\
&+ [A_{ij}E_{ji}]_{12}\cdot[G_{ij}I_{ji}]_3\cdot[\delta_{ij}J_{ji}]_4 - m[A_{ij}F_{ji}]_{12}\cdot[G_{ij}G_{ji}]_4\cdot[\delta_{ij}]_4\\
&+ [\delta_{ij}C_{ji}]_{12}\cdot[H_{ij}G_{ji}]_3\cdot[\delta_{ij}]_4 + [\delta_{ij}E_{ji}]_{12}\cdot[H_{ij}H_{ji}]_4\cdot[\delta_{ij}]_4\\
&+ [\delta_{ij}E_{ji}]_{12}\cdot[H_{ij}I_{ji}]_3\cdot[\delta_{ij}J_{ji}]_4 - m[\delta_{ij}F_{ji}]_{12}\cdot[H_{ij}G_{ji}]_4\cdot[\delta_{ij}]_4\\
&+ [\delta_{ij}C_{ji}]_{12}\cdot[I_{ij}G_{ji}]_3\cdot[J_{ij}\delta_{ji}]_4 + [\delta_{ij}E_{ji}]_{12}\cdot[I_{ij}H_{ji}]_3\cdot[J_{ij}\delta_{ji}]_4\\
&+ [\delta_{ij}E_{ji}]_{12}\cdot[I_{ij}I_{ji}]_3\cdot[J_{ij}J_{ji}]_4 - m[\delta_{ij}F_{ji}]_{12}\cdot[I_{ij}G_{ji}]_3\cdot[J_{ij}\delta_{ij}]_4\\
&- m[B_{ij}C_{ji}]_{12}\cdot[G_{ij}G_{ji}]_3\cdot[\delta_{ij}]_4 - m[B_{ij}E_{ji}]_{12}\cdot[G_{ij}H_{ji}]_3\cdot[\delta_{ij}]_4\\
&- m[B_{ij}E_{ji}]_{12}\cdot[G_{ij}I_{ji}]_3\cdot[\delta_{ij}J_{ji}]_4 + m^2[B_{ij}F_{ji}]_{12}\cdot[G_{ij}G_{ji}]_3\cdot[\delta_{ij}]_4\Big\}\\
&+\frac{1}{N_s^3 N_t}\frac{1}{4}\sum_i\frac{1}{\lambda_i^2+m^2}\\
&\Big\{[C_{ii}]_{12}\cdot[K_{ii}]_3\cdot[\delta_{ii}]_4 + [E_{ii}]_{12}\cdot[N_{ii}]_3\cdot[J_{ii}]_4\\
&+ [E_{ii}]_{12}\cdot[L_{ii}]_3\cdot[\delta_{ii}]_4 - m[F_{ii}]_{12}\cdot[K_{ii}]_3\cdot[\delta_{ii}]_4\Big\},
\end{aligned} \quad (\text{B.29})$$

with

$$\begin{aligned}
A_{ij} &\equiv \rho_i^\dagger(-1)^{n_1+n_2}\slashed{D}_{12}\rho_j\,, & B_{ij} &\equiv \rho_i^\dagger(-1)^{n_1+n_2}\rho_j\,,\\
C_{ij} &\equiv \rho_i^\dagger S_{12}(-1)^{n_2}\slashed{D}_{12}\rho_j\,, & E_{ij} &\equiv \rho_i^\dagger S_{12}(-1)^{n_1}\rho_j\,,\\
F_{ij} &\equiv \rho_i^\dagger S_{12}(-1)^{n_2}\rho_j\,, & J_{ij} &\equiv \xi_i^\dagger D_4\xi_j\,,\\
G_{ij} &\equiv \phi_i^\dagger S_3\phi_j\,, & H_{ij} &\equiv \phi_i^\dagger S_3 D_3\phi_j\,,\\
I_{ij} &\equiv \phi_i^\dagger S_3(-1)^{n_3}\phi_j\,, & K_{ij} &\equiv \phi_i^\dagger D_3\phi_j\,,\\
L_{ij} &\equiv \phi_i^\dagger D_3 D_3\phi_j\,, & N_{ij} &\equiv \phi_i^\dagger D_3(-1)^{n_3}\phi_j\,.
\end{aligned} \quad (\text{B.30})$$

The expressions for the different cases in Fig. 5.3 follow analogously using the same symmetry.

## B.3 Localized Chiral Magnetic Effect

Finally, we can calculate the observable $H$ from Eq. (6.11),

$$H(n_1, n_1') = \frac{1}{4N_s^2 N_t}\left[-\text{Tr}\big(\mathcal{P}_{n_1}\Gamma_3 M^{-1}\mathcal{P}_{n_1'}\Gamma_{45}M^{-1}\big) + \text{Tr}\left(\mathcal{P}_{n_1}\frac{\delta\Gamma_3}{\delta\mu_5(n_1')}M^{-1}\right)\right]. \quad \text{(B.31)}$$

where we have considered $n_1 = x_1/a$. First, we note that the functional derivative in the tadpole term yields a Kronecker delta $\delta_{n_1 n_1'}$, so this term is diagonal in position space. We can expand the trace, obtaining

$$H(n_1, n_1') = \frac{1}{N_s^2 N_t}\left[-\frac{1}{4}\sum_{i,j}\frac{1}{(\lambda_i^2+m^2)(\lambda_j^2+m^2)}(\Psi_i^{n_1})^\dagger \Gamma_3 M^\dagger \Psi_j \left(\Psi_j^{n_1'}\right)^\dagger \Gamma_{45} M^\dagger \Psi_i \right. \\ \left. + \delta_{n_1 n_1'}\frac{1}{4}\sum_i \frac{1}{\lambda_i^2+m^2}(\Psi_i^{n_1})^\dagger\frac{\partial\Gamma_3}{\partial\mu_5}M^\dagger\Psi_i\right], \quad \text{(B.32)}$$

where we have defined

$$\Psi^x_{\{i_{12},i_3,i_4\}}(n_1,n_2,n_3,n_4) = \rho_{i_{12}}(n_1=x,n_2)\,\phi_{i_3}(n_3)\,\xi_{i_4}(n_4) \\ \equiv \rho^x_{i_{12}}(n_1,n_2)\,\phi_{i_3}(n_3)\,\xi_{i_4}(n_4)\,. \quad \text{(B.33)}$$

Now we see that this observable is completely analogous to Eq. (B.29), with one extra subtlety. In this case, we need to consider that

$$(\rho_i^x)^\dagger \rho_j \neq \delta_{ij}\,. \quad \text{(B.34)}$$

Taking this into account, we can get the formula for the observable by defining

$$\begin{aligned}
A^{n_1}_{ij} &\equiv (\rho_i^{n_1})^\dagger (-1)^{n_1+n_2} \slashed{D}_{12}\rho_j\,, & B^{n_1}_{ij} &\equiv (\rho_i^{n_1})^\dagger (-1)^{n_1+n_2}\rho_j\,, \\
C^{n_1}_{ij} &\equiv (\rho_i^{n_1})^\dagger S_{12}(-1)^{n_2}\slashed{D}_{12}\rho_j\,, & E^{n_1}_{ij} &\equiv (\rho_i^{n_1})^\dagger S_{12}(-1)^{n_1}\rho_j\,, \\
F^{n_1}_{ij} &\equiv (\rho_i^{n_1})^\dagger S_{12}(-1)^{n_2}\rho_j\,, & J_{ij} &\equiv \xi_i^\dagger D_4 \xi_j\,, \\
V^{n_1}_{ij} &\equiv (\rho_i^{n_1})^\dagger \rho_j\,, & H_{ij} &\equiv \phi_i^\dagger S_3 D_3 \phi_j\,, \\
G_{ij} &\equiv \phi_i^\dagger S_3 \phi_j\,, & K_{ij} &\equiv \phi_i^\dagger D_3 \phi_j\,, \\
I_{ij} &\equiv \phi_i^\dagger S_3 (-1)^{n_3}\phi_j\,, & N_{ij} &\equiv \phi_i^\dagger D_3(-1)^{n_3}\phi_j\,, \\
L_{ij} &\equiv \phi_i^\dagger D_3 D_3 \phi_j\,,
\end{aligned} \quad \text{(B.35)}$$

so that

$$\begin{aligned}
H(n_1, n_1') = &+ \frac{1}{N_s^2 N_t} \frac{1}{4} \sum_{i,j} \frac{1}{(\lambda_i^2 + m^2)(\lambda_j^2 + m^2)} \\
&\Big\{ [A_{ij}^{n_1} C_{ji}^{n_1'}]_{12} \cdot [G_{ij} G_{ji}]_3 \cdot [\delta_{ij}]_4 + [A_{ij}^{n_1} E_{ji}^{n_1'}]_{12} \cdot [G_{ij} H_{ji}]_3 \cdot [\delta_{ij}]_4 \\
&+ [A_{ij}^{n_1} E_{ji}^{n_1'}]_{12} \cdot [G_{ij} I_{ji}]_3 \cdot [\delta_{ij} J_{ji}]_4 - m[A_{ij}^{n_1} F_{ji}^{n_1'}]_{12} \cdot [G_{ij} G_{ji}]_4 \cdot [\delta_{ij}]_4 \\
&+ [V_{ij}^{n_1} C_{ji}^{n_1'}]_{12} \cdot [H_{ij} G_{ji}]_3 \cdot [\delta_{ij}]_4 + [V_{ij}^{n_1} E_{ji}^{n_1'}]_{12} \cdot [H_{ij} H_{ji}]_4 \cdot [\delta_{ij}]_4 \\
&+ [V_{ij}^{n_1} E_{ji}^{n_1'}]_{12} \cdot [H_{ij} I_{ji}]_3 \cdot [\delta_{ij} J_{ji}]_4 - m[V_{ij}^{n_1} F_{ji}^{n_1'}]_{12} \cdot [H_{ij} G_{ji}]_4 \cdot [\delta_{ij}]_4 \\
&+ [V_{ij}^{n_1} C_{ji}^{n_1'}]_{12} \cdot [I_{ij} G_{ji}]_3 \cdot [J_{ij} \delta_{ji}]_4 + [V_{ij}^{n_1} E_{ji}^{n_1'}]_{12} \cdot [I_{ij} H_{ji}]_3 \cdot [J_{ij} \delta_{ji}]_4 \\
&+ [V_{ij}^{n_1} E_{ji}^{n_1'}]_{12} \cdot [I_{ij} I_{ji}]_3 \cdot [J_{ij} J_{ji}]_4 - m[V_{ij}^{n_1} F_{ji}^{n_1'}]_{12} \cdot [I_{ij} G_{ji}]_3 \cdot [J_{ij} \delta_{ij}]_4 \\
&- m[B_{ij}^{n_1} C_{ji}^{n_1'}]_{12} \cdot [G_{ij} G_{ji}]_3 \cdot [\delta_{ij}]_4 - m[B_{ij}^{n_1} E_{ji}^{n_1'}]_{12} \cdot [G_{ij} H_{ji}]_3 \cdot [\delta_{ij}]_4 \\
&- m[B_{ij}^{n_1} E_{ji}^{n_1'}]_{12} \cdot [G_{ij} I_{ji}]_3 \cdot [\delta_{ij} J_{ji}]_4 + m^2 [B_{ij}^{n_1} F_{ji}^{n_1'}]_{12} \cdot [G_{ij} G_{ji}]_3 \cdot [\delta_{ij}]_4 \Big\} \\
&+ \frac{1}{N_s^3 N_t} \frac{1}{4} \delta_{n_1 n_1'} \sum_i \frac{1}{\lambda_i^2 + m^2} \\
&\Big\{ [C_{ii}^{n_1}]_{12} \cdot [K_{ii}]_3 \cdot [\delta_{ii}]_4 + [E_{ii}^{n_1}]_{12} \cdot [N_{ii}]_3 \cdot [J_{ii}]_4 \\
&+ [E_{ii}^{n_1}]_{12} \cdot [L_{ii}]_3 \cdot [\delta_{ii}]_4 - m[F_{ii}^{n_1}]_{12} \cdot [K_{ii}]_3 \cdot [\delta_{ii}]_4 \Big\}.
\end{aligned}$$

(B.36)

# *Acknowledgements*

First and foremost, I would like to thank Bastian Brandt and Gergely Endrődi for giving me the opportunity to pursue a PhD under their supervision. They have been an example for me, both as scientists and mentors. I have greatly benefited from their knowledge and expertise, and I am very grateful for their continuous support and dedication, even in the busiest times.

I am also indebted to Gergely Markó, for always being happy to help me understand new concepts and discuss. His scientific rigor and ability to convey physics in a simplified form are an example for me. In addition, I would like to thank Laurin Pannullo for his support and advice, especially during the writing process of this thesis. I wish him a lot of success in his future projects. These years, I have also enjoyed the weekly discussion sessions with Rocco Francesco Basta, Volodymyr Chelnokov and Francesca Cuteri. Moreover, I would like to thank Owe Philipsen for being my external supervisor within the HGS-HIRe program and Wolfgang Unger for accepting to be the second referee for this thesis.

I am very thankful to the secretaries of the theoretical physics department in Bielefeld, Irene Kehler and Susette von Reder, who have made living in Germany much easier with their constant help. The other students and postdocs in the department have been a big part of the reason why I have really enjoyed my years in Bielefeld, an incomplete list of these people includes Federico Rocco, Isabel Oldengott, James Keeble, Jérôme Vandecasteele, Oscar Garcia, Philipp Klose, Pratitee Pattanaik, Pavan, Simran Singh, Sipaz Sharma, Shivani Deshmukh, Travis Dore and Yu Zhang. I am particularly thankful to Oscar and Travis for helping me organize a lot of fun game nights. I am also very grateful to Rasmus Nielsen for being a constant source of support, for going to the gym with me and for (almost) going through this thesis. But two people deserve a special mention, my office mates Javier Hernández and Dean Valois. I cannot finish thanking them for coping with me these years, for always being there for me regardless of the situation and for all the great memories I gathered in our trips. From our office, I have taken two friendships for the rest of my life.

I would like to finish by thanking the most important people in my life. A mi padre Fernando, a mi madre Esther y a mi hermana Marta por su apoyo incondicional todos estos años, sin importar el camino que elija. Gracias a su esfuerzo, dedicación y comprensión, puedo estar feliz de la persona que soy hoy. And to Yunxin, even though it is difficult to put into words how much I owe her, I would like to thank her for helping me maintain balance in my life, for making everything brighter with her presence and for completing me.